\documentclass[preprint,aps,prl,superscriptaddress,longbibliography,nobibnotes]{revtex4-1}
\usepackage{amsthm,amsmath,amsfonts,amssymb,verbatim,color}
\usepackage{graphicx,times}
\usepackage[export]{adjustbox}
\usepackage[caption=false]{subfig}
\usepackage{bm}
\usepackage{epsfig,slashed}
\usepackage[colorlinks=true,citecolor=blue,linkcolor=blue,urlcolor=blue]{hyperref}
\DeclareUnicodeCharacter{2009}{\,}

\begin{document}

\newcommand{\tr}{\mathop{\mathrm{tr}}}
\newcommand{\bsigma}{\boldsymbol{\sigma}}
\newcommand{\re}{\mathop{\mathrm{Re}}}
\newcommand{\im}{\mathop{\mathrm{Im}}}
\renewcommand{\b}[1]{{\boldsymbol{#1}}}
\newcommand{\diag}{\mathrm{diag}}
\newcommand{\sign}{\mathrm{sign}}
\newcommand{\sgn}{\mathop{\mathrm{sgn}}}
\renewcommand{\c}[1]{\mathcal{#1}}
\newcommand{\Jac}{\mathop{\mathrm{Jac}}}
\newcommand{\Mob}{\mathop{\textrm{M\"ob}}}
\newcommand{\Aut}{\mathop{\mathrm{Aut}}}
\newcommand{\g}{\mathfrak{g}}
\renewcommand{\geq}{\geqslant}
\renewcommand{\leq}{\leqslant}

\newcommand{\mb}{\bm}
\newcommand{\ua}{\uparrow}
\newcommand{\da}{\downarrow}
\newcommand{\ra}{\rightarrow}
\newcommand{\la}{\leftarrow}
\newcommand{\mc}{\mathcal}                                                                                                                                                                                                                                                                                                                                                                                                                                                                                                                                                                                
\newcommand{\bs}{\boldsymbol}
\newcommand{\lra}{\leftrightarrow}
\newcommand{\nn}{\nonumber}
\newcommand{\half}{{\textstyle{\frac{1}{2}}}}
\newcommand{\mf}{\mathfrak}
\newcommand{\MF}{\text{MF}}
\newcommand{\IR}{\text{IR}}
\newcommand{\UV}{\text{UV}}

\DeclareGraphicsExtensions{.png}

\title{Hyperbolic band theory}

\author{Joseph Maciejko}
\email[Corresponding author. Email: ]{maciejko@ualberta.ca}
\affiliation{Department of Physics \& Theoretical Physics Institute (TPI), University of Alberta, Edmonton, Alberta T6G 2E1, Canada}
\author{Steven Rayan}
\email[Corresponding author. Email: ]{rayan@math.usask.ca}
\affiliation{Department of Mathematics and Statistics \& Centre for Quantum Topology and Its Applications (quanTA), University of Saskatchewan, Saskatoon, Saskatchewan S7N 5E6, Canada}

\date{\today}

\begin{abstract}
The notions of Bloch wave, crystal momentum, and energy bands are commonly regarded as unique features of crystalline materials with commutative translation symmetries. Motivated by the recent realization of hyperbolic lattices in circuit quantum electrodynamics, we exploit ideas from algebraic geometry to construct the first hyperbolic generalization of Bloch theory, despite the absence of commutative translation symmetries. For a quantum particle propagating in a hyperbolic lattice potential, we construct a continuous family of eigenstates that acquire Bloch-like phase factors under a discrete but noncommutative group of hyperbolic translations, the Fuchsian group of the lattice. A hyperbolic analog of crystal momentum arises as the set of Aharonov-Bohm phases threading the cycles of a higher-genus Riemann surface associated with this group. This crystal momentum lives in a higher-dimensional Brillouin zone torus, the Jacobian of the Riemann surface, over which a discrete set of continuous energy bands can be computed.
\end{abstract}

\maketitle

\section{Introduction}
\label{sec:intro}

The concept of \emph{Bloch wave} is a cornerstone of modern physics. Introduced by Felix Bloch in 1928 to describe the quantum-mechanical propagation of electrons in crystalline solids~\cite{bloch1929}, this phenomenon applies generally to the propagation of waves of any kind in periodic media, including atomic matter waves in optical lattices, light in photonic crystals, and sound in acoustic metamaterials. The key condition for the existence of a Bloch wave is periodicity of the underlying medium --- specifically, that the latter be composed of identical unit cells that are repeated under elementary translations. A Bloch wave traveling through such a medium is not itself a periodic function, but acquires predictable phase shifts under those elementary translations. The phase shifts, in turn, define the crystal momentum $\b{k}$ of the wave and its associated reciprocal space. (While our discussion is applicable to wave phenomena in general, for concreteness we will utilize the language of quantum condensed matter physics and employ a units convention familiar in that field, whereas the reduced Planck's constant $\hbar$ is set to one, and the terms ``momentum'' and ``wave vector'', and ``energy'' and ``frequency'', are used interchangeably.) Because the allowed translations are discrete, the crystal momentum is itself a periodic variable, and an irreducible set of inequivalent crystal momenta is given by the (first) Brillouin zone. This basic fact is the foundation upon which the edifice of band theory is built~\cite{SSP}. Energy levels are organized into energy bands, a discrete set $\{E_n(\b{k})\}$ of continuous functions of $\b{k}$ over the Brillouin zone. In $d$ spatial dimensions, the latter is topologically equivalent to a $d$-dimensional torus. Our focus is on $d=2$ spatial dimensions, where this topological space is an ordinary torus, homeomorphic to the surface of a doughnut. The nontrivial topology of the Brillouin zone, stemming from the periodicity of crystalline lattices, is ultimately responsible for the topological revolution in condensed matter physics, initiated by Haldane's discovery of the Chern insulator~\cite{haldane1988} and firmly established through the development of a comprehensive topological band theory~\cite{chiu2016}.

The absence of periodicity, that is, of a discrete translation symmetry in the system's underlying Hamiltonian, significantly complicates the theoretical study of wave propagation. In a limited number of cases, band theory may still serve as a starting point. Localized or weak deviations from strict periodicity can often be successfully modeled as perturbations of a periodic Hamiltonian, as in standard theories of impurity states or randomized impurity scattering in crystals~\cite{SSP}. Incommensurate modulated phases and quasicrystals, while strongly aperiodic, can be described as projections of ordinary periodic lattices in higher dimensions~\cite{QCbook}. Key aspects of wave propagation in such media, such as the existence of sharp Bragg peaks in x-ray diffraction, can thus be understood via analogous projections of a correspondingly higher-dimensional reciprocal space~\cite{janssen2014}. In spite of these cases, the central tenets of band theory --- crystal momentum, the toroidal Brillouin zone, and sharply-defined energy bands --- are expected to fundamentally break down in generic aperiodic media.

Recently, an example of a new class of synthetic aperiodic structures has been engineered using the technology of circuit quantum electrodynamics~\cite{kollar2019}. The structure is an ordered but aperiodic network of microwave resonators that, from the point of view of wave propagation, can be described effectively as a regular heptagonal tessellation of the hyperbolic plane. Such tessellations, also known as hyperbolic tilings, were studied by Coxeter~\cite{coxeter1957} and popularized through M.~C. Escher's now famous ``Circle Limit'' woodcuts~\cite{coxeter1979}. As with an ordinary two-dimensional crystal such as graphene, whose geometry corresponds to a regular tiling of the Euclidean plane, a hyperbolic tiling consists of repeated unit cells that are all geometrically identical, but allows for patterns impossible in Euclidean space, such as a tiling by regular heptagons~\cite{kollar2019}. That such a tiling is only possible in hyperbolic space follows from the fact that the latter is endowed with a uniform negative curvature. As a result, the sum of the interior angles of an $n$-sided polygon is strictly less than $(n-2)\pi$, and repeated unit cells are identical in the sense of non-Euclidean geometry.  Put somewhat differently, using the phrase \emph{geometrically identical} to describe the unit cells depends crucially upon comparing them under the lens of a particular choice of metric, the \emph{hyperbolic} or \emph{Poincar\'e} metric. As such, hyperbolic tilings are also qualitatively distinct from quasicrystalline ones, which tile the Euclidean plane (albeit aperiodically) and in which the unit cells are identical under the standard Euclidean metric. In the experiments of Ref.~\cite{kollar2019}, negative curvature is simulated by artificially engineering the couplings between the resonators, such that resonators that appear closer together from a Euclidean vantage point --- near the circular edge of Escher's artwork~\cite{coxeter1979}, metaphorically speaking --- are in fact coupled with the same strength as resonators near the center of the device, which appear further apart.

Spurred by this experimental breakthrough, recent theoretical studies have explored the propagation of matter waves on hyperbolic lattices. Using graph theory and numerical diagonalization, Ref.~\cite{kollar2019b} obtained general mathematical results concerning the existence of extended degeneracies and gaps in the spectrum of tight-binding Hamiltonians on a variety of discrete hyperbolic lattices. Ref.~\cite{boettcher2019} developed a hyperbolic analog of the effective-mass approximation in solid-state physics, showing that such tight-binding Hamiltonians reduce in the long-distance limit to the hyperbolic Laplacian --- the Laplace-Beltrami operator associated with the Poincar\'e metric on the hyperbolic plane --- and proposing the synthetic structures of Ref.~\cite{kollar2019} as a new platform for the simulation of quantum field theory in curved space. Topological quantum phenomena in hyperbolic lattices were explored using real-space numerical diagonalization in Ref.~\cite{yu2020}. Notwithstanding these significant advances, quoting Ref.~\cite{kollar2019}: ``no hyperbolic equivalent of Bloch theory currently exists, and there is no known general procedure for calculating band structures in either the nearly-free-electron or tight-binding limits.'' The authors have thus concluded that explicit spectra can only be obtained using numerical diagonalization, ``a brute-force method which yields a list of eigenvectors and eigenvalues, but no classification of eigenstates by a momentum quantum number''~\cite{kollar2019}.

In this work, we present the first hyperbolic generalization of Bloch theory. We show that aperiodic Hamiltonians with the symmetry of a particular class of hyperbolic tilings can be described by such a generalization, which we dub {\it hyperbolic band theory}. Despite the absence of a commutative, discrete translation group, we show that a {\it hyperbolic crystal momentum} $\b{k}$ can be suitably defined, but lives in a vector space of dimension higher than two. There exists a corresponding {\it hyperbolic Brillouin zone} that is topologically equivalent to a higher-dimensional, compact torus.  A {\it hyperbolic bandstructure} $\{E_n(\b{k})\}$, a discrete set of continuous functions of $\b{k}$ on this higher-dimensional Brillouin zone, can be defined and explicitly computed.

The higher-dimensional torus that is the hyperbolic Brillouin zone is related to the tessellation of the two-dimensional hyperbolic plane through a particular construction commonly studied in the field of algebraic geometry in mathematics.  It emerges naturally from our setup that the torus always has even dimension $2g$, where $g$ is the genus, or number of holes, of a compact Riemann surface.  The torus and the Riemann surface are related in a precise mathematical way: the Brillouin zone is exactly the \emph{Jacobian}~\cite{Tata} of the surface.  The Riemann surface is itself a minimal representation of the original configuration space, arising after quotienting the hyperbolic plane by a noncommutative translation group $\Gamma$, called a \emph{Fuchsian group}, which amounts to identifying pairs of edges of a $4g$-sided fundamental cell --- see Fig.~\ref{fig:lattice}(c).

From the point of view of the presence of Riemann surfaces, our hyperbolic band theory is simultaneously a \emph{higher-genus band theory}.  In comparison, ordinary band theory is a genus-one theory, where the now standard two-dimensional torus arises as the quotient of the Euclidean plane by a commutative, discrete group of lattice translations, where the lattice is determined by a $(4\times1)$-sided fundamental cell.  The torus serves both as a minimal representation of the configuration space and as reduced momentum space.  Viewed from algebraic geometry, the real space torus is a genus-one Riemann surface known commonly as an \emph{elliptic curve} and the momentum space torus is the Jacobian of the elliptic curve. Indeed, it is a classical fact from algebraic geometry that an elliptic curve and its Jacobian are isomorphic, not only topologically but also as complex manifolds.  The equivalence between them is given by the Abel-Jacobi map~\cite{Tata}, which can be thought of as a geometric Fourier transform. Because of their identical geometry, one can pass easily back and forth between the two tori, blurring the lines between position space and momentum space when convenient.  In our hyperbolic band theory, the Riemann surface and the Jacobian no longer share the same topology, nor even the same dimension.  Still, the passage between them is given by a higher-dimensional Abel-Jacobi map, which can be approximated numerically as required.  The realization of the role of algebraic geometry in what has been, until now, a squarely topological theory of materials anticipates a plethora of new constructions and algebro-geometric invariants for describing and classifying quantum material structures.

\section{Results and Discussion}

\subsection{Euclidean lattices and Bloch phases}

We begin by reviewing Bloch theory for Euclidean lattices. In the absence of a periodic potential, the propagation of electrons on the two-dimensional (2D) Euclidean plane $\mathbb{E}\cong\mathbb{R}^2$ is described by the usual free-particle Hamiltonian $H_0=\b{p}^2/2m$ where $m$ is the electron mass and $\b{p}=-i\nabla$ is the momentum operator. The continuous translation invariance of $H_0$ is expressed mathematically by the fact that it commutes with the operator $T_\b{a}=e^{-i\b{p}\cdot\b{a}}$ for translations of $\mathbb E$ by an arbitrary vector $\b{a}$. Likewise, its continuous $SO(2)$ symmetry under planar rotations corresponds to the fact that $H_0$ commutes with $e^{-i\theta L_z}$, the rotation operator through angle $\theta$, with $L_z=-i\partial/\partial\theta$ the angular momentum operator. Together, translations and rotations form the special Euclidean group $SE(2)$ of rigid motions of $\mathbb{E}$. In the presence of a periodic potential $V(x,y)$, the Hamiltonian is augmented as $H=H_0+V$ and is now only invariant under a discrete subgroup $G\subset SE(2)$.  We shall take the potential to have the symmetry of a square lattice [Fig.~\ref{fig:lattice}(a)], with the lattice constant set to unity. As Bloch's theorem is a consequence of the periodicity of $H$ exclusively~\cite{SSP}, we will ignore point-group operations and take $G$ to be the abelian group of discrete translations on the lattice. The latter is isomorphic as a group to $\mathbb Z\times\mathbb Z$ and is generated by the unit translations in the $x$ and $y$ directions, respectively. Bloch's theorem states that eigenstates of $H$ enjoy the property $\psi(x+1,y)=e^{ik_x}\psi(x,y)$, $\psi(x,y+1)=e^{ik_y}\psi(x,y)$. Since $k_x$ and $k_y$ appear as phase factors, they are determined up to integer multiples of $2\pi$. It follows that $\b{k}=(k_x,k_y)$ lives in the first Brillouin zone, which is the $2$-torus given by the product of two circles, each of unit radius.

Eigenfunctions of $H$ satisfying the Bloch condition can be explicitly constructed as follows. One solves the Schr\"odinger equation in a reference unit cell $\mathcal{D}$, say $[0,1]\times[0,1]$, with the twisted, periodic boundary conditions $\psi(1,y)=e^{ik_x}\psi(0,y)$, $\psi(x,1)=e^{ik_y}\psi(x,0)$, and identical conditions on $\partial_x\psi$ and $\partial_y\psi$, obtained by taking derivatives of the earlier Bloch condition. Since the unit cell is a compact region and the Hamiltonian is self-adjoint on the space of twice-differentiable, square-integrable functions on $\mathcal D$ with such boundary conditions, one obtains a discrete set of real eigenvalues $E_n(\b{k})$ for $H$ on $\mathcal{D}$. Since $H$ is the same in every unit cell, the corresponding solution on the entire Euclidean plane $\mathbb{E}$ is simply obtained by translating the solution in $\c{D}$ in a manner that respects the Bloch condition. The solution at position $\b{r}=(x,y)$ in a unit cell displaced from $\c{D}$ by the lattice translation $\b{R}=(R_x,R_y)\in\mathbb{Z}^2$ is given in terms of the solution in $\mathcal{D}$ by $\psi(\b{r})=e^{i\b{k}\cdot\b{R}}\psi(\b{r}-\b{R})$. This function obeys the Schr\"odinger equation and the Bloch condition everywhere, and the function and its derivatives are manifestly continuous.

To generalize the ideas at play to the hyperbolic case, it will be useful to reinterpret this manner of constructing Bloch waves for $H$ as follows. In reducing the Schr\"odinger problem on $\mathbb{E}$ to its solution on a single unit cell $\c{D}$, we replace $\mathbb E$ with its quotient by $G$.  This action produces a $2$-torus: $\mathbb{E}/G\cong\mathbb{R}^2/\mathbb{Z}^2\cong T^2$ [Fig.~\ref{fig:lattice}(b)]. The Bloch phase factors $e^{ik_x}$ and $e^{ik_y}$ can then be interpreted as Aharonov-Bohm phases produced by fluxes, $k_x=\oint_{C_x}A$ and $k_y=\oint_{C_y}A$, which thread the two noncontractible cycles $C_x,C_y$ of this torus, where $A$ is a flat connection on the torus. Alternatively, each Bloch phase factor can be viewed as a $U(1)$-representation of the fundamental group of the torus, $\pi_1(T^2)$, which is generated by the homotopy classes $C_x$ and $C_y$ and obeys the presentation $C_xC_yC_x^{-1}C_y^{-1}=1$~\cite{Nakahara}. The representation $\chi(C_{x,y})=\chi(C_{x,y}^{-1})^*=e^{ik_{x,y}}\in U(1)$ manifestly obeys this presentation. Note that $\pi_1(T^2)\cong\mathbb{Z}^2$ is in fact isomorphic to $G$. Thus, we recover the usual point of view according to which the Bloch phase factors form a $U(1)$-representation of the discrete translation group.

What may be overlooked is that, strictly speaking, the construction above involves \emph{two} homeomorphic $2$-tori.  The first, which we denote $\Sigma$, is the one obtained by taking the real configuration space $\mathbb{E}$ and quotienting by the symmetry group of the lattice.  The second, which we shall call the \emph{Jacobian} of $\Sigma$ and denote $\Jac(\Sigma)$, is obtained from collecting the Bloch phase factors into a topological space, which naturally has the topology of a $2$-torus.  In fact, we take as a definition that $\Jac(\Sigma)$ is the set of all representations of $\pi_1(\Sigma)$ into $U(1)$, although classically there are several distinct-appearing yet equivalent ways to define the Jacobian~\cite{Tata}. It is also crucial to observe that these two spaces are not simply topological tori but rather complex manifolds.  The torus $\Sigma$ was constructed from orthogonal unit translations, which correspond to the basis vectors $1$ on the real axis and $i$ on the imaginary one.  Their ratio $\tau=i/1=i$ is the value of a parameter in the complex upper half-plane that determines a particular \emph{elliptic curve}, which is a compact Riemann surface of genus $1$.  The Riemann surface structure is extra geometric information on top of the topological structure of the torus.  At the same time, the choice of $\tau$ determines a particular elliptic curve structure on $\Jac(\Sigma)$, which we can take to be identical to that of $\Sigma$.  In general, there is no canonical choice of identification between an elliptic curve and its Jacobian.  Any such identification depends upon a choice of base point, which is the basepoint for the Abel-Jacobi map, and changes of basepoint are simply translations of the lattice.  Another interesting observation is that both $\Sigma$ and $\Jac(\Sigma)$ are algebraic groups, as elliptic curves come equipped with an abelian group law, the existence of which has tremendous implications for number theory and cryptography (e.g., Ref.~\cite{Silverman}).  On the Jacobian side, the group structure manifests in the addition of crystal momenta, $\b{k}+\b{k}'$, modulo the reciprocal lattice.

\subsection{Hyperbolic lattices and automorphic Bloch phases}

We now turn to hyperbolic lattices. By analogy with the Euclidean case, the Hamiltonian of an electron propagating freely in the 2D hyperbolic plane $\mathbb{H}$ should be invariant under the group of rigid motions of this space, which is isomorphic to the projective special unitary group $PSU(1,1)\cong SU(1,1)/\{\pm I\}$. (The key aspects of hyperbolic geometry we will be needing here, such as the Poincar\'e disk model, the compactification of the hyperbolic octagon, and the hyperbolic Laplacian, are reviewed in a language accessible to scientists in Ref.~\cite{BalazsVoros}.) We will be mostly working with the Poincar\'e disk model of the hyperbolic plane, in which $\mathbb{H}$ corresponds to the interior of the complex unit disk $|z|<1$, with the Poincar\'e metric given by the line element $ds^2=4(1-|z|^2)^{-2}(dx^2+dy^2)$, with $z=x+iy$. The Poincar\'e disk model also underlies the effectively non-Euclidean geometry of the engineered structures in Ref.~\cite{kollar2019}. Elements $\gamma$ of the symmetry group act on a point $z\in\mathbb{H}$ by M\"obius transformations: $z\rightarrow \gamma(z)=(\alpha z+\beta)/(\beta^*z+\alpha^*)$, where $|\alpha|^2-|\beta|^2=1$. Since the Euclidean free-particle Hamiltonian $H_0$ is proportional to the Euclidean Laplacian $\nabla^2$, its natural generalization to the hyperbolic case is $H_0=-\Delta$, where
\begin{align}\label{laplacian}
\Delta=\frac{1}{4}(1-|z|^2)^2\left(\frac{\partial^2}{\partial x^2}+\frac{\partial^2}{\partial y^2}\right),
\end{align}
is the \emph{Laplace-Beltrami operator} on the Poincar\'e disk $\mathbb{H}$, which we will refer to as the ``hyperbolic Laplacian''. One then explicitly checks that $\Delta$ commutes with M\"obius transformations.

\begin{figure}[t]
\centering
\begin{tabular}[c]{cc}
\subfloat{\includegraphics[width=0.37\columnwidth,valign=c]{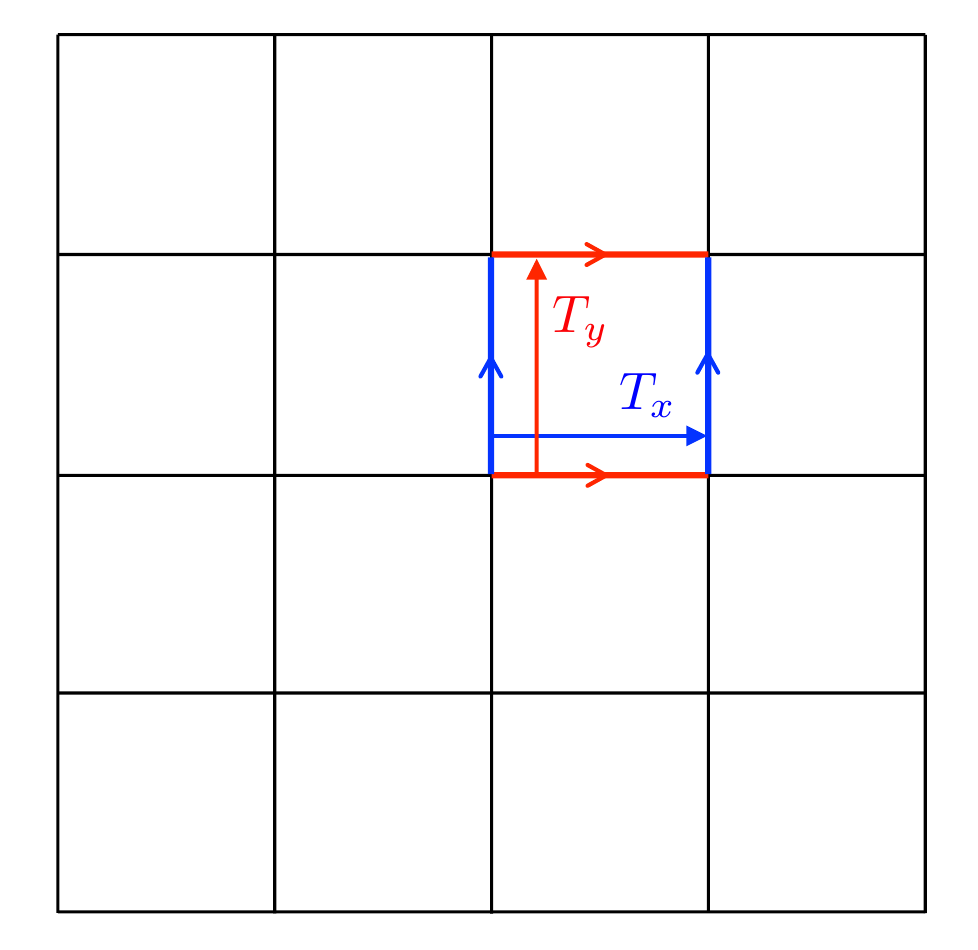}}
&
\subfloat{\includegraphics[width=0.25\columnwidth,valign=c]{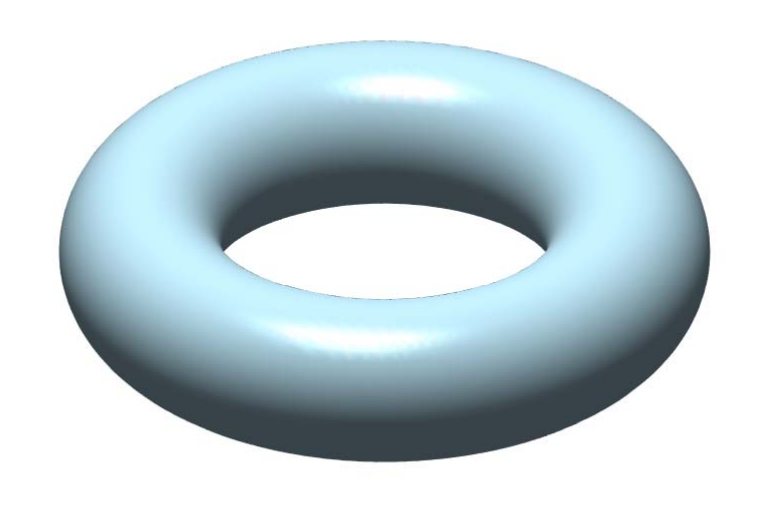}} \\
(a) & (b) \\
\subfloat{\includegraphics[width=0.4\columnwidth,valign=c]{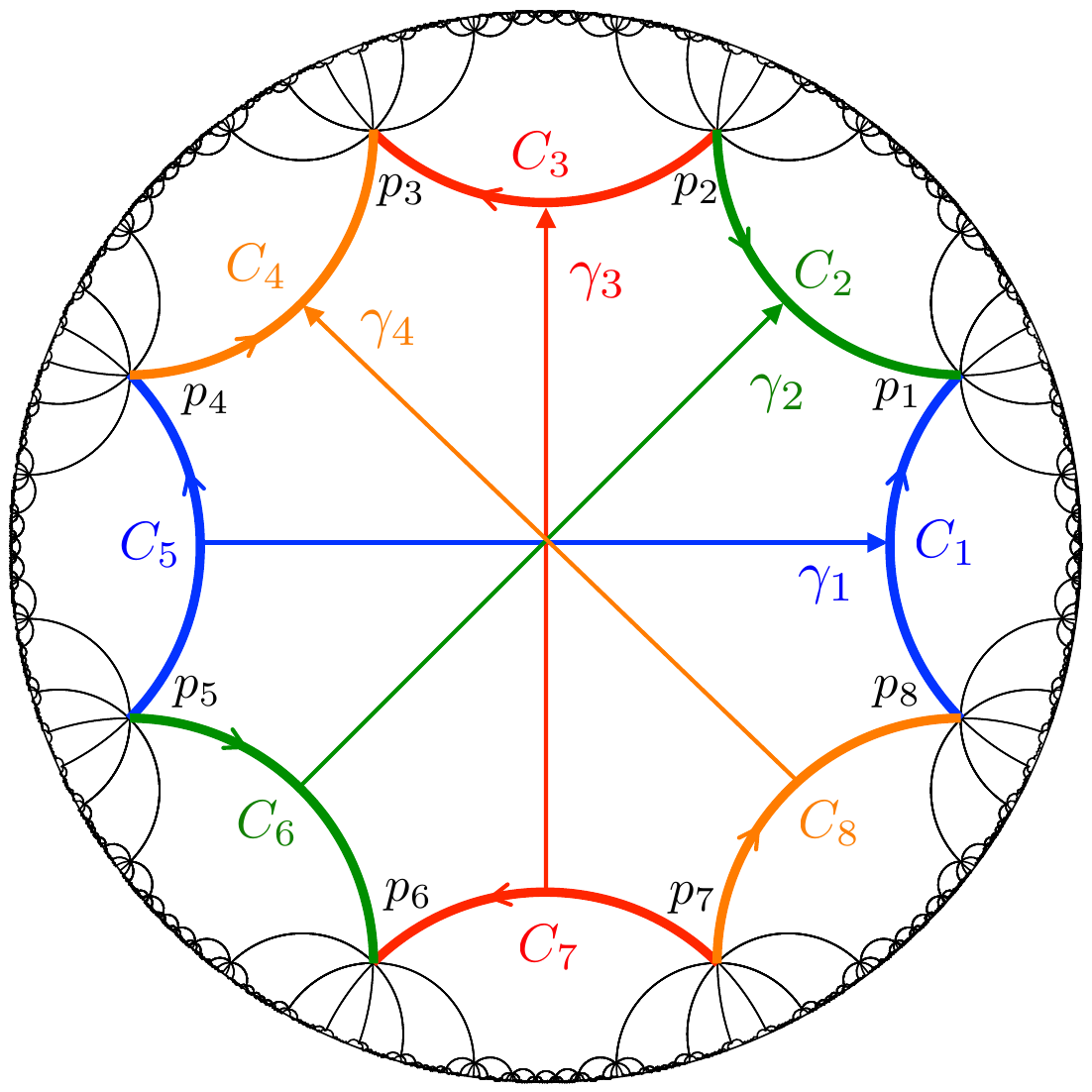}}
&
\subfloat{\includegraphics[width=0.4\columnwidth,valign=c]{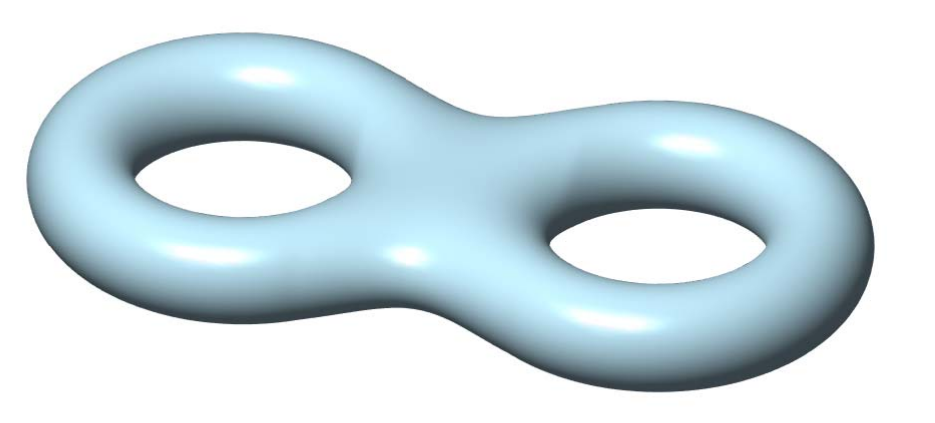}} \\
(c) & (d)
\end{tabular}
 \caption{Euclidean vs hyperbolic lattices. For the Euclidean lattice (a), the unit translations, denoted $T_x$, $T_y$ here, identify pairwise the four sides of the unit cell, which gives the ordinary torus (b). On the hyperbolic lattice (c), Fuchsian group transformations $\gamma_1$, $\gamma_2$, $\gamma_3$, $\gamma_4$ identify pairwise the eight sides of the hyperbolic unit cell, which gives the genus-$2$ surface (d). In both (a) and (c), the identifications preserve the orientations of the sides, which are indicated by arrows.}
  \label{fig:lattice}
\end{figure}

To introduce a hyperbolic lattice, we consider a potential $V(x,y)$ with the symmetry of a $\{4g,4g\}$ hyperbolic tiling with $g\geq 2$. The unit cell of such tilings --- which are impossible in Euclidean space --- is a hyperbolic $4g$-gon, $4g$ of which meet at each vertex of the lattice. We first outline the key steps of our construction for general tilings of this type, and later proceed with detailed calculations for a specific example: the Poincar\'e regular octagonal $\{8,8\}$ tiling ($g=2$) illustrated in Fig.~\ref{fig:lattice}(c).

The full Hamiltonian $H=H_0+V$ is now not invariant under continuous $PSU(1,1)$ transformations, but rather under the discrete Fuchsian subgroup $\Gamma$ determined by the tiling. While nonabelian in general, this group behaves as a hyperbolic analog of a discrete translation group: it acts properly discontinuously on the hyperbolic plane $\mathbb{H}$, meaning that its repeated action on a single fundamental region or reference unit cell $\c{D}$ in $\mathbb{H}$ tiles all of $\mathbb{H}$ with geometrically identical copies of $\c{D}$, with neither gaps nor overlaps~\cite{Katok}. We will focus on the case where $\Gamma$ is co-compact and strictly hyperbolic, in which case the unit cell $\c{D}$ is compact and has finite area under the Poincar\'e metric. For the $\{4g,4g\}$ tiling, $\c{D}$ is a hyperbolic $4g$-gon, meaning a polygon whose $4g$ sides are geodesic segments under the metric.  The uniformization theorem, an important result appearing in algebraic geometry, differential geometry, and number theory, states that the quotient $\mathbb{H}/\Gamma$ is a smooth, compact Riemann surface $\Sigma_g$ of genus $g\geq 2$~\cite{FarkasKra}. Topologically, this surface originates from $2g$ pairwise identifications of the sides of $\c{D}$ under the action of $\Gamma$. Such a surface has $2g$ noncontractible cycles, corresponding to homotopy classes $a_1,b_1,\ldots,a_g,b_g$ through a common basepoint $p_0\in\Sigma_g$ and through which $2g$ Aharonov-Bohm fluxes $k_a^{(1)},k_b^{(1)},\ldots,k_a^{(g)},k_b^{(g)}\in[0,2\pi)$ can be threaded, each of which can again be interpreted as the integral of a flat connection around the corresponding cycle. What also persists is that the $2g$ phase factors $e^{ik_a^{(1)}},e^{ik_b^{(1)}},\ldots,e^{ik_a^{(g)}},e^{ik_b^{(g)}}$ form a $U(1)$-representation $\chi$ of the fundamental group $\pi_1(\Sigma_g)$ of $\Sigma_g$, which is generated by the homotopy classes of $\Sigma_g$, as seen in the relation $a_1^{\phantom{}}b_1^{\phantom{}}a_1^{-1}b_1^{-1}\cdots a_g^{\phantom{}} b_g^{\phantom{}} a_g^{-1} b_g^{-1}=1$ that defines $\pi_1(\Sigma_g)$. We define $\chi$ in terms of the Aharonov-Bohm fluxes by $\chi(a_i)=\chi(a_i^{-1})^*=e^{ik_a^{(i)}}$, $\chi(b_i)=\chi(b_i^{-1})^*=e^{ik_b^{(i)}}$, $i=1,\ldots,g$. In analogy with the Euclidean case, the Fuchsian group $\Gamma$ is in fact isomorphic to $\pi_1(\Sigma_g)$.  Isomorphic subgroups $\Gamma\subset PSU(1,1)$ that generate the same $\{4g,4g\}$ hyperbolic tiling are the analog of distinct choices of basis vectors for the same periodic lattice in the Euclidean case.

In this geometric picture, we again have two complex manifolds, although they are no longer isomorphic --- not as complex manifolds and not even topologically.  One is the Riemann surface $\Sigma_g$, which is a minimal domain for the real configuration space.  As in the Euclidean case, there is a particular complex manifold structure on $\Sigma_g$ that is inherited from the quotient by $\Gamma$ and, hence, from the particular choice of tessellation.  The other manifold is $\Jac(\Sigma_g)$, the Jacobian of $\Sigma_g$, which parametrizes distinct $U(1)$-representations $\chi$ of $\pi_1(\Sigma_g)$.  The manifold $\Sigma_g$ is $2$-dimensional just as in the Euclidean case, although it is no longer homeomorphic to a $2$-torus.  On the other hand, $\Jac(\Sigma_g)$ is $2g$-dimensional and is homeomorphic to the $2g$-torus $T^{2g}=(S^1)^{2g}$.  Yet another difference is that while $\Jac(\Sigma_g)$ remains a group under addition of phases, $\Sigma_g$ does not admit a group law.

From these observations, we propose that despite the absence of an abelian translation group, the choice of a $\{4g,4g\}$ hyperbolic lattice induces naturally a notion of crystal momentum: a $2g$-dimensional hyperbolic crystal  momentum, $\b{k}=\left(k_a^{(1)},k_b^{(1)},\ldots,k_a^{(g)},k_b^{(g)}\right)\in T^{2g}\cong\Jac(\Sigma_g)$.  In other words, we propose that $\Jac(\Sigma_g)$ plays the role of a \emph{hyperbolic Brillouin zone}. By analogy with the Euclidean case described earlier, the notion of hyperbolic crystal momentum can be used to construct eigenfunctions $\psi$ of $H$ starting from a single reference unit cell $\c{D}$. For $z=x+iy$ in the Poincar\'e disk, we generalize the Bloch condition to
\begin{align}\label{automorphic}
\psi(\gamma(z))=\chi(\gamma)\psi(z),
\end{align}
where $\gamma\in\Gamma$ acts by M\"obius transformations and where $\chi$ is the map discussed earlier.  Appearing as early as works of Poincar\'e, functions obeying the condition (\ref{automorphic}) are known as \emph{automorphic functions} with \emph{factor of automorphy} $\chi$~\cite{Gunning,Venkov}, and can be seen as hyperbolic analogs of periodic functions. (We use a convention standard in the number-theory literature; from the point of view of representation theory, a more natural but physically equivalent convention is $\psi(\gamma^{-1}(z))=\chi(\gamma)\psi(z)$, which simply amounts to replacing $\chi(\gamma)$ by $\chi(\gamma)^{-1}$.) The factor of automorphy here is the simplest possible type --- that of weight $0$, also known as a \emph{multiplier system}.  More generally, one may consider factors of automorphy that depend holomorphically on $z$ --- that is, weight-$k$ factors of automorphy $\hat\chi(\gamma,z)=\chi(\gamma)(cz+d)^k$ for some real numbers $c$ and $d$, and where $\chi:\Gamma\to U(1)$ and $z\in\mathbb H$.  We consider only unitary automorphic factors in our Bloch condition, as a direct generalization of the Euclidean situation.

By assumption, the potential $V$ itself is an automorphic function with trivial automorphy factor, $V(\gamma(z))=V(z)$.  Accordingly, we shall refer to such a potential as an \emph{automorphic potential}.  Again, we solve the Schr\"odinger equation
\begin{align}\label{schrodinger}
(-\Delta+V)\psi = E\psi,
\end{align}
on the single reference unit cell $\c{D}$, with the boundary conditions specified by Eq.~(\ref{automorphic}). By analogy with the Euclidean or genus-$1$ case, there are now $2g$ linearly independent boundary conditions to apply, corresponding to the $2g$ generators of $\pi_1(\Sigma_g)$. In practice, one requires an explicit representation of those generators as $PSU(1,1)$ matrices. The potential $V$ does not involve derivatives and is thus trivially self-adjoint. With the boundary conditions (\ref{automorphic}), the hyperbolic Laplacian $\Delta$ can be shown to be self-adjoint on $\c{D}$ as well~\cite{HejhalVol1}. Since the region is compact, we obtain a discrete set of real eigenvalues $\{E_n(\b{k})\}$ for each value of the hyperbolic crystal momentum $\b{k}$. Since $H$ is the same in every unit cell, i.e., it is invariant under the action of $\Gamma$, the solution on $\mathcal{D}$ can be extended to the entire Poincar\'e disk $\mathbb{H}$ by $\Gamma$-translating it in a manner that respects the generalized Bloch condition (\ref{automorphic}). In other words, the solution in any fundamental domain $\c{D}'\subset\mathbb{H}$, that is necessarily the image of $\c{D}$ under the action of a particular element $\gamma\in\Gamma$, is given by $\psi(z)=\chi(\gamma)\psi(\gamma^{-1}(z))$, where $z\in\mathcal{D}'$ and $\gamma^{-1}(z)\in\mathcal{D}$.  This construction ensures that, as in the Euclidean case, $\psi$ obeys the Schr\"odinger equation (\ref{schrodinger}) and the generalized Bloch condition (\ref{automorphic}) everywhere, and $\psi$ and its derivatives are continuous in the entire Poincar\'e disk.

With these observations in hand, we have the desired identifications: $\Jac(\Sigma_g)$ is indeed our hyperbolic momentum space and we may describe each factor of automorphy $\chi$ as a \emph{hyperbolic Bloch phase}.

\subsection{Particle-wave duality and the Abel-Jacobi map}

The geometry emerging from our construction is a pair of complex manifolds, $\Sigma_g$ and $\Jac(\Sigma_g)$.  In the Euclidean case, these manifolds manifest as a pair of essentially indistinguishable elliptic curves.  One can ask whether we retain a direct passage from one to the other in the hyperbolic, or $g\geq2$, case.

To this end, we shall very briefly review the complex manifold structure of the Riemann surface $\Sigma_g$.  The surface $\Sigma_g$ is covered by overlapping open patches $U$, each of which is homeomorphic under a map $\psi_U$ to an open set of the complex plane $\mathbb C$.  This allows us to assign to a point $p\in U$ a coordinate, which is the corresponding complex number $\psi_U(p)$.  When $p$ is in the intersection of two open patches $U$ and $V$, we can translate from one coordinate system to another via $\psi_V\circ\psi_U^{-1}$.  The fact that $\Sigma_g$ is a Riemann surface implies that there exists such an atlas of coordinate charts in which each and every composition $\psi_V\circ\psi_U^{-1}$ is holomorphic as a map between two open sets in $\mathbb C$, meaning that the composite functions satisfy the standard Cauchy-Riemann equations of complex analysis.  (The complex manifold structure of Fuchsian quotients such as $\Sigma_g$ is further discussed in \cite{GK}.)

The fact that $\Sigma_g$ looks locally like an open set in $\mathbb C$ means that it is also possible to discuss complex-valued holomorphic functions $f$ on $\Sigma_g$; locally, they are holomorphic functions from $U$ to $\mathbb C$.  Moreover, we have an associated notion of holomorphic one-form: these are the one-forms on $\Sigma_g$ that can be written locally as $\theta=fdz$, where $f$ is a holomorphic function on $\Sigma_g$.

A well-known result in algebraic geometry that follows from the Riemann-Roch theorem and Serre duality (e.g.,~\cite{Forster,Miranda}) is that the global holomorphic one-forms on $\Sigma_g$ constitute a vector space of complex dimension $g$. In other words, there are $g$-many global, linearly independent, holomorphic one-forms $\theta_1,\dots,\theta_g$ on the Riemann surface.  This is an algebraic interpretation of the genus that complements the topological one: rather than counting the number of holes, we think of $g$ as counting the number of independent one-forms --- a fact consistent with the reality that there is no global holomorphic one-form on the Riemann sphere other than $\theta=0$.  

Now, recall that we chose $2g$ cycles with a common basepoint $p_0$ via which we defined Aharonov-Bohm fluxes, leading to $U(1)$-representations $\chi$ of $\pi_1(\Sigma_g)$.  These cycles provide a basis for the first homology group of the surface.  At this point, we will replace these cycles with a \emph{symplectic basis}, which is a collection of loops $a_i,b_i$, $i=1,\dots,g$, such that $a_i$ and $b_i$ intersect in exactly one point and all other intersections are empty.  At the same time, we choose a basis $\theta_1,\dots,\theta_g$ of holomorphic one-forms in such a way that they are ``dual'' to the $a$ loops, meaning that $\oint_{a_i}\theta_j\;=\;\delta_{ij}$. The remaining integrals, which form $g$-many $g$-tuples $\left(\oint_{b_1}\theta_j,\dots,\oint_{b_g}\theta_j\right)$, produce a nondegenerate $g\times g$ matrix $\Omega$, the \emph{period matrix} of $\Sigma_g$.  The full rank of $\Omega$ follows from the Riemann bilinear relations~\cite{Tata}.  As such, the columns are a basis for $\mathbb C^g$, giving us a lattice structure on the underlying $\mathbb R^{2g}$, known as the \emph{period lattice}.  Let us denote this lattice by $\Lambda$.  The quotient $\mathbb R^{2g}/\Lambda$ is precisely $\Jac(\Sigma_g)$.  The matrix $\Omega$ can be shown to be always symmetric with positive-definite imaginary part.  The space of all such matrices is called the \emph{Siegel upper half-space}.  Note that in the $g=1$ or elliptic curve case, the period matrix is $1\times 1$ --- it is precisely the modular parameter $\tau$ in the upper half-plane.

Now, let $p$ be any point in $\Sigma_g$ and let $c_p$ be a continuous path from $p_0$ to $p$, where $p_0$ is a basepoint (not necessarily the one we chose earlier).  We can define a map $a:\Sigma_g\to\Jac(\Sigma_g)$ by setting
\begin{align}\label{AJMap}
a(p)=\left(\int_{c_p}\theta_1,\dots,\int_{c_p}\theta_g\right)\,\mbox{mod}\,\Lambda.
\end{align}
Here, the integral yields a vector in $\mathbb C^g\cong\mathbb R^{2g}$.   We then translate the output to the fundamental unit cell in $\mathbb C^g\cong\mathbb R^{2g}$ of the lattice $\Lambda$, thus producing a point in $\Jac(\Sigma_g)$ --- equivalently, a crystal momentum $\b k$.  It is readily apparent that the map is independent of both the specific basepoint, as well as the chosen path to $p$.  Changing the path perturbs the calculation by an integral over a cycle, which can be written in the basis $(a_i,b_i)$, and so the difference that we pick up is precisely an element of $\Lambda$.  This difference is killed by the quotient. Changing the basepoint simply translates the torus. Finally, when we take $p=p_0$, we are integrating only over cycles, which again are killed by the quotient, and so $a(p_0)=\b k=0$ is the identity in the Jacobian as a group.

The map defined here is the \emph{Abel-Jacobi map}.  As it maps a $1$-dimensional space (over $\mathbb C$) to a $g$-dimensional space, it is only an isomorphism in the genus-$1$ case, where it provides the familiar particle-wave duality of Euclidean quantum mechanics and, hence, conventional band theory.  Intuitively, the line integrals in Eq.~(\ref{AJMap}) can be interpreted as the set of topologically distinct contributions to the geometric phase accumulated under adiabatic motion of a quantum particle from a reference point $p_0$ to a given point $p$ inside the unit cell. In the Euclidean case, the unit cell is geometrically flat, and the two contributions to the geometric phase are linear functions of two linearly independent displacements, producing an isomorphism between real and momentum spaces. Apart from the obvious dimensional differences inherent in the hyperbolic case, the nontrivial negative curvature of the unit cell required by the Gauss-Bonnet theorem renders such a linear mapping impossible.

To counter the difference in dimension between the configuration and momentum spaces for $g\geq2$, we can ask about the effect of applying the map in Eq.~(\ref{AJMap}) to $g$-tuples of points from $\Sigma_g$, by defining $a(p_1,\dots,p_g)=\sum_{i=1}^ga(p_i)$. One immediate observation is that the order of the $g$ inputs has no effect on the output, and so the map is well-defined on the symmetric product of the Riemann surface with itself $g$ times (rather than simply the Cartesian product). It is a classical fact from algebraic geometry, e.g.,~\cite{Macdonald}, that this map $a$ from the $g$-fold symmetric product of $\Sigma_g$ to $\Jac(\Sigma_g)$ is \emph{almost} an isomorphism of complex manifolds.  The map is only \emph{birational}, which means that a certain submanifold of the symmetric product must be ``blown down'' in order to recover $\Jac(\Sigma_g)$.  This submanifold is an example of a ``high-symmetry'' region, related to the so-called \emph{theta divisor} in $\Jac(\Sigma_g)$~\cite{Tata}.  While involving some technical aspects of algebraic geometry, this construction exhibits the Jacobian as a particular complex manifold arising from the data of our Riemann surface in a direct way, providing an algebraic particle-wave duality that exists in spite of dimensional and curvature differences.

The aforementioned high-symmetry region is worthy of further investigation, as it suggests the existence of a special set of points in the hyperbolic unit cell whose physical relevance is not yet appreciated. The map further suggests that an ideal Liouville-Arnol'd-type phase space for this physical system in which the configuration space and momentum space have equal dimension might be given by a fibration of Jacobian tori over a $g$-dimensional complex space associated to $\Sigma_g$.  We leave the formalization of this dynamical system to forthcoming work.  In the meantime, we now proceed with a concrete example of our hyperbolic band theory in genus $g=2$.

\subsection{The Bolza lattice}
\label{sec:bolza}

Having outlined the key ideas of our general theory, we now apply it to the simplest hyperbolic analog of the Euclidean square lattice: the regular octagonal $\{8,8\}$ tiling depicted in Fig.~\ref{fig:lattice}(c). This tiling is generated by the action of a Fuchsian group $\Gamma$ on a reference unit cell $\c{D}$, which can be taken to be the regular hyperbolic octagon centered at the origin $z=0$ of the Poincar\'e disk, that is, the region bounded by the colored geodesic segments $C_1,\ldots,C_8$ in Fig.~\ref{fig:lattice}(c); see also the Supplementary Materials. We define the hyperbolic Bloch factor $\chi(\gamma)$ in Eq.~(\ref{automorphic}) by its action on the Fuchsian group generators $\gamma_j$: $\chi(\gamma_j)=\chi(\gamma_j^{-1})^*=e^{ik_j}$, $j=1,\ldots,4$, writing the hyperbolic crystal momentum as $\b{k}=(k_1,k_2,k_3,k_4)\in\Jac(\Sigma_2)$. Indeed, in this case, the underlying topology of the hyperbolic Brillouin zone $\Jac(\Sigma_2)$ is a four-dimensional torus $T^4$. Since we require $\chi$ to be a representation of $\Gamma$, we further define $\chi(\gamma_i\gamma_j)=\chi(\gamma_i)\chi(\gamma_j)$, for any $i,j=1,\ldots,4$.  Since $\Gamma$ is finitely generated by the $\gamma_j$, this is sufficient to define $\chi(\gamma)$ for any $\gamma\in\Gamma$. Combining the definition of the hyperbolic Bloch factor with the automorphic Bloch condition (\ref{automorphic}) and the pairwise identifications imposed by $\Gamma$, the four boundary conditions we impose when solving the hyperbolic Bloch problem (\ref{schrodinger}) on the hyperbolic octagon $\c{D}$ become: $\psi(C_j)=e^{ik_j}\psi(C_{j+4})$, $j=1,\ldots,4$.

In ordinary band theory, the simplest problem that illustrates many salient features of generic bandstructures, including zone folding and symmetry-protected or accidental degeneracies, is the {\it empty-lattice approximation}~\cite{SSP}. In this approximation, the potential is taken to be constant, and thus necessarily periodic; without loss of generality, one can further choose $V=0$. As we then have $H=H_0=-\nabla^2/2m$, the problem thus reduces to finding the eigenvalues and eigenfunctions of the Euclidean Laplacian with Bloch (twisted) boundary conditions. One easily finds $E_\b{n}(\b{k})=\frac{1}{2m}(\b{k}+2\pi\b{n})^2$ and $\psi_{\b{n}\b{k}}(\b{r})\propto e^{i(\b{k}+2\pi\b{n})\cdot\b{r}}$, with $\b{k}\in T^2$ as the crystal momentum and $\b{n}=(n_x,n_y)\in\mathbb{Z}^2$ as a discrete band index.

\begin{figure}[t]
\centering
\begin{tabular}[c]{cc}
\subfloat{\includegraphics[width=0.37\columnwidth,valign=c]{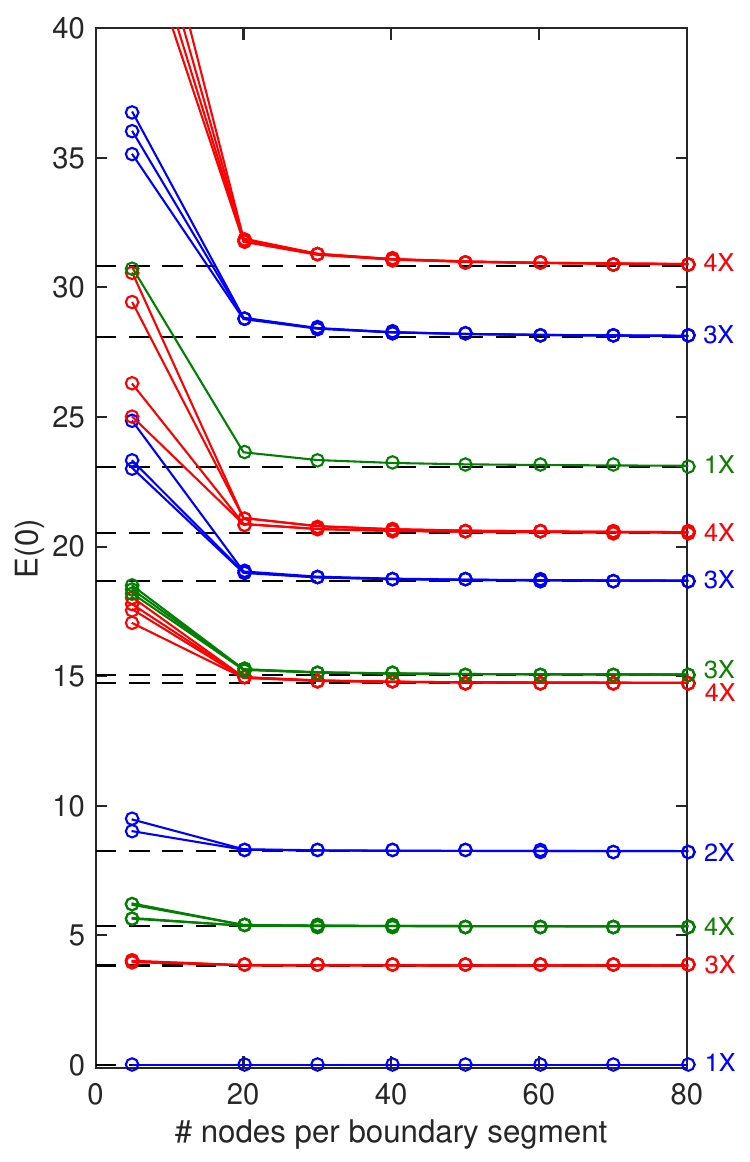}}
&
\subfloat{\includegraphics[width=0.54\columnwidth,valign=c]{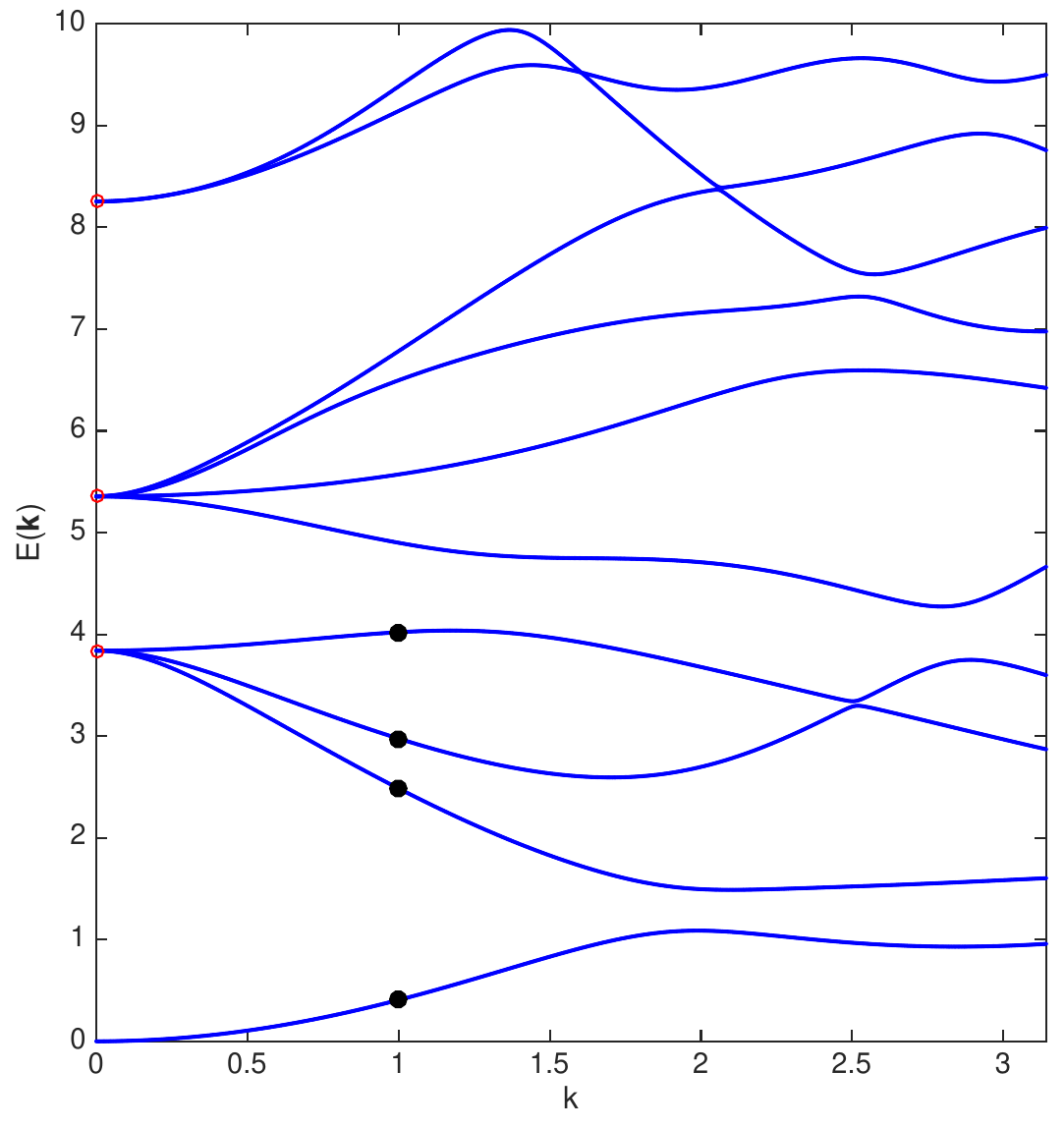}} \\
(a) & (b) 
\end{tabular}
 \caption{Hyperbolic bandstructure of the Bolza lattice in the empty-lattice approximation. (a) $\b{k}=0$ eigenenergies computed using the finite element method (colored plots) vs eigenvalues of the Laplacian on the Bolza surface taken from Ref.~\cite{strohmaier2013} (dashed lines); only the lowest eleven distinct eigenvalues are shown (degeneracies from Ref.~\cite{strohmaier2013} shown on the right). The total number of mesh nodes grows approximately quadratically with the number of boundary nodes (see Supplementary Materials). (b) Hyperbolic bandstructure plotted along a generic direction in the hyperbolic Brillouin zone: $\b{k}=(k_1,k_2,k_3,k_4)=(0.8,0.3,1.2,1.7)k$. Red circles: eigenvalues of the Laplacian on the Bolza surface taken from Ref.~\cite{strohmaier2013}; black dots: eigenstates whose probabilities densities are plotted in Fig.~\ref{fig:wavef}(e-h).}
  \label{fig:band1}
\end{figure}

In the hyperbolic case, we wish to find the eigenvalues $E$ and eigenfunctions $\psi$ of the hyperbolic Laplacian $-\Delta$ on the hyperbolic octagon $\c{D}$ with the twisted boundary conditions mentioned earlier. At the origin of the hyperbolic Brillouin zone, $\b{k}=0$, those boundary conditions reduce to the condition that the solutions be strictly automorphic, $\psi(\gamma(z))=\psi(z)$, the case usually considered in mathematics~\cite{Buser}.

While exact analytical solutions for the eigenfunctions and eigenenergies are unavailable, this problem can be studied numerically. Motivated by questions in the theory of quantum chaos, approximate eigenenergies and eigenfunctions of the hyperbolic Laplacian on hyperbolic octagons with strictly automorphic boundary conditions were first obtained in Ref.~\cite{aurich1989} using the finite element method. Subsequent work studied this problem using the boundary element method~\cite{aurich1993}, quantization via the Selberg trace formula~\cite{ninnemann1995}, time-dependent methods~\cite{bachelot-motet2010}, and an algorithm based on the method of particular solutions~\cite{strohmaier2013}. In accordance with our previous expectations, the spectrum $\{E_n(0)\}$ of $-\Delta$ with strictly automorphic boundary conditions is indeed found to be real and discrete. For the Bolza surface of interest to us, the lowest eigenvalue is $E_0(0)=0$, corresponding to a constant eigenfunction over $\c{D}$, and the next three eigenvalues are given approximately by $E_1(0)\approx 3.839$, $E_2(0)\approx 5.354$, and $E_3(0)\approx 14.726$~\cite{strohmaier2013}.

Here, we study the general case $\b{k}\neq 0$ for $\Sigma_2$ using the finite element method. We use a freely available software package, FreeFEM++~\cite{hecht2012}, which was used successfully to study the spectrum of the Bolza surface with strictly automorphic boundary conditions~\cite{CookThesis}. Our implementation of the twisted boundary conditions is discussed in the Supplementary Materials. As a check on our calculations, we first compute the spectrum $\{E_n(0)\}$, i.e., with strictly automorphic boundary conditions [colored plots in Fig.~\ref{fig:band1}(a)]. With increased refinement of the finite element mesh, the $\b{k}=0$ spectrum gradually converges to previously obtained results~\cite{strohmaier2013}. In particular, the degeneracies found in Ref.~\cite{strohmaier2013,CookThesis} are correctly reproduced with a sufficiently fine mesh. Such degeneracies are a consequence of the large symmetry group (automorphism group) of the Bolza surface~\cite{CookThesis} which, as will be seen later, can be thought of as the hyperbolic analog of a point group. We use a mesh with 70 nodes per boundary segment in all remaining plots, which achieves satisfactory accuracy at reasonable computational cost. Since the spectrum is unbounded, we only compute a small number of low-lying eigenvalues using standard numerical linear algebra techniques.

A well-known result in conventional band theory is that, ignoring spin degrees of freedom, degeneracies at high-symmetry points fully split as one moves away from such points along a generic direction in reciprocal space~\cite{SSP}.  An example of a high-symmetry point is the origin $\b{k}=0$ of the Brillouin zone --- equivalently, the group identity in $\Jac(\Sigma_2)$.  To ascertain whether this behavior holds in the hyperbolic case, we compute the hyperbolic bandstructure for $\b{k}\neq 0$ along a generic direction in the hyperbolic Brillouin zone [Fig.~\ref{fig:band1}(b)]. The lowest eigenvalue $E_0(0)=0$ is nondegenerate at $\b{k}=0$ and thus does not split. As in the Euclidean case, the energy $E_0(\b{k})$ of the lowest band increases with the magnitude of $\b{k}$ at small $\b{k}$, in accordance with the intuitive expectation that (kinetic) energy increases with crystal momentum in the long-wavelength limit. The next three eigenvalues $E_1(0)$, $E_2(0)$, $E_3(0)$ are three-, four-, and two-fold degenerate, respectively~\cite{strohmaier2013,CookThesis}, but this degeneracy is completely lifted as $\b{k}$ moves away from zero, as in conventional band theory. We also observe linear crossings between some of the bands emanating from $E_2(0)$ and $E_3(0)$. According to the von Neumann-Wigner theorem~\cite{vonneumann1929}, only codimension-3 level crossings are perturbatively stable in the absence of symmetries other than translational. Thus by contrast with 2D (or 3D) Euclidean lattices, we expect generically stable nodal-line crossings~\cite{fang2016} in the hyperbolic bandstructures of $\{8,8\}$ tessellations, and for general $\{4g,4g\}$ tessellations, stable crossings forming $(2g-3)$-dimensional submanifolds of $\Jac(\Sigma_g)$.

In algebraic geometry, the points of degeneracy are known as \emph{ramification points} while the splitting-off of eigensheets is known as \emph{branching}.  From this point of view, the total energy manifold $E_n(\b k)$, for all $n$ and for all $\b k$, is a \emph{branched cover} of $\Jac(\Sigma_2)$, although not one of finite type, as there are countably- but not finitely-many levels $n$.  Finite-type branched covers arising from eigenvalues of finite-rank linear operators, known as \emph{spectral covers}, are studied frequently in algebraic geometry, especially in connection with gauge theories, integrable systems, and high-energy physics, e.g., Ref.~\cite{Donagi}.

\begin{figure}[t]
\centering
\begin{tabular}[c]{cccc}
\subfloat{\includegraphics[width=0.235\columnwidth,valign=c]{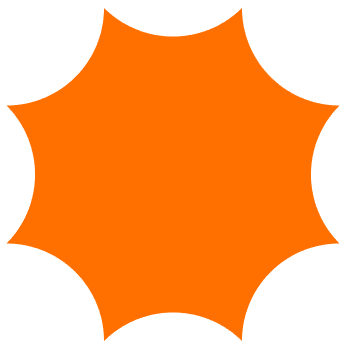}}
&
\subfloat{\includegraphics[width=0.235\columnwidth,valign=c]{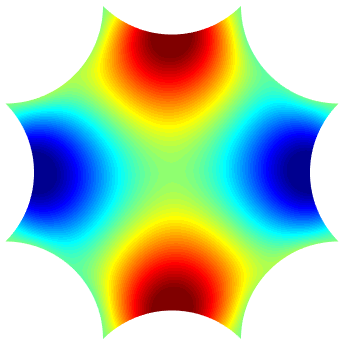}} 
&
\subfloat{\includegraphics[width=0.235\columnwidth,valign=c]{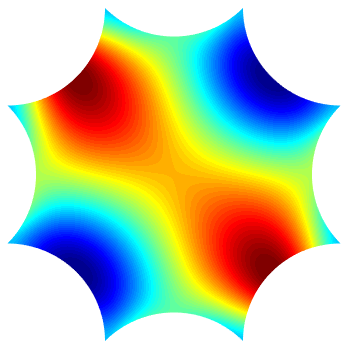}}
&
\subfloat{\includegraphics[width=0.295\columnwidth,valign=c]{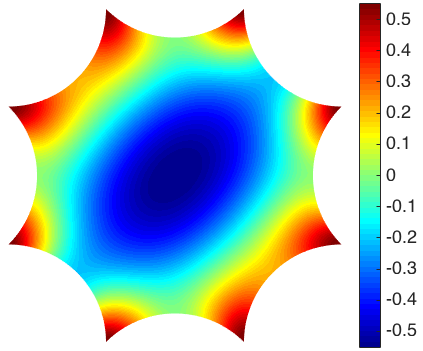}}  
\\
(a) & (b) & (c) & (d)\\
\subfloat{\includegraphics[width=0.235\columnwidth,valign=c]{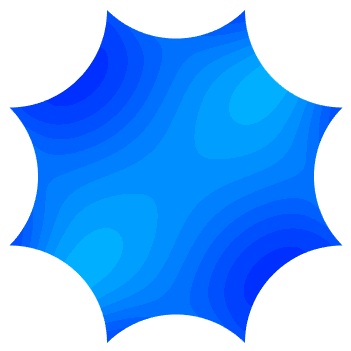}}
&
\subfloat{\includegraphics[width=0.235\columnwidth,valign=c]{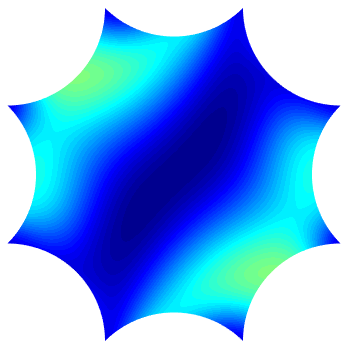}} 
&
\subfloat{\includegraphics[width=0.235\columnwidth,valign=c]{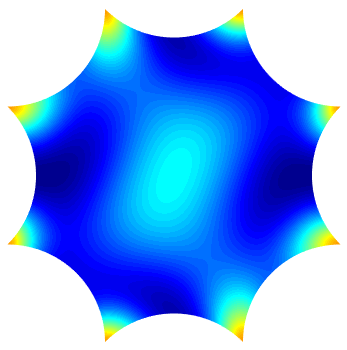}}
&
\subfloat{\includegraphics[width=0.295\columnwidth,valign=c]{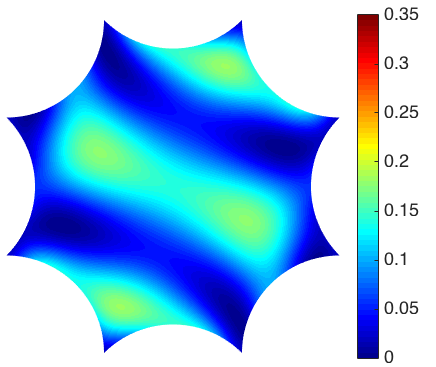}}  
\\
(e) & (f) & (g) & (h)
\end{tabular}
 \caption{Hyperbolic Bloch eigenstates in the empty-lattice approximation. Wavefunction $\psi_\b{k}(z)$ for the (a) ground state and (b-d) degenerate first excited states at $\b{k}=0$; modulus squared $|\psi_\b{k}(z)|^2$ for the (e) ground and (f-h) first three excited states corresponding to the black dots in Fig.~\ref{fig:band1}(b), i.e., at $\b{k}=(0.8,0.3,1.2,1.7)$, in order of increasing eigenenergy.}
  \label{fig:wavef}
\end{figure}

Our finite element calculation also gives us access to the detailed spatial profile of the hyperbolic Bloch wavefunctions $\psi_{n\b{k}}(z)$. Since these wavefunctions obey the automorphic Bloch condition (\ref{automorphic}) by construction, it is sufficient to plot them for $z$ in the central hyperbolic octagon $\c{D}$ [Fig.~\ref{fig:wavef}]. At $\b{k}=0$, the wavefunctions are purely real. The ground state [Fig.~\ref{fig:wavef}(a)] is nodeless and perfectly uniform, while the excited states [Fig.~\ref{fig:wavef}(b-d)] acquire nodes. For $\b{k}\neq 0$, the wavefunctions are in general complex, as in the Euclidean case. The probability densities for ground and excited states [Fig.~\ref{fig:wavef}(e-h)] are modulated by the $\b{k}$ vector with respect to their $\b{k}=0$ counterparts. (Note however that the three excited states in Fig.~\ref{fig:wavef}(b-d) are degenerate, and only represent one possible basis of the degenerate subspace, which is split at $\b{k}\neq 0$; thus one cannot directly match Fig.~\ref{fig:wavef}(b-d) and Fig.~\ref{fig:wavef}(f-h).)

\subsection{A particle in an automorphic potential}
\label{sec:sec:automorphicV}

We now consider turning on a nonzero automorphic potential $V$. Such a potential can be constructed by summing over all $\Gamma$-translates of a localized potential $U(z)$,
\begin{align}\label{AutomV}
V(z)=\sum_{\gamma\in\Gamma}U(\gamma(z)),
\end{align}
which is a kind of generalized theta series~\cite{Tata}. To ensure this series converges everywhere, we choose $U(z)$ with compact support in $\c{D}$, for instance, a circular well of radius $R$ and depth $V_0$. Since the full Hamiltonian $H=-\Delta+V$ is invariant under $\Gamma$-translations, it is sufficient to solve the Schr\"odinger equation (\ref{schrodinger}) with the automorphic Bloch boundary conditions (\ref{automorphic}) on $\c{D}$.

\begin{figure}[t]
\centering
\begin{tabular}[c]{cc}
\subfloat{\includegraphics[width=0.5\columnwidth,valign=c]{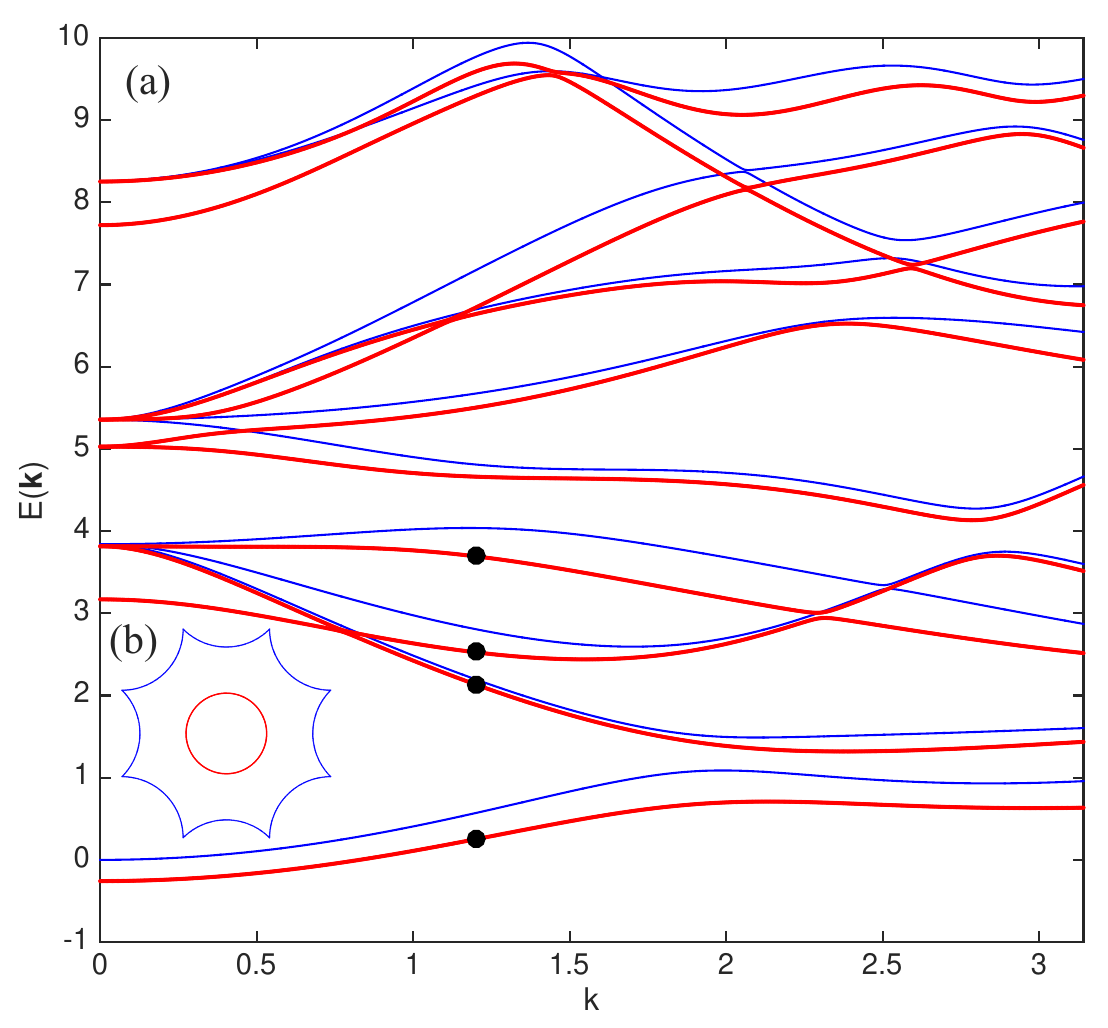}}
&
\subfloat{\includegraphics[width=0.5\columnwidth,valign=c]{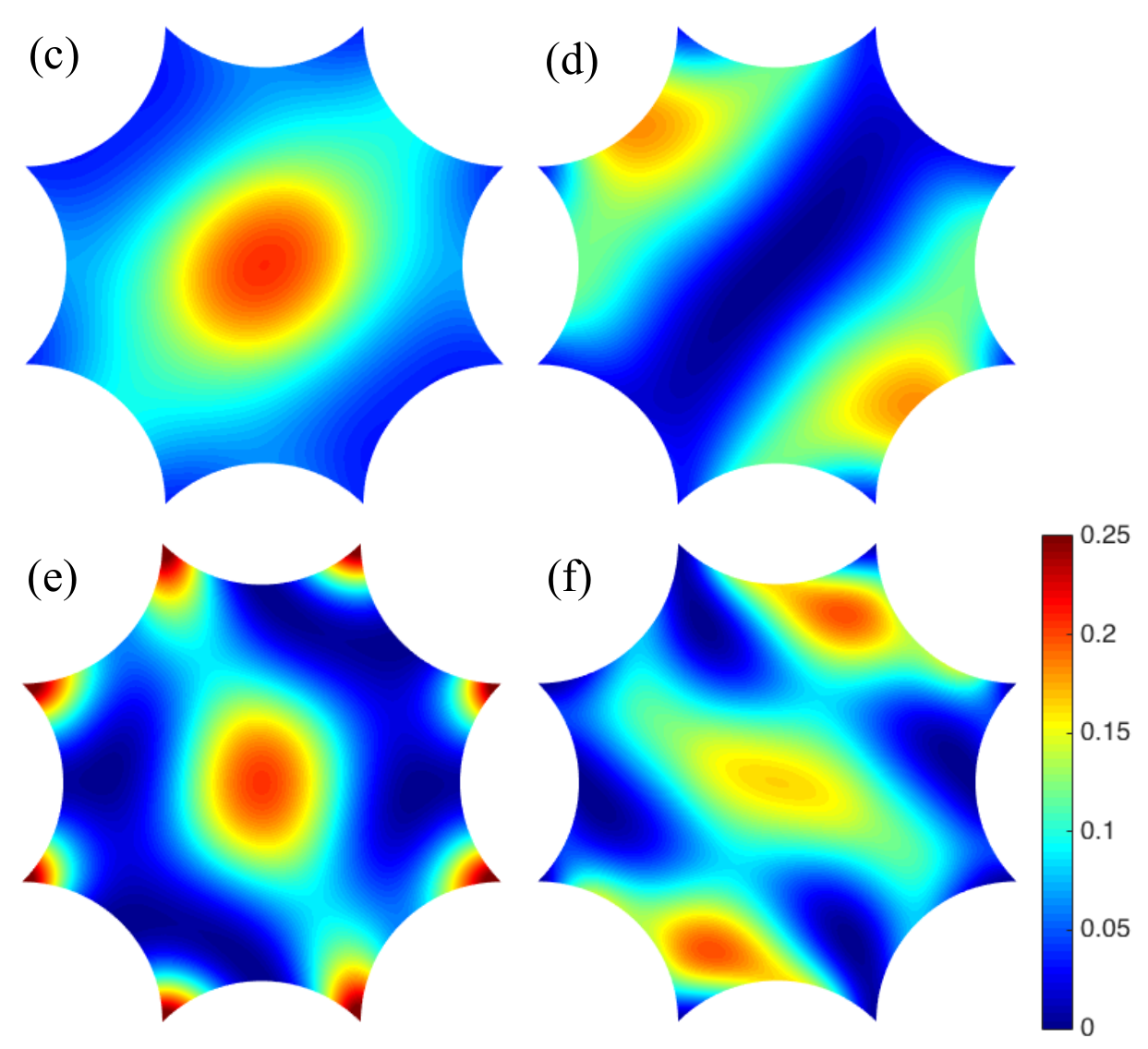}}
\end{tabular}
 \caption{Hyperbolic Bloch problem with nontrivial automorphic potential of the form (\ref{AutomV}). (a) Bandstructure with (red) and without (blue) automorphic potential, along the same direction in the hyperbolic Brillouin zone as in Fig.~\ref{fig:band1}; (b) circularly symmetric potential $U(z)$ in the octagonal unit cell, with $R=0.3$ and $V_0=2$; (c-f) modulus squared $|\psi_\b{k}(z)|^2$ of the eigenstates corresponding to the black dots in (a), i.e., at $\b{k}=(0.8,0.3,1.2,1.7)k$ with $k=1.2$, in increasing order of eigenenergy.}
  \label{fig:pot}
\end{figure}

In Fig.~\ref{fig:pot}(a), we plot the hyperbolic bandstructure for the potential (\ref{AutomV}) with $R=0.3$ and $V_0=2$, illustrated schematically in the inset [Fig.~\ref{fig:pot}(b)]. Focusing first on the $\b{k}=0$ eigenenergies, we find that the ground-state energy is lowered from $E_0(0)=0$ to a negative value, as expected for an attractive potential. We observe a (partial) lifting of the $\b{k}=0$ degeneracies: the $3$-fold degeneracy of $E_1(0)$ is split as $2\oplus 1$; the $4$-fold degeneracy of $E_2(0)$, as $2\oplus 2$; and the $2$-fold degeneracy of $E_3(0)$ is lifted. For both the first and second excited spectral manifolds, we find the energy of one of the doublets is virtually unchanged from the original unperturbed eigenvalue. For the first excited manifold, this can be understood from the fact that for two of the three unperturbed eigenstates [Fig.~\ref{fig:wavef}(b-c)], most of the probability density is concentrated near the boundary segments, with very little near the center of the octagon. From the perspective of degenerate perturbation theory, the average of the potential $U(z)$ over the appropriate linear combinations would yield a small correction to the eigenenergies. By contrast, the third unperturbed eigenfunction [Fig.~\ref{fig:wavef}(d)] has modulus squared peaked near the center of the octagon, and also at its corners: it registers the potential more, and the correction to its eigenenergy is correspondingly greater.

In Fig.~\ref{fig:pot}(c-f) we plot the modulus squared of the hyperbolic Bloch wavefunctions corresponding to the (nondegenerate) levels indicated by black dots in Fig.~\ref{fig:pot}(a). For the lowest band [Fig.~\ref{fig:pot}(c)], due to the attractive potential the probability density is much more concentrated near the center of the unit cell, as compared to the empty-lattice approximation [e.g., Fig.~\ref{fig:wavef}(e)], although the value of $\b{k}$ is not exactly the same]. The eigenfunctions for the next three bands [Fig.~\ref{fig:pot}(d-f)] are also distorted with respect to their empty-lattice counterparts [Fig.~\ref{fig:wavef}(f-h)]. The observation that the probability density in Fig.~\ref{fig:pot}(e) is peaked near the center and at the corners of the octagonal unit cell, combined with the fact that at $\b{k}=0$ this same hyperbolic Bloch state belongs to the singlet in the splitting $3\rightarrow 2\oplus 1$ discussed above for the $E_1(0)$ spectral manifold, confirms our earlier speculation concerning the qualitative reason for this splitting.

\subsection{Hyperbolic point-group symmetries}
\label{sec:sec:pointgroup}

We have so far only discussed the hyperbolic analog of lattice translations, namely, elements of a co-compact, strictly hyperbolic Fuchsian group $\Gamma\subset PSU(1,1)$. Like Euclidean translations, these elements act on the hyperbolic plane without fixed points, and are essentially 2D Lorentz boosts~\cite{BalazsVoros}. Also akin to Euclidean lattices, hyperbolic lattices admit the analog of point-group symmetries, which are discrete symmetries that leave at least one point of the lattice fixed.  A complete hyperbolic band theory must also include a discussion of these, with particular attention paid to how such point-group symmetries manifest in $\b{k}$-space.

For a 2D Euclidean lattice, the point group $G$ is a finite subgroup of the orthogonal group $O(2)$, which includes $SO(2)$ rotations but also orientation-reversing transformations, that is, reflections. Point-group symmetries imply that if $\psi_\b{k}(\b{r})$ is a Bloch eigenstate for such a lattice with energy $E(\b{k})$, the transformed state $\psi_\b{k}^{h}(\b{r})\equiv\psi_\b{k}(h\b{r})$, with ${h}\in G$, is also an eigenstate with the same energy. By elementary properties of Fourier transforms, this transformed state is in fact a Bloch state with wavevector $\b{k}^{h}\equiv{h}\b{k}$, which implies that the bandstructure must obey $E({h}\b{k})=E(\b{k})$.

In the absence of an abelian translation group, Fourier transforms cannot be directly used to generalize these ideas to hyperbolic lattices. Furthermore, since for a $\{4g,4g\}$ hyperbolic lattice $\b{k}$-space is in fact $2g$-dimensional, the very question of how non-translational discrete symmetries in 2D hyperbolic space act in a higher-dimensional $\b{k}$-space is a deep conceptual one.  That said, given that the Abel-Jacobi map provides an algebraic replacement for the Fourier transform, this duality provides a potentially lucrative route for exploring the effect of point-group symmetries. As the group acts on $\Sigma_g$, it acts on the symmetric product of $\Sigma_g$ with itself and, hence, on $\Jac(\Sigma_g)$ via Abel-Jacobi.  As the action moves the points $p_1,\dots,p_g$ in $\Sigma_g$, it moves the end points of the paths of integration in the definition (\ref{AJMap}) of the map $a$, which is the induced action on $\Jac(\Sigma_g)$. We recall that there is a high-symmetry region within the Jacobian --- in this region, the action may have more fixed points.  We aim to utilize this point of view in further work.

For the specific case of the Bolza curve, we are able to examine the point-group action directly. Via concrete calculations for the Bolza lattice, we argue in the Supplementary Materials that the proper generalization of point group for $\{4g,4g\}$ hyperbolic lattices is the finite group $G\cong\Aut(\Sigma_g)$ of automorphisms (i.e., self-maps) of the genus-$g$ Riemann surface~\cite{FarkasKra} associated with the compactified $4g$-gonal unit cell. For the Bolza surface, it is a nonabelian group of order 96 generated by four M\"obius transformations~\cite{CookThesis}: an eightfold rotation ($R$) around the center of the octagon and a threefold rotation-like operation ($U$), both orientation-preserving, and two reflection-like operations ($S$ and $T$), both orientation-reversing. 
Furthermore, as in the Euclidean case, we find this hyperbolic point group acts linearly on hyperbolic $\b{k}$-space: $\b{k}^{h}=M(h)\b{k}$, $h\in G$, where the $4\times 4$ matrices $M(h)$, $h\in G$, form an $SL(4,\mathbb{Z})$ representation of $G$. Explicit representation matrices for the generators $h=R,S,T,U$, from which the representation matrix of any element of $G$ can be constructed by matrix multiplication, are given in the Supplementary Materials. For a $\{4g,4g\}$ hyperbolic lattice, we conjecture that $\b{k}$ transforms similarly, with $M$ a representation of $G$ valued in $SL(2g,\mathbb{Z})$. By contrast with the Euclidean case however, the matrices $M(h)$ are in general not orthogonal, and thus do not simply correspond to the action of a Euclidean point group in $2g$ dimensions.

\begin{figure}[t]
\centering
\begin{tabular}[c]{cc}
\subfloat{\includegraphics[width=0.5\columnwidth,valign=c]{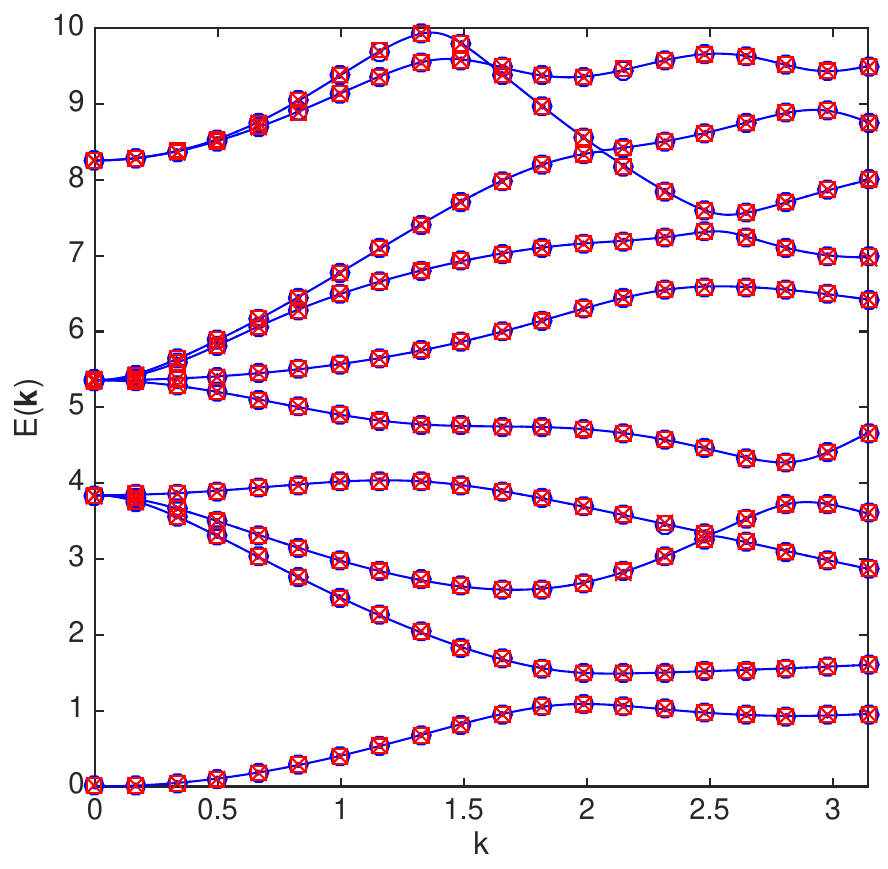}}
&
\subfloat{\includegraphics[width=0.5\columnwidth,valign=c]{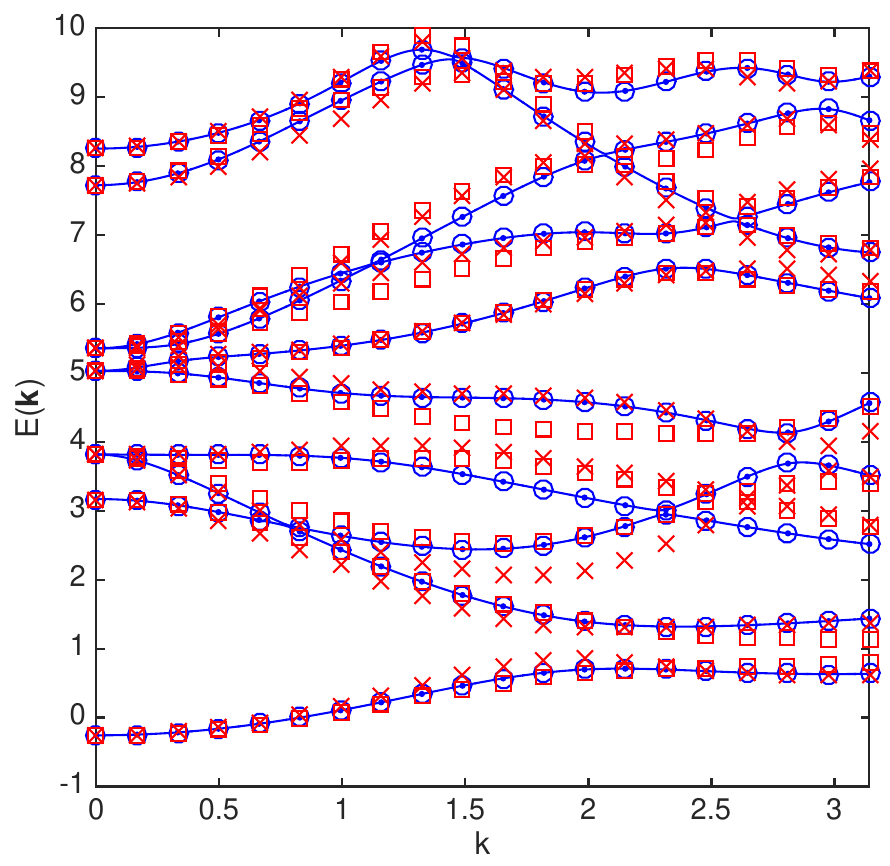}} \\
(a) & (b) 
\end{tabular}
 \caption{Point-group symmetries in hyperbolic $\b{k}$-space. (a) corresponds to the empty-lattice approximation, and (b) to the automorphic potential of Fig.~\ref{fig:pot}. Blue lines: hyperbolic bandstructure $E_n(\b{k})$ along the direction of Figs.~\ref{fig:band1} and \ref{fig:pot}, blue dots: $E_n(\b{k}^R)$, blue circles: $E_n(\b{k}^S)$, red crosses: $E_n(\b{k}^T)$, red squares: $E_n(\b{k}^U)$.}
  \label{fig:pointgroup}
\end{figure}

In Fig.~\ref{fig:pointgroup}(a), we verify numerically that the hyperbolic bandstructure in the empty-lattice approximation is invariant under the full hyperbolic point group $G$ of the Bolza lattice, meaning that $E_n(\b{k}^h)=E_n(\b{k})$ for all $h\in G$. We choose $\b{k}$ along the generic direction already considered in Fig.~\ref{fig:band1} and plot both $E_n(\b{k})$ (blue lines) and $E_n(\b{k}^h)$ (colored symbols), where $\b{k}^h$ is the direction related to $\b{k}$ by point-group symmetry $h$. We verify that the bandstructure is left unchanged under the action of all four generators $h=R,S,T,U$ of $G$, thus establishing invariance under the full point group.

Fig.~\ref{fig:pointgroup}(a) is to be contrasted with Fig.~\ref{fig:pointgroup}(b), which illustrates that the automorphic potential chosen in Eq.~(\ref{AutomV}) breaks at least some of the hyperbolic point-group symmetries of the Bolza lattice. The $R$ operation is a $\pi/4$ rotation about the origin, and the $S$ operation is a reflection across the $x$ axis followed by a $\pi/4$ rotation. Though formally defined as M\"obius transformations, they reduce to simple Euclidean isometries that are obvious symmetries of a circular potential well. As a result, the bandstructure is left unchanged under $\b{k}\rightarrow\b{k}^h$ with $h=R,S$. By contrast, the $T$ and $U$ operations are genuine non-Euclidean isometries involving boosts (see the Supplementary Materials) that do not leave the potential invariant. Correspondingly, the bandstructure does not exhibit invariance under $\b{k}\rightarrow\b{k}^h$ with $h=T,U$.

\subsection{The tight-binding limit}

In conventional band theory, the tight-binding method is a commonly-used approximation scheme to analyze the Schr\"odinger equation in the limit of deep periodic potentials~\cite{SSP}. While inexact, it provides a conceptually important, and often sufficiently accurate, framework to study the Bloch problem in this limit. The tight-binding method starts from the discrete spectrum and localized eigenstates of isolated potential wells, and builds on the idea that propagation throughout the crystal proceeds via weak quantum tunneling between those localized states. Our hyperbolic band theory described so far is based on the full Schr\"odinger equation, and applies to arbitrary $\{4g,4g\}$ automorphic potentials, including deep ones. However, to further develop our generalization of band theory and in light of the experiments of Ref.~\cite{kollar2019}, which are most simply modelled using the tight-binding method, it is natural to ask whether an explicit tight-binding formulation of hyperbolic band theory can be devised. In the Supplementary Materials, we show this is indeed possible. In the limit of a deep localized potential $U(z)$, approximate eigenstates that obey the automorphic Bloch condition (\ref{automorphic}) can be constructed as linear combinations of eigenstates of the ``atomic'' problem $-\Delta+U$ and their $\Gamma$-translates, in the spirit of the linear combination of atomic orbitals (LCAO) familiar in solid-state physics~\cite{SSP}. The coefficients of this expansion are eigenvectors of a finite-dimensional $\b{k}$-dependent matrix eigenvalue problem, the dimension of which is equal to the number of atomic eigenstates kept in the expansion, and the eigenvalues produce an approximate hyperbolic bandstructure. A hyperbolic analog of Wannier functions can likewise be constructed (see the Supplementary Materials).

Our work opens up several exciting avenues of research. While we have shown how to construct a continuous family of Bloch eigenstates for a large class of Hamiltonians with the symmetry of a hyperbolic tessellation, we have {\it not} provided a hyperbolic equivalent of Bloch's theorem --- that is, a statement that {\it all} eigenstates of the Hamiltonian are hyperbolic Bloch eigenstates. What precise fraction of the full spectrum is captured by the hyperbolic Bloch family of eigenstates, and the nature of those eigenstates that may not be of hyperbolic Bloch form, are thus important questions for future research. One obvious line of attack is to attempt to match our predictions with those obtained from numerical diagonalization on $\{4g,4g\}$ lattices, keeping in mind possible subtle issues related to the implementation of automorphic boundary conditions in finite lattices, especially given the different relative importance of bulk versus boundary in Euclidean versus hyperbolic geometries. It may also be possible to approach those spectral questions using number-theoretic tools such as the Selberg trace formula and associated zeta function~\cite{selberg1956,HejhalVol1}. Even within the Bloch condition, the role of factors of automorphy of nonzero weight is an intriguing question in the hyperbolic setting. 

In further pursuing the connections to algebraic geometry and number theory, we note that higher-dimensional versions of our construction may be produced now for K3 surfaces and Calabi-Yau manifolds (e.g.,~\cite{Fields}), which generalize elliptic curves, and for Shimura varieties (e.g.,~\cite{Shimura}), which generalize modular curves.  Working over Calabi-Yau manifolds is especially tantalizing as a potential pathway for novel connections between high-energy physics and condensed matter, which may offer new tools to the latter from string theory and mirror symmetry (e.g.,~\cite{Clay}).  In three spatial dimensions specifically, we also anticipate connections with the work of Thurston~\cite{thurston1982}, whereby hyperbolic bandstructures may arise in connection with three-dimensional hyperbolic tessellations, their (Kleinian) groups of discrete translations, and the geometry and topology of compact three-manifolds produced by the quotienting of three-dimensional hyperbolic space by Kleinian translations. Finally, on the experimental side, we advocate the fabrication and characterization of $\{4g,4g\}$ lattices using circuit QED~\cite{kollar2019}, photonic~\cite{yu2020}, or other metamaterial platforms.

Our construction carries with it a realization that our topological understanding of condensed matter is a small corner of a theory that is perhaps, by and large, algebro-geometric in nature. Indeed, our construction anticipates the emergence of algebro-geometric invariants alongside topological ones, such as Donaldson-Thomas invariants~\cite{DT} of higher-dimensional complex varieties.

\nocite{bolza1887,DM,hladkyhennion1991,langlet1995,ballandras2002,Liboff,carinena2007}

\section{Supplementary Materials}

\noindent Supplementary material for this article is available at XXX\\
Section S1. Geometry of the Bolza lattice\\
Section S2. Fundamental group of the Bolza surface\\
Section S3. Twisted automorphic boundary conditions in the finite element method\\
Section S4. Point-group symmetries of the Bolza lattice\\
Section S5. The tight-binding approximation\\
Section S6. Hyperbolic Wannier functions\\
Fig. S1. Finite element triangulation of the hyperbolic octagon\\
Fig. S2. Automorphisms of the Bolza surface\\
References (43-49)

\bigskip

{\bf Acknowledgements:} The authors wish to thank M. Protter and G. Shankar for assisting with some of the derivations in the Supplementary Materials. The authors also acknowledge helpful conversations with M. Berg, V. Bouchard, R. Boyack, T. Creutzig, C. Doran, D. Freed, T. Gannon, C. Mahadeo, R. Mazzeo, A. Neitzke, A. Stevens, J. Szmigielski, and K. Tanaka, and thank the anonymous reviewers for very useful remarks and suggestions. {\bf Funding:} While working on the research reported in this manuscript, J.M. was supported by NSERC Discovery Grants \#RGPIN-2014-4608, \#RGPIN-2020-06999, and \#RGPAS-2020-00064; the Canada Research Chair (CRC) Program; CIFAR; the Government of Alberta's Major Innovation Fund (MIF); and the University of Alberta.  S.R. was supported by NSERC Discovery Grant \#RGPIN-2017-04520; the Canada Foundation for Innovation John R. Evans Leaders Fund; the GEAR Network (National Science Foundation grants DMS 1107452, 1107263, 1107367 \emph{RNMS: Geometric Structures and Representation Varieties}); the School of Mathematics and Physics at the University of Queensland (through the Ethel Raybould Fellowship program); and the University of Saskatchewan.  Both J.M. and S.R. were supported by the Tri-Agency New Frontiers in Research Fund (NFRF, Exploration Stream) and the Pacific Institute for the Mathematical Sciences (PIMS) Collaborative Research Group program. {\bf Author contributions:} J.M. and S.R. have contributed equally to this paper. {\bf Competing interests:} The authors declare that they have no competing interests. {\bf Data and materials availability:} All data needed to evaluate the conclusions in the paper are present in the paper and/or the Supplementary Materials.

\bibliography{hbt}

\begin{thebibliography}{49}%
\makeatletter
\providecommand \@ifxundefined [1]{%
 \@ifx{#1\undefined}
}%
\providecommand \@ifnum [1]{%
 \ifnum #1\expandafter \@firstoftwo
 \else \expandafter \@secondoftwo
 \fi
}%
\providecommand \@ifx [1]{%
 \ifx #1\expandafter \@firstoftwo
 \else \expandafter \@secondoftwo
 \fi
}%
\providecommand \natexlab [1]{#1}%
\providecommand \enquote  [1]{``#1''}%
\providecommand \bibnamefont  [1]{#1}%
\providecommand \bibfnamefont [1]{#1}%
\providecommand \citenamefont [1]{#1}%
\providecommand \href@noop [0]{\@secondoftwo}%
\providecommand \href [0]{\begingroup \@sanitize@url \@href}%
\providecommand \@href[1]{\@@startlink{#1}\@@href}%
\providecommand \@@href[1]{\endgroup#1\@@endlink}%
\providecommand \@sanitize@url [0]{\catcode `\\12\catcode `\$12\catcode
  `\&12\catcode `\#12\catcode `\^12\catcode `\_12\catcode `\%12\relax}%
\providecommand \@@startlink[1]{}%
\providecommand \@@endlink[0]{}%
\providecommand \url  [0]{\begingroup\@sanitize@url \@url }%
\providecommand \@url [1]{\endgroup\@href {#1}{\urlprefix }}%
\providecommand \urlprefix  [0]{URL }%
\providecommand \Eprint [0]{\href }%
\providecommand \doibase [0]{http://dx.doi.org/}%
\providecommand \selectlanguage [0]{\@gobble}%
\providecommand \bibinfo  [0]{\@secondoftwo}%
\providecommand \bibfield  [0]{\@secondoftwo}%
\providecommand \translation [1]{[#1]}%
\providecommand \BibitemOpen [0]{}%
\providecommand \bibitemStop [0]{}%
\providecommand \bibitemNoStop [0]{.\EOS\space}%
\providecommand \EOS [0]{\spacefactor3000\relax}%
\providecommand \BibitemShut  [1]{\csname bibitem#1\endcsname}%
\let\auto@bib@innerbib\@empty
\bibitem [{\citenamefont {Bloch}(1929)}]{bloch1929}%
  \BibitemOpen
  \bibfield  {author} {\bibinfo {author} {\bibfnamefont {F.}~\bibnamefont
  {Bloch}},\ }\bibfield  {title} {\enquote {\bibinfo {title} {{\"Uber} die
  {Quantenmechanik} der {Elektronen} in {Kristallgittern}},}\ }\href {\doibase
  10.1007/BF01339455} {\bibfield  {journal} {\bibinfo  {journal} {Z. Phys.}\
  }\textbf {\bibinfo {volume} {52}},\ \bibinfo {pages} {555--600} (\bibinfo
  {year} {1929})}\BibitemShut {NoStop}%
\bibitem [{\citenamefont {Ashcroft}\ and\ \citenamefont {Mermin}(1976)}]{SSP}%
  \BibitemOpen
  \bibfield  {author} {\bibinfo {author} {\bibfnamefont {N.~W.}\ \bibnamefont
  {Ashcroft}}\ and\ \bibinfo {author} {\bibfnamefont {N.~D.}\ \bibnamefont
  {Mermin}},\ }\href@noop {} {\emph {\bibinfo {title} {Solid State Physics}}}\
  (\bibinfo  {publisher} {Saunders College},\ \bibinfo {address}
  {Philadelphia},\ \bibinfo {year} {1976})\BibitemShut {NoStop}%
\bibitem [{\citenamefont {Haldane}(1988)}]{haldane1988}%
  \BibitemOpen
  \bibfield  {author} {\bibinfo {author} {\bibfnamefont {F.~D.~M.}\
  \bibnamefont {Haldane}},\ }\bibfield  {title} {\enquote {\bibinfo {title}
  {Model for a {quantum} {Hall} {effect} without {Landau} {levels}:
  {condensed}-{matter} {realization} of the `{parity} {anomaly}'},}\ }\href
  {\doibase 10.1103/PhysRevLett.61.2015} {\bibfield  {journal} {\bibinfo
  {journal} {Phys. Rev. Lett.}\ }\textbf {\bibinfo {volume} {61}},\ \bibinfo
  {pages} {2015--2018} (\bibinfo {year} {1988})}\BibitemShut {NoStop}%
\bibitem [{\citenamefont {Chiu}\ \emph {et~al.}(2016)\citenamefont {Chiu},
  \citenamefont {Teo}, \citenamefont {Schnyder},\ and\ \citenamefont
  {Ryu}}]{chiu2016}%
  \BibitemOpen
  \bibfield  {author} {\bibinfo {author} {\bibfnamefont {C.-K.}\ \bibnamefont
  {Chiu}}, \bibinfo {author} {\bibfnamefont {J.~C.~Y.}\ \bibnamefont {Teo}},
  \bibinfo {author} {\bibfnamefont {A.~P.}\ \bibnamefont {Schnyder}}, \ and\
  \bibinfo {author} {\bibfnamefont {S.}~\bibnamefont {Ryu}},\ }\bibfield
  {title} {\enquote {\bibinfo {title} {Classification of topological quantum
  matter with symmetries},}\ }\href {\doibase 10.1103/RevModPhys.88.035005}
  {\bibfield  {journal} {\bibinfo  {journal} {Rev. Mod. Phys.}\ }\textbf
  {\bibinfo {volume} {88}},\ \bibinfo {pages} {035005} (\bibinfo {year}
  {2016})}\BibitemShut {NoStop}%
\bibitem [{\citenamefont {Steinhardt}\ and\ \citenamefont
  {Ostlund}(1987)}]{QCbook}%
  \BibitemOpen
  \bibfield  {author} {\bibinfo {author} {\bibfnamefont {P.~J.}\ \bibnamefont
  {Steinhardt}}\ and\ \bibinfo {author} {\bibfnamefont {S.}~\bibnamefont
  {Ostlund}},\ }\href@noop {} {\emph {\bibinfo {title} {The Physics of
  Quasicrystals}}}\ (\bibinfo  {publisher} {World Scientific},\ \bibinfo
  {address} {Singapore},\ \bibinfo {year} {1987})\BibitemShut {NoStop}%
\bibitem [{\citenamefont {Janssen}\ and\ \citenamefont
  {Janner}(2014)}]{janssen2014}%
  \BibitemOpen
  \bibfield  {author} {\bibinfo {author} {\bibfnamefont {T.}~\bibnamefont
  {Janssen}}\ and\ \bibinfo {author} {\bibfnamefont {A.}~\bibnamefont
  {Janner}},\ }\bibfield  {title} {\enquote {\bibinfo {title} {Aperiodic
  crystals and superspace concepts},}\ }\href {\doibase
  10.1107/S2052520614014917} {\bibfield  {journal} {\bibinfo  {journal} {Acta
  Crystallogr. Sect. B}\ }\textbf {\bibinfo {volume} {70}},\ \bibinfo {pages}
  {617--651} (\bibinfo {year} {2014})}\BibitemShut {NoStop}%
\bibitem [{\citenamefont {Koll\'ar}\ \emph {et~al.}(2019)\citenamefont
  {Koll\'ar}, \citenamefont {Fitzpatrick},\ and\ \citenamefont
  {Houck}}]{kollar2019}%
  \BibitemOpen
  \bibfield  {author} {\bibinfo {author} {\bibfnamefont {A.~J.}\ \bibnamefont
  {Koll\'ar}}, \bibinfo {author} {\bibfnamefont {M.}~\bibnamefont
  {Fitzpatrick}}, \ and\ \bibinfo {author} {\bibfnamefont {A.~A.}\ \bibnamefont
  {Houck}},\ }\bibfield  {title} {\enquote {\bibinfo {title} {Hyperbolic
  lattices in circuit quantum electrodynamics},}\ }\href {\doibase
  10.1038/s41586-019-1348-3} {\bibfield  {journal} {\bibinfo  {journal}
  {Nature}\ }\textbf {\bibinfo {volume} {571}},\ \bibinfo {pages} {45--50}
  (\bibinfo {year} {2019})}\BibitemShut {NoStop}%
\bibitem [{\citenamefont {Coxeter}(1957)}]{coxeter1957}%
  \BibitemOpen
  \bibfield  {author} {\bibinfo {author} {\bibfnamefont {H.~S.~M.}\
  \bibnamefont {Coxeter}},\ }\bibfield  {title} {\enquote {\bibinfo {title}
  {Crystal symmetry and its generalizations},}\ }\href@noop {} {\bibfield
  {journal} {\bibinfo  {journal} {Trans. Royal Soc. Canada}\ }\textbf {\bibinfo
  {volume} {51}},\ \bibinfo {pages} {1--13} (\bibinfo {year}
  {1957})}\BibitemShut {NoStop}%
\bibitem [{\citenamefont {Coxeter}(1979)}]{coxeter1979}%
  \BibitemOpen
  \bibfield  {author} {\bibinfo {author} {\bibfnamefont {H.~S.~M.}\
  \bibnamefont {Coxeter}},\ }\bibfield  {title} {\enquote {\bibinfo {title}
  {The {non}-{Euclidean} {symmetry} of {Escher}'s {picture} `{Circle} {Limit}
  {III}'},}\ }\href {\doibase 10.2307/1574078} {\bibfield  {journal} {\bibinfo
  {journal} {Leonardo}\ }\textbf {\bibinfo {volume} {12}},\ \bibinfo {pages}
  {19--25} (\bibinfo {year} {1979})}\BibitemShut {NoStop}%
\bibitem [{\citenamefont {{Koll\'ar}}\ \emph {et~al.}(2020)\citenamefont
  {{Koll\'ar}}, \citenamefont {Fitzpatrick}, \citenamefont {Sarnak},\ and\
  \citenamefont {Houck}}]{kollar2019b}%
  \BibitemOpen
  \bibfield  {author} {\bibinfo {author} {\bibfnamefont {A.~J.}\ \bibnamefont
  {{Koll\'ar}}}, \bibinfo {author} {\bibfnamefont {M.}~\bibnamefont
  {Fitzpatrick}}, \bibinfo {author} {\bibfnamefont {P.}~\bibnamefont {Sarnak}},
  \ and\ \bibinfo {author} {\bibfnamefont {A.~A.}\ \bibnamefont {Houck}},\
  }\bibfield  {title} {\enquote {\bibinfo {title} {Line-{Graph} {Lattices}:
  {Euclidean} and {Non}-{Euclidean} {Flat} {Bands}, and {Implementations} in
  {Circuit} {Quantum} {Electrodynamics}},}\ }\href {\doibase
  10.1007/s00220-019-03645-8} {\bibfield  {journal} {\bibinfo  {journal}
  {Commun. Math. Phys.}\ }\textbf {\bibinfo {volume} {376}},\ \bibinfo {pages}
  {1909--1956} (\bibinfo {year} {2020})}\BibitemShut {NoStop}%
\bibitem [{\citenamefont {Boettcher}\ \emph {et~al.}(2020)\citenamefont
  {Boettcher}, \citenamefont {Bienias}, \citenamefont {Belyansky},
  \citenamefont {{Koll\'ar}},\ and\ \citenamefont {Gorshkov}}]{boettcher2019}%
  \BibitemOpen
  \bibfield  {author} {\bibinfo {author} {\bibfnamefont {I.}~\bibnamefont
  {Boettcher}}, \bibinfo {author} {\bibfnamefont {P.}~\bibnamefont {Bienias}},
  \bibinfo {author} {\bibfnamefont {R.}~\bibnamefont {Belyansky}}, \bibinfo
  {author} {\bibfnamefont {A.~J.}\ \bibnamefont {{Koll\'ar}}}, \ and\ \bibinfo
  {author} {\bibfnamefont {A.~V.}\ \bibnamefont {Gorshkov}},\ }\bibfield
  {title} {\enquote {\bibinfo {title} {Quantum simulation of hyperbolic space
  with circuit quantum electrodynamics: {From} graphs to geometry},}\ }\href
  {\doibase 10.1103/PhysRevA.102.032208} {\bibfield  {journal} {\bibinfo
  {journal} {Phys. Rev. A}\ }\textbf {\bibinfo {volume} {102}},\ \bibinfo
  {pages} {032208} (\bibinfo {year} {2020})}\BibitemShut {NoStop}%
\bibitem [{\citenamefont {Yu}\ \emph {et~al.}(2020)\citenamefont {Yu},
  \citenamefont {Piao},\ and\ \citenamefont {Park}}]{yu2020}%
  \BibitemOpen
  \bibfield  {author} {\bibinfo {author} {\bibfnamefont {S.}~\bibnamefont
  {Yu}}, \bibinfo {author} {\bibfnamefont {X.}~\bibnamefont {Piao}}, \ and\
  \bibinfo {author} {\bibfnamefont {N.}~\bibnamefont {Park}},\ }\bibfield
  {title} {\enquote {\bibinfo {title} {Topological {hyperbolic} {lattices}},}\
  }\href {\doibase 10.1103/PhysRevLett.125.053901} {\bibfield  {journal}
  {\bibinfo  {journal} {Phys. Rev. Lett.}\ }\textbf {\bibinfo {volume} {125}},\
  \bibinfo {pages} {053901} (\bibinfo {year} {2020})}\BibitemShut {NoStop}%
\bibitem [{\citenamefont {Mumford}(1983)}]{Tata}%
  \BibitemOpen
  \bibfield  {author} {\bibinfo {author} {\bibfnamefont {D.}~\bibnamefont
  {Mumford}},\ }\href@noop {} {\emph {\bibinfo {title} {Tata Lectures on Theta
  I}}}\ (\bibinfo  {publisher} {Birkh\"auser},\ \bibinfo {address} {Boston},\
  \bibinfo {year} {1983})\BibitemShut {NoStop}%
\bibitem [{\citenamefont {Nakahara}(1990)}]{Nakahara}%
  \BibitemOpen
  \bibfield  {author} {\bibinfo {author} {\bibfnamefont {M.}~\bibnamefont
  {Nakahara}},\ }\href@noop {} {\emph {\bibinfo {title} {Geometry, Topology,
  and Physics}}}\ (\bibinfo  {publisher} {IOP Publishing},\ \bibinfo {address}
  {Bristol},\ \bibinfo {year} {1990})\BibitemShut {NoStop}%
\bibitem [{\citenamefont {Silverman}(2005)}]{Silverman}%
  \BibitemOpen
  \bibfield  {author} {\bibinfo {author} {\bibfnamefont {J.~H.}\ \bibnamefont
  {Silverman}},\ }\bibfield  {title} {\enquote {\bibinfo {title} {Elliptic
  {c}urves and {c}ryptography},}\ }\href {\doibase 10.1090/psapm/062/2211873}
  {\bibfield  {journal} {\bibinfo  {journal} {Proc. Sympos. Appl. Math.}\
  }\textbf {\bibinfo {volume} {62}},\ \bibinfo {pages} {91--112} (\bibinfo
  {year} {2005})}\BibitemShut {NoStop}%
\bibitem [{\citenamefont {Balazs}\ and\ \citenamefont
  {Voros}(1986)}]{BalazsVoros}%
  \BibitemOpen
  \bibfield  {author} {\bibinfo {author} {\bibfnamefont {N.~L.}\ \bibnamefont
  {Balazs}}\ and\ \bibinfo {author} {\bibfnamefont {A.}~\bibnamefont {Voros}},\
  }\bibfield  {title} {\enquote {\bibinfo {title} {Chaos on the
  pseudosphere},}\ }\href {\doibase 10.1016/0370-1573(86)90159-6} {\bibfield
  {journal} {\bibinfo  {journal} {Phys. Rep.}\ }\textbf {\bibinfo {volume}
  {143}},\ \bibinfo {pages} {109--240} (\bibinfo {year} {1986})}\BibitemShut
  {NoStop}%
\bibitem [{\citenamefont {Katok}(1992)}]{Katok}%
  \BibitemOpen
  \bibfield  {author} {\bibinfo {author} {\bibfnamefont {S.}~\bibnamefont
  {Katok}},\ }\href@noop {} {\emph {\bibinfo {title} {Fuchsian Groups}}}\
  (\bibinfo  {publisher} {The University of Chicago Press},\ \bibinfo {address}
  {Chicago},\ \bibinfo {year} {1992})\BibitemShut {NoStop}%
\bibitem [{\citenamefont {Farkas}\ and\ \citenamefont {Kra}(1992)}]{FarkasKra}%
  \BibitemOpen
  \bibfield  {author} {\bibinfo {author} {\bibfnamefont {H.~M.}\ \bibnamefont
  {Farkas}}\ and\ \bibinfo {author} {\bibfnamefont {I.}~\bibnamefont {Kra}},\
  }\href@noop {} {\emph {\bibinfo {title} {Riemann Surfaces}}},\ \bibinfo
  {edition} {2nd}\ ed.\ (\bibinfo  {publisher} {Springer},\ \bibinfo {address}
  {New York},\ \bibinfo {year} {1992})\BibitemShut {NoStop}%
\bibitem [{\citenamefont {Gunning}(1956)}]{Gunning}%
  \BibitemOpen
  \bibfield  {author} {\bibinfo {author} {\bibfnamefont {R.~C.}\ \bibnamefont
  {Gunning}},\ }\bibfield  {title} {\enquote {\bibinfo {title} {The structure
  of factors of automorphy},}\ }\href {http://www.jstor.org/stable/2372521}
  {\bibfield  {journal} {\bibinfo  {journal} {Amer. J. Math.}\ }\textbf
  {\bibinfo {volume} {78}},\ \bibinfo {pages} {357--382} (\bibinfo {year}
  {1956})}\BibitemShut {NoStop}%
\bibitem [{\citenamefont {Venkov}(1990)}]{Venkov}%
  \BibitemOpen
  \bibfield  {author} {\bibinfo {author} {\bibfnamefont {A.~B.}\ \bibnamefont
  {Venkov}},\ }\href@noop {} {\emph {\bibinfo {title} {Spectral Theory of
  Automorphic Functions and Its Applications}}}\ (\bibinfo  {publisher} {Kluwer
  Academic Publishers},\ \bibinfo {address} {Dordrecht},\ \bibinfo {year}
  {1990})\BibitemShut {NoStop}%
\bibitem [{\citenamefont {Hejhal}(1976)}]{HejhalVol1}%
  \BibitemOpen
  \bibfield  {author} {\bibinfo {author} {\bibfnamefont {D.~A.}\ \bibnamefont
  {Hejhal}},\ }\href@noop {} {\emph {\bibinfo {title} {The Selberg Trace
  Formula for $PSL(2,\mathbb{R})$}}},\ Vol.~\bibinfo {volume} {1}\ (\bibinfo
  {publisher} {Springer-Verlag},\ \bibinfo {address} {Berlin},\ \bibinfo {year}
  {1976})\BibitemShut {NoStop}%
\bibitem [{\citenamefont {Kiremidjian}(1972)}]{GK}%
  \BibitemOpen
  \bibfield  {author} {\bibinfo {author} {\bibfnamefont {G.}~\bibnamefont
  {Kiremidjian}},\ }\bibfield  {title} {\enquote {\bibinfo {title} {Complex
  structures on {R}iemann surfaces},}\ }\href {\doibase 10.2307/1996246}
  {\bibfield  {journal} {\bibinfo  {journal} {Trans. Amer. Math. Soc.}\
  }\textbf {\bibinfo {volume} {169}},\ \bibinfo {pages} {317--336} (\bibinfo
  {year} {1972})}\BibitemShut {NoStop}%
\bibitem [{\citenamefont {Forster}(1981)}]{Forster}%
  \BibitemOpen
  \bibfield  {author} {\bibinfo {author} {\bibfnamefont {O.}~\bibnamefont
  {Forster}},\ }\href@noop {} {\emph {\bibinfo {title} {Lectures on {R}iemann
  {S}urfaces}}}\ (\bibinfo  {publisher} {Springer-Verlag, New York-Berlin},\
  \bibinfo {year} {1981})\BibitemShut {NoStop}%
\bibitem [{\citenamefont {Miranda}(1995)}]{Miranda}%
  \BibitemOpen
  \bibfield  {author} {\bibinfo {author} {\bibfnamefont {R.}~\bibnamefont
  {Miranda}},\ }\href@noop {} {\emph {\bibinfo {title} {Algebraic {C}urves and
  {R}iemann {S}urfaces}}}\ (\bibinfo  {publisher} {American Mathematical
  Society, Providence},\ \bibinfo {year} {1995})\BibitemShut {NoStop}%
\bibitem [{\citenamefont {Macdonald}(1962)}]{Macdonald}%
  \BibitemOpen
  \bibfield  {author} {\bibinfo {author} {\bibfnamefont {I.~G.}\ \bibnamefont
  {Macdonald}},\ }\bibfield  {title} {\enquote {\bibinfo {title} {Symmetric
  products of an algebraic curve},}\ }\href {\doibase
  10.1016/0040-9383(62)90019-8} {\bibfield  {journal} {\bibinfo  {journal}
  {Topology}\ }\textbf {\bibinfo {volume} {1}},\ \bibinfo {pages} {319--343}
  (\bibinfo {year} {1962})}\BibitemShut {NoStop}%
\bibitem [{\citenamefont {Strohmaier}\ and\ \citenamefont
  {Uski}(2013)}]{strohmaier2013}%
  \BibitemOpen
  \bibfield  {author} {\bibinfo {author} {\bibfnamefont {A.}~\bibnamefont
  {Strohmaier}}\ and\ \bibinfo {author} {\bibfnamefont {V.}~\bibnamefont
  {Uski}},\ }\bibfield  {title} {\enquote {\bibinfo {title} {An {algorithm} for
  the {computation} of {eigenvalues}, {spectral} {zeta} {functions} and
  {zeta}-{determinants} on {hyperbolic} {surfaces}},}\ }\href {\doibase
  10.1007/s00220-012-1557-1} {\bibfield  {journal} {\bibinfo  {journal}
  {Commun. Math. Phys.}\ }\textbf {\bibinfo {volume} {317}},\ \bibinfo {pages}
  {827--869} (\bibinfo {year} {2013})}\BibitemShut {NoStop}%
\bibitem [{\citenamefont {Buser}(1992)}]{Buser}%
  \BibitemOpen
  \bibfield  {author} {\bibinfo {author} {\bibfnamefont {P.}~\bibnamefont
  {Buser}},\ }\href@noop {} {\emph {\bibinfo {title} {Geometry and Spectra of
  Compact Riemann Surfaces}}}\ (\bibinfo  {publisher} {Birkh\"auser},\ \bibinfo
  {address} {Boston},\ \bibinfo {year} {1992})\BibitemShut {NoStop}%
\bibitem [{\citenamefont {Aurich}\ and\ \citenamefont
  {Steiner}(1989)}]{aurich1989}%
  \BibitemOpen
  \bibfield  {author} {\bibinfo {author} {\bibfnamefont {R.}~\bibnamefont
  {Aurich}}\ and\ \bibinfo {author} {\bibfnamefont {F.}~\bibnamefont
  {Steiner}},\ }\bibfield  {title} {\enquote {\bibinfo {title} {Periodic-orbit
  sum rules for the {Hadamard}-{Gutzwiller} model},}\ }\href {\doibase
  10.1016/0167-2789(89)90003-1} {\bibfield  {journal} {\bibinfo  {journal}
  {Physica D}\ }\textbf {\bibinfo {volume} {39}},\ \bibinfo {pages} {169--193}
  (\bibinfo {year} {1989})}\BibitemShut {NoStop}%
\bibitem [{\citenamefont {Aurich}\ and\ \citenamefont
  {Steiner}(1993)}]{aurich1993}%
  \BibitemOpen
  \bibfield  {author} {\bibinfo {author} {\bibfnamefont {R.}~\bibnamefont
  {Aurich}}\ and\ \bibinfo {author} {\bibfnamefont {F.}~\bibnamefont
  {Steiner}},\ }\bibfield  {title} {\enquote {\bibinfo {title} {Statistical
  properties of highly excited quantum eigenstates of a strongly chaotic
  system},}\ }\href {\doibase 10.1016/0167-2789(93)90255-Y} {\bibfield
  {journal} {\bibinfo  {journal} {Physica D}\ }\textbf {\bibinfo {volume}
  {64}},\ \bibinfo {pages} {185--214} (\bibinfo {year} {1993})}\BibitemShut
  {NoStop}%
\bibitem [{\citenamefont {Ninnemann}(1995)}]{ninnemann1995}%
  \BibitemOpen
  \bibfield  {author} {\bibinfo {author} {\bibfnamefont {H.}~\bibnamefont
  {Ninnemann}},\ }\bibfield  {title} {\enquote {\bibinfo {title} {Gutzwiller's
  octagon and the triangular billiard {$T^*(2,3,8)$} as models for the
  quantization of chaotic systems by {Selberg}'s trace formula},}\ }\href
  {\doibase 10.1142/S0217979295000719} {\bibfield  {journal} {\bibinfo
  {journal} {Int. J. Mod. Phys. B}\ }\textbf {\bibinfo {volume} {09}},\
  \bibinfo {pages} {1647--1753} (\bibinfo {year} {1995})}\BibitemShut {NoStop}%
\bibitem [{\citenamefont {Bachelot-Motet}(2010)}]{bachelot-motet2010}%
  \BibitemOpen
  \bibfield  {author} {\bibinfo {author} {\bibfnamefont {A.}~\bibnamefont
  {Bachelot-Motet}},\ }\bibfield  {title} {\enquote {\bibinfo {title} {Wave
  computation on the hyperbolic double doughnut},}\ }\href
  {https://www.jstor.org/stable/43693616} {\bibfield  {journal} {\bibinfo
  {journal} {J. Comput. Math.}\ }\textbf {\bibinfo {volume} {28}},\ \bibinfo
  {pages} {790--806} (\bibinfo {year} {2010})}\BibitemShut {NoStop}%
\bibitem [{\citenamefont {Hecht}(2012)}]{hecht2012}%
  \BibitemOpen
  \bibfield  {author} {\bibinfo {author} {\bibfnamefont {F.}~\bibnamefont
  {Hecht}},\ }\bibfield  {title} {\enquote {\bibinfo {title} {New development
  in freefem++},}\ }\href {\doibase 10.1515/jnum-2012-0013} {\bibfield
  {journal} {\bibinfo  {journal} {J. Numer. Math.}\ }\textbf {\bibinfo {volume}
  {20}},\ \bibinfo {pages} {251--266} (\bibinfo {year} {2012})}\BibitemShut
  {NoStop}%
\bibitem [{\citenamefont {Cook}(2018)}]{CookThesis}%
  \BibitemOpen
  \bibfield  {author} {\bibinfo {author} {\bibfnamefont {J.}~\bibnamefont
  {Cook}},\ }\emph {\bibinfo {title} {Properties of Eigenvalues on Riemann
  Surfaces with Large Symmetry Groups}},\ \href@noop {} {Ph.D. thesis},\
  \bibinfo  {school} {Loughborough University} (\bibinfo {year}
  {2018})\BibitemShut {NoStop}%
\bibitem [{\citenamefont {{von Neumann}}\ and\ \citenamefont
  {Wigner}(1929)}]{vonneumann1929}%
  \BibitemOpen
  \bibfield  {author} {\bibinfo {author} {\bibfnamefont {J.}~\bibnamefont {{von
  Neumann}}}\ and\ \bibinfo {author} {\bibfnamefont {E.~P.}\ \bibnamefont
  {Wigner}},\ }\bibfield  {title} {\enquote {\bibinfo {title} {{\"Uber} das
  {Verhalten} von {Eigenwerten} bei adiabatischen {Prozessen}},}\ }\href@noop
  {} {\bibfield  {journal} {\bibinfo  {journal} {Z. Phys.}\ }\textbf {\bibinfo
  {volume} {30}},\ \bibinfo {pages} {467--470} (\bibinfo {year}
  {1929})}\BibitemShut {NoStop}%
\bibitem [{\citenamefont {Fang}\ \emph {et~al.}(2016)\citenamefont {Fang},
  \citenamefont {Weng}, \citenamefont {Dai},\ and\ \citenamefont
  {Fang}}]{fang2016}%
  \BibitemOpen
  \bibfield  {author} {\bibinfo {author} {\bibfnamefont {C.}~\bibnamefont
  {Fang}}, \bibinfo {author} {\bibfnamefont {H.}~\bibnamefont {Weng}}, \bibinfo
  {author} {\bibfnamefont {X.}~\bibnamefont {Dai}}, \ and\ \bibinfo {author}
  {\bibfnamefont {Z.}~\bibnamefont {Fang}},\ }\bibfield  {title} {\enquote
  {\bibinfo {title} {Topological nodal line semimetals},}\ }\href {\doibase
  10.1088/1674-1056/25/11/117106} {\bibfield  {journal} {\bibinfo  {journal}
  {Chinese Phys. B}\ }\textbf {\bibinfo {volume} {25}},\ \bibinfo {pages}
  {117106} (\bibinfo {year} {2016})}\BibitemShut {NoStop}%
\bibitem [{\citenamefont {Donagi}(1995)}]{Donagi}%
  \BibitemOpen
  \bibfield  {author} {\bibinfo {author} {\bibfnamefont {R.}~\bibnamefont
  {Donagi}},\ }\bibfield  {title} {\enquote {\bibinfo {title} {Spectral
  covers},}\ }in\ \href@noop {} {\emph {\bibinfo {booktitle} {Current {T}opics
  in {C}omplex {A}lgebraic {G}eometry ({B}erkeley, {CA}, 1992/93)}}},\ \bibinfo
  {series} {Math. Sci. Res. Inst. Publ.}, Vol.~\bibinfo {volume} {28}\
  (\bibinfo  {publisher} {Cambridge Univ. Press, Cambridge},\ \bibinfo {year}
  {1995})\ pp.\ \bibinfo {pages} {65--86}\BibitemShut {NoStop}%
\bibitem [{\citenamefont {Selberg}(1956)}]{selberg1956}%
  \BibitemOpen
  \bibfield  {author} {\bibinfo {author} {\bibfnamefont {A.}~\bibnamefont
  {Selberg}},\ }\bibfield  {title} {\enquote {\bibinfo {title} {Harmonic
  analysis and discontinuous groups in weakly symmetric {Riemannian} spaces
  with applications to {Dirichlet} series},}\ }\href@noop {} {\bibfield
  {journal} {\bibinfo  {journal} {J. Indian Math. Soc.}\ }\textbf {\bibinfo
  {volume} {20}},\ \bibinfo {pages} {47--87} (\bibinfo {year}
  {1956})}\BibitemShut {NoStop}%
\bibitem [{\citenamefont {Laza}\ \emph {et~al.}(2013)\citenamefont {Laza},
  \citenamefont {Sch\"{u}tt},\ and\ \citenamefont {Yui}}]{Fields}%
  \BibitemOpen
  \bibinfo {editor} {\bibfnamefont {R.}~\bibnamefont {Laza}}, \bibinfo {editor}
  {\bibfnamefont {M.}~\bibnamefont {Sch\"{u}tt}}, \ and\ \bibinfo {editor}
  {\bibfnamefont {N.}~\bibnamefont {Yui}},\ eds.,\ \href@noop {} {\emph
  {\bibinfo {title} {Arithmetic and {G}eometry of {K}3 {S}urfaces and
  {C}alabi-{Y}au {T}hreefolds}}},\ \bibinfo {series} {Proceedings of the
  {W}orkshop {H}eld at the {F}ields {I}nstitute and {U}niversity of {T}oronto,
  {T}oronto, {ON}, {A}ugust 16--25, 2011, {F}ields {I}nstitute
  {C}ommunications}, Vol.~\bibinfo {volume} {67}\ (\bibinfo  {publisher}
  {Springer, New York; Fields Institute for Research in Mathematical Sciences,
  Toronto, ON},\ \bibinfo {year} {2013})\BibitemShut {NoStop}%
\bibitem [{\citenamefont {Deligne}\ \emph {et~al.}(1982)\citenamefont
  {Deligne}, \citenamefont {Milne}, \citenamefont {Ogus},\ and\ \citenamefont
  {Shih}}]{Shimura}%
  \BibitemOpen
  \bibfield  {author} {\bibinfo {author} {\bibfnamefont {P.}~\bibnamefont
  {Deligne}}, \bibinfo {author} {\bibfnamefont {J.~S.}\ \bibnamefont {Milne}},
  \bibinfo {author} {\bibfnamefont {A.}~\bibnamefont {Ogus}}, \ and\ \bibinfo
  {author} {\bibfnamefont {K.-Y.}\ \bibnamefont {Shih}},\ }\href@noop {} {\emph
  {\bibinfo {title} {Hodge {C}ycles, {M}otives, and {S}himura {V}arieties}}}\
  (\bibinfo  {publisher} {Springer-Verlag, Berlin-New York},\ \bibinfo {year}
  {1982})\BibitemShut {NoStop}%
\bibitem [{\citenamefont {Hori}\ \emph {et~al.}(2003)\citenamefont {Hori},
  \citenamefont {Katz}, \citenamefont {Klemm}, \citenamefont {Pandharipande},
  \citenamefont {Thomas}, \citenamefont {Vafa}, \citenamefont {Vakil},\ and\
  \citenamefont {Zaslow}}]{Clay}%
  \BibitemOpen
  \bibfield  {author} {\bibinfo {author} {\bibfnamefont {K.}~\bibnamefont
  {Hori}}, \bibinfo {author} {\bibfnamefont {S.}~\bibnamefont {Katz}}, \bibinfo
  {author} {\bibfnamefont {A.}~\bibnamefont {Klemm}}, \bibinfo {author}
  {\bibfnamefont {R.}~\bibnamefont {Pandharipande}}, \bibinfo {author}
  {\bibfnamefont {R.~P.}\ \bibnamefont {Thomas}}, \bibinfo {author}
  {\bibfnamefont {C.}~\bibnamefont {Vafa}}, \bibinfo {author} {\bibfnamefont
  {R.}~\bibnamefont {Vakil}}, \ and\ \bibinfo {author} {\bibfnamefont
  {E.}~\bibnamefont {Zaslow}},\ }\href@noop {} {\emph {\bibinfo {title} {Mirror
  {S}ymmetry}}}\ (\bibinfo  {publisher} {American Mathematical Society,
  Providence, RI; Clay Mathematics Institute, Cambridge, MA},\ \bibinfo {year}
  {2003})\BibitemShut {NoStop}%
\bibitem [{\citenamefont {Thurston}(1982)}]{thurston1982}%
  \BibitemOpen
  \bibfield  {author} {\bibinfo {author} {\bibfnamefont {W.~P.}\ \bibnamefont
  {Thurston}},\ }\bibfield  {title} {\enquote {\bibinfo {title} {Three
  dimensional manifolds, {Kleinian} groups and hyperbolic geometry},}\ }\href
  {https://projecteuclid.org/euclid.bams/1183548782} {\bibfield  {journal}
  {\bibinfo  {journal} {Bull. Amer. Math. Soc. (N.S.)}\ }\textbf {\bibinfo
  {volume} {6}},\ \bibinfo {pages} {357--381} (\bibinfo {year}
  {1982})}\BibitemShut {NoStop}%
\bibitem [{\citenamefont {Donaldson}\ and\ \citenamefont {Thomas}(1998)}]{DT}%
  \BibitemOpen
  \bibfield  {author} {\bibinfo {author} {\bibfnamefont {S.~K.}\ \bibnamefont
  {Donaldson}}\ and\ \bibinfo {author} {\bibfnamefont {R.~P.}\ \bibnamefont
  {Thomas}},\ }\bibfield  {title} {\enquote {\bibinfo {title} {Gauge theory in
  higher dimensions},}\ }in\ \href@noop {} {\emph {\bibinfo {booktitle} {The
  {G}eometric {U}niverse ({O}xford, 1996)}}}\ (\bibinfo  {publisher} {Oxford
  Univ. Press, Oxford},\ \bibinfo {year} {1998})\ pp.\ \bibinfo {pages}
  {31--47}\BibitemShut {NoStop}%
\bibitem [{\citenamefont {Bolza}(1887)}]{bolza1887}%
  \BibitemOpen
  \bibfield  {author} {\bibinfo {author} {\bibfnamefont {O.}~\bibnamefont
  {Bolza}},\ }\bibfield  {title} {\enquote {\bibinfo {title} {On {binary}
  {sextics} with {linear} {transformations} into {themselves}},}\ }\href
  {\doibase 10.2307/2369402} {\bibfield  {journal} {\bibinfo  {journal} {Am. J.
  Math.}\ }\textbf {\bibinfo {volume} {10}},\ \bibinfo {pages} {47--70}
  (\bibinfo {year} {1887})}\BibitemShut {NoStop}%
\bibitem [{\citenamefont {Deligne}\ and\ \citenamefont {Mumford}(1969)}]{DM}%
  \BibitemOpen
  \bibfield  {author} {\bibinfo {author} {\bibfnamefont {P.}~\bibnamefont
  {Deligne}}\ and\ \bibinfo {author} {\bibfnamefont {D.}~\bibnamefont
  {Mumford}},\ }\bibfield  {title} {\enquote {\bibinfo {title} {The
  irreducibility of the space of curves of given genus},}\ }\href
  {http://www.numdam.org/item?id=PMIHES_1969__36__75_0} {\bibfield  {journal}
  {\bibinfo  {journal} {Inst. Hautes \'{E}tudes Sci. Publ. Math.}\ }\textbf
  {\bibinfo {volume} {36}},\ \bibinfo {pages} {75--109} (\bibinfo {year}
  {1969})}\BibitemShut {NoStop}%
\bibitem [{\citenamefont {{Hladky-Hennion}}\ and\ \citenamefont
  {Decarpigny}(1991)}]{hladkyhennion1991}%
  \BibitemOpen
  \bibfield  {author} {\bibinfo {author} {\bibfnamefont {A.-C.}\ \bibnamefont
  {{Hladky-Hennion}}}\ and\ \bibinfo {author} {\bibfnamefont {J.-N.}\
  \bibnamefont {Decarpigny}},\ }\bibfield  {title} {\enquote {\bibinfo {title}
  {Analysis of the scattering of a plane acoustic wave by a doubly periodic
  structure using the finite element method: {Application} to {Alberich}
  anechoic coatings},}\ }\href {\doibase 10.1121/1.401395} {\bibfield
  {journal} {\bibinfo  {journal} {J. Acoust. Soc. Am.}\ }\textbf {\bibinfo
  {volume} {90}},\ \bibinfo {pages} {3356--3367} (\bibinfo {year}
  {1991})}\BibitemShut {NoStop}%
\bibitem [{\citenamefont {Langlet}\ \emph {et~al.}(1995)\citenamefont
  {Langlet}, \citenamefont {{Hladky-Hennion}},\ and\ \citenamefont
  {Decarpigny}}]{langlet1995}%
  \BibitemOpen
  \bibfield  {author} {\bibinfo {author} {\bibfnamefont {P.}~\bibnamefont
  {Langlet}}, \bibinfo {author} {\bibfnamefont {A.-C.}\ \bibnamefont
  {{Hladky-Hennion}}}, \ and\ \bibinfo {author} {\bibfnamefont {J.-N.}\
  \bibnamefont {Decarpigny}},\ }\bibfield  {title} {\enquote {\bibinfo {title}
  {Analysis of the propagation of plane acoustic waves in passive periodic
  materials using the finite element method},}\ }\href {\doibase
  10.1121/1.413244} {\bibfield  {journal} {\bibinfo  {journal} {J. Acoust. Soc.
  Am.}\ }\textbf {\bibinfo {volume} {98}},\ \bibinfo {pages} {2792--2800}
  (\bibinfo {year} {1995})}\BibitemShut {NoStop}%
\bibitem [{\citenamefont {Ballandras}\ \emph {et~al.}(2002)\citenamefont
  {Ballandras}, \citenamefont {Wilm}, \citenamefont {Edoa}, \citenamefont
  {Soufyane}, \citenamefont {Laude}, \citenamefont {Steichen},\ and\
  \citenamefont {Lardat}}]{ballandras2002}%
  \BibitemOpen
  \bibfield  {author} {\bibinfo {author} {\bibfnamefont {S.}~\bibnamefont
  {Ballandras}}, \bibinfo {author} {\bibfnamefont {M.}~\bibnamefont {Wilm}},
  \bibinfo {author} {\bibfnamefont {P.-F.}\ \bibnamefont {Edoa}}, \bibinfo
  {author} {\bibfnamefont {A.}~\bibnamefont {Soufyane}}, \bibinfo {author}
  {\bibfnamefont {V.}~\bibnamefont {Laude}}, \bibinfo {author} {\bibfnamefont
  {W.}~\bibnamefont {Steichen}}, \ and\ \bibinfo {author} {\bibfnamefont
  {R.}~\bibnamefont {Lardat}},\ }\bibfield  {title} {\enquote {\bibinfo {title}
  {Finite-element analysis of periodic piezoelectric transducers},}\ }\href
  {\doibase 10.1063/1.1524711} {\bibfield  {journal} {\bibinfo  {journal} {J.
  Appl. Phys.}\ }\textbf {\bibinfo {volume} {93}},\ \bibinfo {pages} {702--711}
  (\bibinfo {year} {2002})}\BibitemShut {NoStop}%
\bibitem [{\citenamefont {Liboff}(2004)}]{Liboff}%
  \BibitemOpen
  \bibfield  {author} {\bibinfo {author} {\bibfnamefont {R.~L.}\ \bibnamefont
  {Liboff}},\ }\href@noop {} {\emph {\bibinfo {title} {Primer for Point and
  Space Groups}}}\ (\bibinfo  {publisher} {Springer-Verlag},\ \bibinfo
  {address} {New York},\ \bibinfo {year} {2004})\BibitemShut {NoStop}%
\bibitem [{\citenamefont {Cari\~nena}\ \emph {et~al.}(2007)\citenamefont
  {Cari\~nena}, \citenamefont {Ra\~nada},\ and\ \citenamefont
  {Santander}}]{carinena2007}%
  \BibitemOpen
  \bibfield  {author} {\bibinfo {author} {\bibfnamefont {J.~F.}\ \bibnamefont
  {Cari\~nena}}, \bibinfo {author} {\bibfnamefont {M.~F.}\ \bibnamefont
  {Ra\~nada}}, \ and\ \bibinfo {author} {\bibfnamefont {M.}~\bibnamefont
  {Santander}},\ }\bibfield  {title} {\enquote {\bibinfo {title} {The quantum
  harmonic oscillator on the sphere and the hyperbolic plane},}\ }\href
  {\doibase 10.1016/j.aop.2006.10.010} {\bibfield  {journal} {\bibinfo
  {journal} {Ann. Phys. (N.Y.)}\ }\textbf {\bibinfo {volume} {322}},\ \bibinfo
  {pages} {2249--2278} (\bibinfo {year} {2007})}\BibitemShut {NoStop}%
\end{thebibliography}%

\end{document}


\newcommand{\tr}{\mathop{\mathrm{tr}}}
\newcommand{\bsigma}{\boldsymbol{\sigma}}
\newcommand{\re}{\mathop{\mathrm{Re}}}
\newcommand{\im}{\mathop{\mathrm{Im}}}
\renewcommand{\b}[1]{{\boldsymbol{#1}}}
\newcommand{\diag}{\mathrm{diag}}
\newcommand{\sign}{\mathrm{sign}}
\newcommand{\sgn}{\mathop{\mathrm{sgn}}}
\renewcommand{\c}[1]{\mathcal{#1}}
\newcommand{\Jac}{\mathop{\mathrm{Jac}}}
\newcommand{\Mob}{\mathop{\textrm{M\"ob}}}
\newcommand{\Aut}{\mathop{\mathrm{Aut}}}
\newcommand{\g}{\mathfrak{g}}
\renewcommand{\geq}{\geqslant}
\renewcommand{\leq}{\leqslant}

\newcommand{\mb}{\bm}
\newcommand{\ua}{\uparrow}
\newcommand{\da}{\downarrow}
\newcommand{\ra}{\rightarrow}
\newcommand{\la}{\leftarrow}
\newcommand{\mc}{\mathcal}                                                                                                                                                                                                                                                                                                                                                                                                                                                                                                                                                                                
\newcommand{\bs}{\boldsymbol}
\newcommand{\lra}{\leftrightarrow}
\newcommand{\nn}{\nonumber}
\newcommand{\half}{{\textstyle{\frac{1}{2}}}}
\newcommand{\mf}{\mathfrak}
\newcommand{\MF}{\text{MF}}
\newcommand{\IR}{\text{IR}}
\newcommand{\UV}{\text{UV}}

\renewcommand{\thetable}{S\arabic{table}}
\renewcommand{\thefigure}{S\arabic{figure}}
\renewcommand{\thesection}{S\arabic{section}}
\renewcommand{\thesubsection}{\arabic{subsection}}
\renewcommand{\theequation}{S\arabic{equation}}

\DeclareGraphicsExtensions{.png}

\title{Supplementary Materials for ``Hyperbolic band theory''}

\author{Joseph Maciejko}
\email{maciejko@ualberta.ca}
\affiliation{Department of Physics \& Theoretical Physics Institute (TPI), University of Alberta, Edmonton, Alberta T6G 2E1, Canada}
\author{Steven Rayan}
\email{rayan@math.usask.ca}
\affiliation{Department of Mathematics and Statistics\\\& Centre for Quantum Topology and Its Applications (quanTA), University of Saskatchewan, Saskatoon, Saskatchewan S7N 5E6, Canada}

\maketitle


\section{Geometry of the Bolza lattice}
\label{app:geom}

In this Section, we detail the geometry of the regular $\{8,8\}$ hyperbolic tessellation illustrated in Fig.~1(c) of the main text. This tessellation is composed of repeated regular hyperbolic octagons, eight of which meet at each vertex of the lattice. To fully describe the tessellation, it is sufficient to give the geometry of a reference unit cell $\c{D}$, here chosen to be the octagon centered at the origin $z=0$ of the Poincar\'e disk, and the $PSU(1,1)$ matrix representation of the noncommutative Fuchsian group $\Gamma$ that translates $\c{D}$ to cover the entire Poincar\'e disk.

To describe $\c{D}$, we give its boundary segments. These are (circular) geodesic arcs normal to the boundary at infinity $|z|=1$, and can be parametrized as follows:
\begin{align}\label{Cj}
C_j=\left\{z=e^{i(j-1)\frac{\pi}{4}}(c+re^{i\theta}),\frac{3\pi}{4}<\theta<\frac{5\pi}{4}\right\},\hspace{10mm}
j=1,\ldots,8,
\end{align}
where
\begin{align}\label{defcr}
c=\sqrt{\frac{3+2\sqrt{2}}{2+2\sqrt{2}}},\hspace{10mm}
r=\frac{1}{\sqrt{2+2\sqrt{2}}}.
\end{align}
The vertices of $\c{D}$ are given by
\begin{align}
p_j=2^{-1/4}e^{i(2j-1)\frac{\pi}{8}},\hspace{10mm}
j=1,\ldots,8.
\end{align}
An explicit $PSU(1,1)$ matrix representation of the four generators $\gamma_1$, $\gamma_2$, $\gamma_3$, $\gamma_4$ which generate $\Gamma$ is given in Ref.~[30]:
\begin{align}\label{BolzaFuchsian}
\gamma_j=\left(
\begin{array}{cc}
1+\sqrt{2} & (2+\sqrt{2})\lambda e^{i(j-1)\pi/4} \\
(2+\sqrt{2})\lambda e^{-i(j-1)\pi/4} & 1+\sqrt{2}
\end{array}\right),\hspace{10mm}
j=1,\ldots,4,
\end{align}
where $\lambda=\sqrt{\sqrt{2}-1}$.   One can check by explicit computation that the action of those generators identifies the boundary segments pairwise and in an orientation-preserving manner, as depicted in Fig.~1(c) of the main text:
\begin{align}\label{BolzaIdentifications}
\gamma_1(C_5)=C_1,\hspace{10mm}
\gamma_2(C_6)=C_2,\hspace{10mm}
\gamma_3(C_7)=C_3,\hspace{10mm}
\gamma_4(C_8)=C_4.
\end{align}
The inverse M\"obius transformations $\gamma_j^{-1}$ correspond simply to the matrix inverse of Eq.~(\ref{BolzaFuchsian}). The generators of $\Gamma$ obey the relation
\begin{align}\label{relator}
\gamma_1^{\phantom{}}\gamma_2^{-1}\gamma_3^{\phantom{}}\gamma_4^{-1}\gamma_1^{-1}\gamma_2^{\phantom{}}\gamma_3^{-1}\gamma_4^{\phantom{}}=\mathbb{I},
\end{align}
where $\mathbb{I}$ again denotes the identity. As shown in Sec.~\ref{app:bolza}, Eq.~(\ref{relator}) is precisely the presentation of the fundamental group $\pi_1(\Sigma_2)$ of a smooth, compact genus-2 surface, illustrated schematically in Fig.~1(d) of the main text, that is obtained by gluing together the opposite sides of the hyperbolic octagon $\c{D}$. This establishes the group isomorphism $\pi_1(\Sigma_2)\cong\Gamma$.  We emphasize that while all genus-$2$ topological surfaces share this fundamental group, the symbol $\Sigma_2$ refers not to \emph{any} genus-$2$ surface but specifically to the quotient $\mathbb H/\Gamma$ for the $\Gamma$ generated above. This quotient is precisely the side identification illustrated in Fig.~1(c) of the main text.  To use our earlier language, $\Sigma_2$ is a particular Riemann surface structure defined upon a topological genus-$2$ surface.  This Riemann surface is known traditionally as the \emph{Bolza surface}~[43].  Accordingly, we shall refer to the regular $\{8,8\}$ tessellation as the \emph{Bolza lattice}.  The Bolza surface is but one Riemann surface in an entire \emph{moduli space} of distinct Riemann surfaces of genus $2$~[44].  In general, there are $3g-3$ complex degrees of freedom to vary the Riemann surface structure whenever $g\geq2$---this is the dimension of the moduli space---while the underlying topological class has no freedom.  Of the points in the moduli space, the Bolza surface is the Riemann surface of genus $2$ with the largest possible automorphism group, reflecting the high degree of symmetry in the original lattice and its edge identifications. (We use automorphisms of Riemann surfaces to generalize the notion of point-group symmetries to hyperbolic lattices in Sec.~\ref{app:pointgroup}.)

\section{Fundamental group of the Bolza surface}
\label{app:bolza}

The Bolza surface $\Sigma_2$ is a Riemann surface of genus 2~[43] obtained by identifying the opposite sides of the hyperbolic octagon in Fig.~1(c) of the main text. Under this identification, the eight vertices $p_1,\ldots,p_8$ of the octagon are mapped to a single point $p_0$ which we can take as the base point for loops~[14]. Under the identification, each of the eight geodesic boundary segments $C_1,\ldots,C_8$ starts and ends at $p_0$ and is thus a closed loop; thus the $C_j$, $j=1,\ldots,8$ are elements of the fundamental group $\pi_1(\Sigma_2,p_0)$ based at $p_0$. Consider a closed path $C$ that starts at $p_1\sim p_0$ and goes around the boundary of the octagon counterclockwise. Denoting by $C_1,\ldots,C_8$ the oriented paths with orientations indicated by the arrows in Fig.~1(c) of the main text, and by $C_1^{-1},\ldots,C_8^{-1}$ the same paths traversed in reverse, $C$ is given by
\begin{align}
C=C_1C_2^{-1}C_3C_4^{-1}C_5^{-1}C_6C_7^{-1}C_8.
\end{align}
Now, $C$ can be continuously deformed to a point inside the octagon, thus it is homotopic to the trivial path: $C=1$. This remains true after identification. After identification, however, $C_j$ and $C_{j+4}$, $j=1,\ldots,4$ are identified in an orientation-preserving manner. Therefore, the unique relation satisfied by the distinct generators $C_1,C_2,C_3,C_4$ of $\pi(\Sigma_2,p_0)$ is:
\begin{align}\label{app:rel}
C_1C_2^{-1}C_3C_4^{-1}C_1^{-1}C_2C_3^{-1}C_4=1.
\end{align}
Since fundamental groups with different base points are isomorphic, we can simply write:
\begin{align}\label{app:pi1Sigma2}
\pi_1(\Sigma_2)=\{C_1,C_2,C_3,C_4:C_1C_2^{-1}C_3C_4^{-1}C_1^{-1}C_2C_3^{-1}C_4=1\}.
\end{align}

Isomorphic groups may have different, but equivalent, presentations. For example, one can give a different presentation for (\ref{app:pi1Sigma2}) as follows~[30]. Define new generators as $a_1=C_3$, $b_1=C_4^{-1}$, $a_2=C_1C_2^{-1}$, and $b_2=C_3C_4^{-1}C_1^{-1}$. Then using (\ref{app:rel}), one finds that the relation satisfied by these new generators is that quoted in the main text, i.e.,
\begin{align}
\pi_1(\Sigma_g)\cong\{a_1,b_1,\ldots,a_g,b_g:a_1b_1a_1^{-1}b_1^{-1}\cdots a_gb_ga_g^{-1}b_g^{-1}=1\},
\end{align}
with $g=2$. Correspondingly, the original generators can be obtained from the new ones by $C_1=b_2^{-1}a_1b_1$, $C_2=a_2^{-1}b_2^{-1}a_1b_1$, $C_3=a_1$, and $C_4=b_1^{-1}$. Since products of homotopy classes correspond to composition of loops, different choices of generators correspond to different choices of closed loops on the Riemann surface. In the section ``The Bolza lattice'' in the main text, our choice of representation $\chi:\pi_1(\Sigma_2)\rightarrow U(1)$,
\begin{align}\label{BlochFactorBolza}
\chi(\gamma_j)=\chi(\gamma_j^{-1})^*=e^{ik_j},\hspace{10mm}j=1,\ldots,4,
\end{align}
associates $k_1,k_2,k_3,k_4$ with the Aharonov-Bohm phase acquired upon traversing $C_1,C_2,C_3,C_4$, but the choice 
\begin{align}
\chi(a_i)=\chi(a_i^{-1})^*=e^{ik_a^{(i)}},\hspace{10mm}
\chi(b_i)=\chi(b_i^{-1})^*=e^{ik_b^{(i)}},\hspace{10mm}
i=1,\ldots,g,
\end{align}
with $a_1,b_1,a_2,b_2$ defined above, is equally valid. It can simply be thought of as the choice of a different basis for the hyperbolic reciprocal lattice, i.e., the $2g$-dimensional lattice $\Lambda$ such that the quotient $\mathbb{R}^{2g}\cong\mathbb{C}^g$ by it gives $\Jac(\Sigma_g)\cong T^{2g}$. We also note that since the representation $\chi$ is abelian, it automatically satisfies the group relation in $\pi_1(\Sigma_2)$.

\section{Twisted automorphic boundary conditions in the finite element method}
\label{app:FEM}

The FreeFEM++ package~[32] allows for Dirichlet, Neumann, or strictly periodic/automorphic boundary conditions, but not directly for the twisted automorphic boundary conditions required for nonzero $\b{k}$, i.e.,
\begin{align}\label{BolzaPBC}
\psi(C_j)=e^{ik_j}\psi(C_{j+4}),\hspace{10mm}j=1,\ldots,4.
\end{align}
To remedy this problem, we follow the approach of Refs.~[45--47], in which the stiffness and overlap matrices of the weak (variational) formulation of the hyperbolic Schr\"odinger equation [Eq.~(3) in the main text] are first computed using FreeFEM++ with unconstrained boundary conditions, and the number of physical degrees of freedom is subsequently reduced using simple matrix operations before proceeding to numerical diagonalization. The generalized Bloch phases are easily introduced at this second stage, as we now explain.

\begin{figure}[t]
\includegraphics[width=0.35\columnwidth]{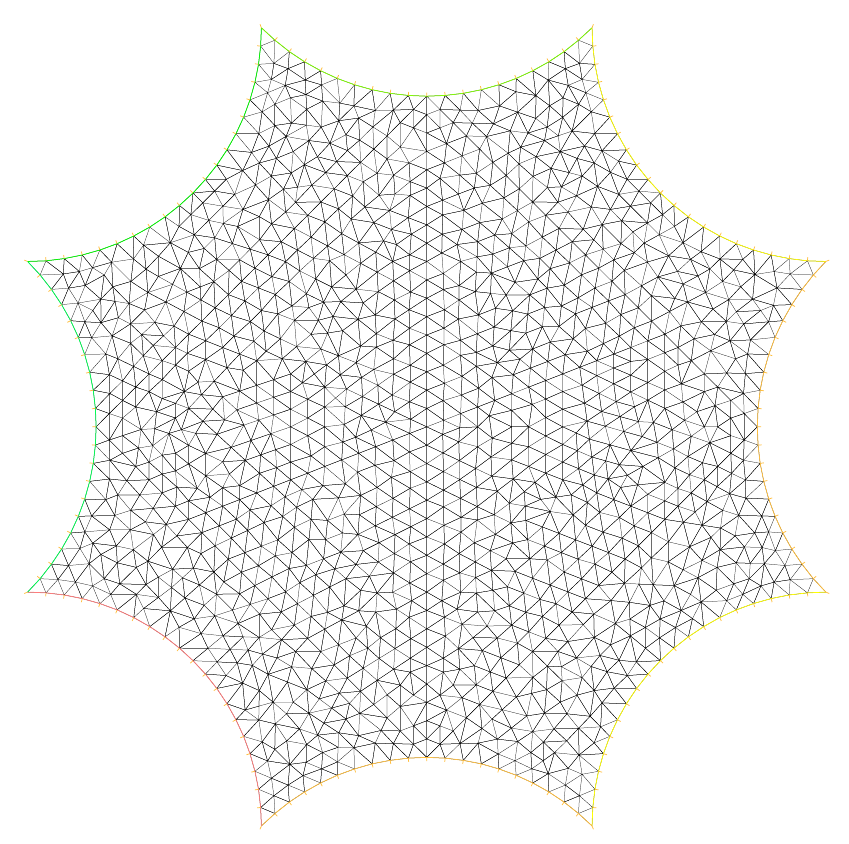}
\caption{A finite element triangulation of the hyperbolic octagon $\c{D}$ with 20 nodes per boundary segment.}
\label{fig:fem}
\end{figure}

Consider two functions $\phi,\psi$ obeying the twisted automorphic condition [Eq.~(2) in the main text]. The weak form of the Schr\"odinger equation is obtained by multiplying this equation by $\phi^*$ and integrating over the domain $\c{D}$,
\begin{align}\label{weak1}
\int_\c{D}d^2r\sqrt{g}\left(-\phi^*\Delta\psi+\phi^*V\psi\right)=E\int_\c{D}d^2r\sqrt{g}\phi^*\psi,
\end{align}
where $\sqrt{g}=4/(1-|z|^2)^2$ is the square root of the determinant of the Poincar\'e metric tensor $g_{\mu\nu}=4\delta_{\mu\nu}/(1-|z|^2)^2$, and $d^2r=dx\,dy$ is the usual Euclidean integration measure. Using Green's theorem and the automorphic Bloch condition, one can show that Eq.~(\ref{weak1}) becomes~[21]
\begin{align}
\int_\c{D}d^2r\sqrt{g}\left(g^{\mu\nu}\partial_\mu\phi^*\partial_\nu\psi+\phi^*V\psi\right)=E\int_\c{D}d^2r\sqrt{g}\phi^*\psi.
\end{align}
Using the inverse metric tensor $g^{\mu\nu}=\frac{1}{4}\delta^{\mu\nu}(1-|z|^2)^2$, we obtain
\begin{align}\label{weak2}
\int_\c{D}d^2r\left(\partial_x\phi^*\partial_x\psi+\partial_y\phi^*\partial_y\psi+\frac{4\phi^*V\psi}{(1-|z|^2)^2}\right)=E\int_\c{D} d^2r\frac{4\phi^*\psi}{(1-|z|^2)^2}. 
\end{align}
In the finite element method, one triangulates the region $\c{D}$ (e.g., Fig.~\ref{fig:fem}) and expands the solution $\psi$ on a basis of functions $u_i$, $i=1,\ldots,M$ with compact support on each finite element (i.e., triangle) in the triangulation. While $M$ is formally infinite, in practice, we truncate $\{u_i\}$ to the set of linear Lagrangian shape functions (P1 elements in the notation of Ref.~[32]). $M$ then equals the finite number of vertices in the triangulation, and the piecewise-linear basis function $u_i$ equals one on vertex $i$ and vanishes on all other vertices. While simple, the choice of P1 elements allows us to achieve satisfactory accuracy with a sufficiently fine triangulation. Expanding $\psi=\sum_{j=1}^M\psi_ju_j$ and taking $\phi=u_i$, Eq.~(\ref{weak2}) becomes
\begin{align}\label{ABeq}
\sum_{j=1}^M A_{ij}\psi_j = E\sum_{j=1}^M B_{ij}\psi_j,\hspace{10mm}
i=1,\ldots,M,
\end{align}
where the $M\times M$ Hermitian matrices $A$ (stiffness matrix) and $B$ (overlap matrix) are
\begin{align}
A_{ij}=\int_\c{D}d^2r\left(\partial_x u_i^* \partial_x u_j+\partial_y u_i^* \partial_y u_j
+\frac{4u_i^*Vu_j}{(1-|z|^2)^2}\right),\hspace{10mm}
B_{ij}=\int_\c{D}d^2r\frac{4u_i^*u_j}{(1-|z|^2)^2}.
\end{align}
The solution of the Schr\"odinger equation in continuous space is thus approximated by the solution of a finite-dimensional generalized eigenvalue problem, Eq.~(\ref{ABeq}).

We now explain how to impose the boundary conditions (\ref{BolzaPBC}). Since $i,j$ are vertex indices, the solution vector $\b{\psi}=(\psi_1,\ldots,\psi_M)$ can be written in block form as
\begin{align}
\b{\psi}=\left(\psi_{p_1},\ldots,\psi_{p_8};\b{\psi}_{C_1},\ldots,\b{\psi}_{C_8};\b{\psi}_\mathrm{bulk}\right),
\end{align}
in the notation of Fig.~1(c), where $p_1,\ldots,p_8$ denote vertices at the eight corners of the hyperbolic octagon; $\b{\psi}_{C_j}$, $j=1,\ldots,8,$ is a vector containing the solution on the set of vertices belonging to the boundary segment $C_j$ of the octagon, excluding corners; and $\b{\psi}_\mathrm{bulk}$ is a vector containing the solution on the interior vertices. The boundary conditions, which identify boundary segments (including corners) pairwise, imply that only $N<M$ degrees of freedom are in fact independent. There thus exists an $M\times N$ matrix $U$ such that $\b{\psi}=U\tilde{\b{\psi}}$, where $\tilde{\b{\psi}}$ is an $N$-dimensional vector containing only the independent degrees of freedom. Substituting this equation inside the $M$-dimensional generalized eigenvalue problem $A\b{\psi}=EB\b{\psi}$ in (\ref{ABeq}), and left-multiplying by $U^\dag$, we obtain the reduced, $N$-dimensional generalized eigenvalue problem:
\begin{align}\label{redABeq}
\tilde{A}\tilde{\b{\psi}}=E\tilde{B}\tilde{\b{\psi}},
\end{align}
where
\begin{align}
\tilde{A}=U^\dag AU,\hspace{10mm}
\tilde{B}=U^\dag BU,
\end{align}
are again Hermitian.

The $M\times N$ matrix $U$ is constructed as follows. First, the bulk vertices are unaffected by the boundary conditions, thus $\b{\psi}_\mathrm{bulk}$ appears in full in $\tilde{\b{\psi}}$. Next, out of the eight boundary vectors $\b{\psi}_{C_1},\ldots,\b{\psi}_{C_8}$, only four are linearly independent due to the boundary conditions. Using Eq.~(\ref{BolzaPBC}), we can thus express all eight boundary vectors in terms of $\b{\psi}_{C_5},\ldots,\b{\psi}_{C_8}$:
\begin{align}\label{reduce1}
\left(\begin{array}{c}
\b{\psi}_{C_1} \\
\b{\psi}_{C_2} \\
\b{\psi}_{C_3} \\
\b{\psi}_{C_4} \\
\b{\psi}_{C_5} \\
\b{\psi}_{C_6} \\
\b{\psi}_{C_7} \\
\b{\psi}_{C_8}
\end{array}\right)=\left(\begin{array}{cccc}
e^{ik_1} & & & \\
& e^{ik_2} & & \\
& & e^{ik_3} & \\
& & & e^{ik_4} \\
1 & & & \\
& 1 & & \\
& & 1 & \\
& & & 1
\end{array}\right)
\left(\begin{array}{c}
\b{\psi}_{C_5} \\
\b{\psi}_{C_6} \\
\b{\psi}_{C_7} \\
\b{\psi}_{C_8}
\end{array}\right),
\end{align}
making also sure that the components of the respective vectors are ordered so as the preserve orientation under pairwise identification [see Fig.~1(c) in the main text]. Finally, all eight corner vertices $p_1,\ldots,p_8$ collapse under this identification to a single point, which can be chosen as $p_5$. Using the action of the Fuchsian group generators depicted in Fig.~1(c) of the main text, we can express the remaining vertices as:
\begin{align}
p_1&=\gamma_1\gamma_4\gamma_3^{-1}\gamma_2(p_5),\hspace{10mm}
p_2=\gamma_2(p_5),\hspace{10mm}
p_3=\gamma_4\gamma_1(p_5),\hspace{10mm}
p_4=\gamma_4\gamma_3^{-1}\gamma_2(p_5),\nn\\
p_6&=\gamma_3^{-1}\gamma_4\gamma_1(p_5),\hspace{10mm}
p_7=\gamma_3^{-1}\gamma_2(p_5),\hspace{10mm}
p_8=\gamma_1(p_5).
\end{align}
Note that this choice of representation is not unique, but different representations can be shown to be equivalent using the relation (\ref{relator}). Using the generalized Bloch factor (\ref{BlochFactorBolza}), we can thus write
\begin{align}\label{reduce2}
\left(\begin{array}{c}
\psi_{p_1} \\
\psi_{p_2} \\
\psi_{p_3} \\
\psi_{p_4} \\
\psi_{p_5} \\
\psi_{p_6} \\
\psi_{p_7} \\
\psi_{p_8}
\end{array}\right)=\left(\begin{array}{c}
e^{i(k_1+k_2-k_3+k_4)} \\
e^{ik_2} \\
e^{i(k_1+k_4)} \\
e^{i(k_2-k_3+k_4)} \\
1 \\
e^{i(k_1-k_3+k_4)} \\
e^{i(k_2-k_3)} \\
e^{ik_1}
\end{array}\right)\psi_{p_5}.
\end{align}
Defining the reduced vector $\tilde{\b{\psi}}$ as
\begin{align}
\tilde{\b{\psi}}=\left(\psi_{p_5};\b{\psi}_{C_5},\ldots,\b{\psi}_{C_8};\b{\psi}_\mathrm{bulk}\right),
\end{align}
and using Eqs.~(\ref{reduce1}) and (\ref{reduce2}), as well as an identity matrix for $\b{\psi}_\mathrm{bulk}$, the matrix $U$ can be easily constructed. Solving the generalized eigenvalue problem (\ref{redABeq}) numerically for each $\b{k}$ in the hyperbolic Brillouin zone, using standard linear algebra techniques, we obtain the hyperbolic bandstructure $\{E_n(\b{k})\}$ and hyperbolic Bloch wavefunctions $\psi_{n\b{k}}(x,y)$.

\section{Point-group symmetries of the Bolza lattice}
\label{app:pointgroup}

In ordinary crystallography, the group of all discrete symmetries of a Euclidean (periodic) lattice constitutes the space group $\c{G}$. The translation group $\c{T}$ is a normal subgroup of $\c{G}$, that is, if $h\in\c{G}$, then $h\c{T}h^{-1}=\c{T}$. The point group $\c{P}$ is the factor group $\c{P}\cong\c{G}/\c{T}$, i.e., space group operations with translations factored out~[48]. Similarly, the group $G=\Aut(X)$ of automorphisms of a compact Riemann surface $X\cong\mathbb{H}/\Gamma$ with $\Gamma$ a co-compact, strictly hyperbolic Fuchsian group is isomorphic to the factor group~[18]
\begin{align}\label{FactorGroup}
G\cong N(\Gamma)/\Gamma,
\end{align}
where $N(\Gamma)$ is the normalizer of $\Gamma$ in $PSL(2,\mathbb{R})$:
\begin{align}\label{normalizer}
N(\Gamma)=\{h\in PSL(2,\mathbb{R}):h\Gamma h^{-1}=\Gamma\}.
\end{align}
By analogy with Euclidean crystallography, it is natural to intepret $N(\Gamma)$ as a hyperbolic space group, its normal subgroup $\Gamma$ as a (nonabelian) translation group, and the factor group (\ref{FactorGroup}) as a point group. Note that in Riemann surface theory~[18], one is typically interested in orientation-preserving automorphisms, thus $N(\Gamma)$ is defined as the normalizer of $\Gamma$ in {$\Mob^+\cong PSL(2,\mathbb{R})$}, the group of orientation-preserving {M\"obius} transformations. Here we consider all automorphisms~[33], both {$\Mob^+$} and orientation-reversing ones ({$\Mob^-$}), and $N(\Gamma)$ is more properly defined as the normalizer of $\Gamma$ in the general {M\"obius} group {$\Mob$}.

In this Section, we describe the group $G\cong\Aut(\Sigma_2)$ of automorphisms of the Bolza surface (Sec.~\ref{app:Aut}), interpreting it as a point group, and construct its linear action
\begin{align}\label{kg}
\b{k}\rightarrow\b{k}^{h}=M(h)\b{k},\hspace{10mm}{h}\in G,
\end{align}
on the Jacobian, i.e., hyperbolic $\b{k}$-space (Sec.~\ref{app:AutJac}).

\subsection{Automorphisms of the Bolza surface}
\label{app:Aut}

The automorphism group $G$ of the Bolza surface is a finite nonabelian group with 96 elements, of the form~[33]
\begin{align}
R^iS^jT^kU^l,\hspace{10mm}i=0,\ldots,7,\hspace{10mm}j,k=0,1,\hspace{10mm}l=0,1,2,
\end{align}
where $R,S,T,U$ are four generators. $R$ and $U$ are orientation-preserving hyperbolic isometries, i.e., M\"obius transformations of the form
\begin{align}\label{mobius}
z\rightarrow \gamma(z)=\frac{\alpha z+\beta}{\beta^*z+\alpha^*}.
\end{align}
Working in the Poincar\'e disk, the group $\Mob^+$ of orientation-preserving M\"obius transformations is isomorphic to $PSU(1,1)$. $S$ and $T$ are reflections, and thus orientation-reversing isometries, which are again $PSU(1,1)$ transformations, but of the form~[16]
\begin{align}\label{mobiusConj}
z\mapsto\gamma(z)=\frac{\alpha z^*+\beta}{\beta^*z^*+\alpha^*},
\end{align}
which is sometimes called an anti-M\"obius or $\Mob^-$ transformation, and is the composition of an ordinary M\"obius transformation with complex conjugation. Together, $\Mob^+$ and $\Mob^-$ transformations form the general M\"obius group, $\Mob$, which is the full group $\mathrm{Isom}(\mathbb{H})$ of isometries of the hyperbolic plane. In the next subsections we figure out the explicit form of the generators $R,S,T,U$ of $G$ as $PSU(1,1)$ matrices. We first consider the $\Mob^+$ generators $R$ and $U$, then the $\Mob^-$ generators $S$ and $T$.

\begin{figure}[t]
\includegraphics[width=0.5\columnwidth]{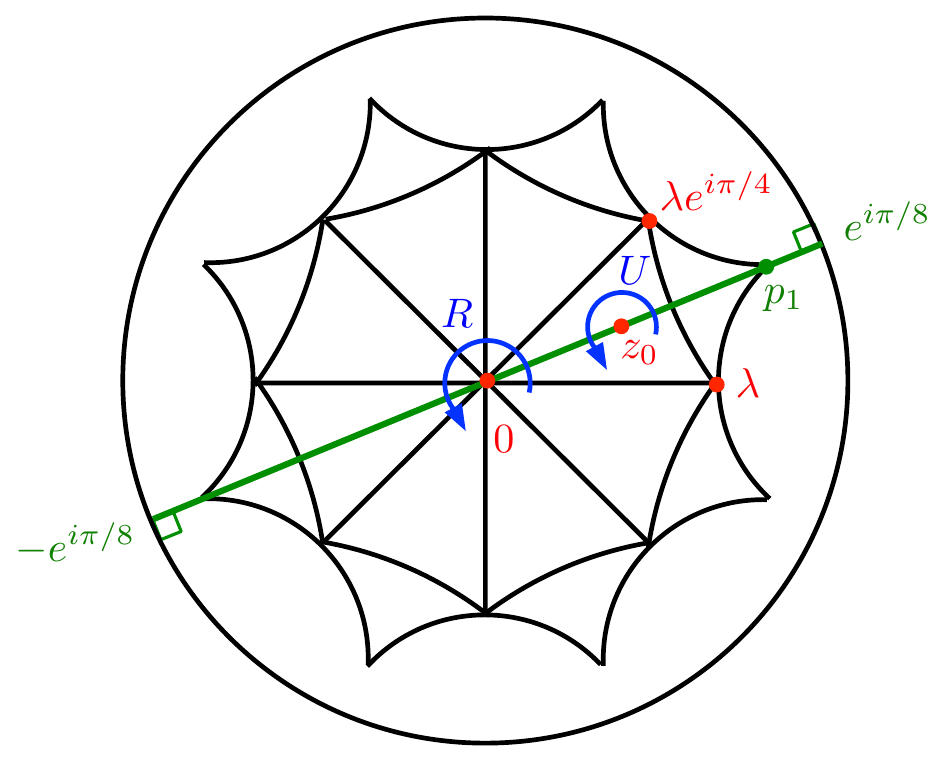}
\caption{Automorphisms of the Bolza surface (adapted from Ref.~[33] with permission of the author).}
\label{fig:aut}
\end{figure}

\emph{Transformation $R$.} The transformation $R$ is a $C_8$ rotation $z\mapsto R(z)=e^{i2\pi/8}z$ about the center of the octagonal unit cell (Fig.~\ref{fig:aut}), which corresponds to the $SU(1,1)$ matrix
\begin{align}
R=\left(\begin{array}{cc}
e^{i\pi/8} & 0 \\
0 & e^{-i\pi/8}
\end{array}\right),
\end{align}
up to an overall sign.

\emph{Transformation $U$.} This is a $C_3$ rotation around the center $z_0$ of a hyperbolic triangle, e.g., that formed by $0$, $\lambda$, and $\lambda e^{i\pi/4}$ in Fig.~\ref{fig:aut}, where $\lambda=\sqrt{\sqrt{2}-1}$ (see Sec.~\ref{app:geom}). The latter two vertices are found by considering that they are the midpoint of the boundary segments $C_1$ and $C_2$, respectively, which corresponds to the point $\theta=\pi$ in the parametrization (\ref{Cj}). The midpoints of $C_1$ and $C_2$ are thus $c-r$ and $e^{i\pi/4}(c-r)$, respectively, with $c$ and $r$ defined in Eq.~(\ref{defcr}), and $c-r=\lambda$.

We first need to find the center $z_0$ of that triangle. By symmetry, we expect $z_0$ to lie on the straight geodesic (green line in Fig.~\ref{fig:aut}) bisecting the triangle, i.e., $z_0=ae^{i\pi/8}$ with $0<a<\lambda$. We determine $z_0$ by requiring that it is the point of equal geodesic distance $d(z,z')$ from the three vertices of the triangle: $d(z_0,0)=d(z_0,\lambda)=d(z_0,\lambda e^{i\pi/8})$. By symmetry, we see that the last equality is automatically satisfied, thus we only need to impose the first one to find $a$. The geodesic distance on the Poincar\'e disk is given by~[16]
\begin{align}
\cosh d(z,z')=1+\frac{2|z-z'|^2}{(1-|z|^2)(1-|z'|^2)}.
\end{align}
Imposing the condition $d(z_0,z_1)=d(z_0,z_2)$ is thus the same as imposing
\begin{align}
\frac{|z_0-z_1|^2}{1-|z_1|^2}=\frac{|z_0-z_2|^2}{1-|z_2|^2}.
\end{align}
Setting $z_1=0$ and $z_2=\lambda$, and after a bit of algebra, we obtain
\begin{align}
a^2-2^{1/4}(1+\sqrt{2})a+1=0.
\end{align}
Keeping the only root satisfying $0<a<\lambda\approx 0.64$, we obtain
\begin{align}
z_0=\left(\frac{1+\sqrt{2}-\sqrt{3}}{2^{3/4}}\right)e^{i\pi/8}.
\end{align}

Now, $U$ is a rotation by $2\pi/3$ around $z_0$. We do not directly know what this looks like, but we know that a rotation around $z=0$ is $C_3:z\mapsto e^{i2\pi/3}z$. To obtain $U$, we simply have to ``translate'' (boost) $z_0$ to the origin, perform $C_3$, then boost back to $z_0$. In other words, $U=\gamma_\eta\circ C_3\circ\gamma_\eta^{-1}$ where $\gamma_\eta$ is a boost from $z=0$ to $z=z_0$, i.e., along the green geodesic joining $-e^{i\pi/8}$ to $e^{i\pi/8}$. To find $\gamma_\eta$, first consider a boost $\tilde{\gamma}_\eta$ along the $x$ axis. Such a boost by a quantity in $-\infty<\eta<\infty$ is given by~[16]
\begin{align}
\tilde{\gamma}_\eta:z\mapsto\tilde{\gamma}_\eta(z)=\frac{(\cosh\eta)z+\sinh\eta}{(\sinh\eta)z+\cosh\eta},
\end{align}
corresponding to a $PSU(1,1)$ transformation with $\alpha=\cosh\eta$, $\beta=\sinh\eta$. The fixed points of this transformation are the points at infinity $\pm 1$, and the origin is boosted to $\tilde{\gamma}_\eta(0)=\tanh\eta$. For $\eta>0$ this is a boost towards the positive $x$ axis. To boost along the green line, one should rotate down to the $x$ axis, boost along the $x$ axis, then rotate back to the green line. In other words, $\gamma_\eta=R_{\pi/8}\circ\tilde{\gamma}_\eta\circ R_{\pi/8}^{-1}$, where $R_{\pi/8}:z\mapsto e^{i\pi/8}z$ is a counterclockwise ($C_{16}$) rotation by $\pi/8$. We thus obtain
\begin{align}
\gamma_\eta(z)=(R_{\pi/8}\circ\tilde{\gamma}_\eta\circ R_{\pi/8}^{-1})(z)
=e^{i\pi/8}\left(\frac{(\cosh\eta)e^{-i\pi/8}z+\sinh\eta}{(\sinh\eta)e^{-i\pi/8}z+\cosh\eta}\right)
=\frac{(\cosh\eta)z+e^{i\pi/8}\sinh\eta}{(e^{-i\pi/8}\sinh\eta)z+\cosh\eta},
\end{align}
a $\Mob^+$ transformation with $\alpha=\cosh\eta$, $\beta=e^{i\pi/8}\sinh\eta$. The points at infinity $\pm e^{i\pi/8}$ are fixed points of this transformation.

We now need to find the value of $\eta$ such that $\gamma_\eta(0)=z_0$. This implies
\begin{align}
\tanh\eta=\frac{1+\sqrt{2}-\sqrt{3}}{2^{3/4}}.
\end{align}
Writing $\sinh\eta=(1+\sqrt{2}-\sqrt{3})b$ and $\cosh\eta=2^{3/4}b$, and finding $b$ using $\cosh^2\eta-\sinh^2\eta=1$, we find that $\gamma_\eta$ is a $\Mob^+$ transformation with
\begin{align}
\alpha=\sqrt{\frac{\sqrt{3}+\sqrt{6}+3}{6}}>1,\hspace{5mm}\beta=e^{i\pi/8}\sqrt{\alpha^2-1}.
\end{align}
Finally, to obtain $U$ we conjugate $C_3$ by $\gamma_\eta$. In $SU(1,1)$ this is simply a product of matrices, and we find that $U$ is a $\Mob^+$ transformation with
\begin{align}
\alpha=e^{i3\pi/8}\sqrt{1+\frac{1}{\sqrt{2}}},\hspace{5mm}\beta=2^{-1/4}e^{-i3\pi/8}.
\end{align}

\emph{Transformation $S$.} The transformation $S$ is defined in Ref.~[33] as a reflection across the green line in Fig.~\ref{fig:aut}. Define $S'$ to be the reflection across the real axis, which sends $(x,y)\mapsto(x,-y)$ and is thus simply complex conjugation: $S':z\mapsto z^*$, the simplest $\Mob^-$ transformation. A reflection $S$ across the green line can be obtained by first rotating by $\pi/8$ clockwise to the real axis, performing $S'$, and then rotating back to the green line. We thus have:
\begin{align}
S(z)=(R_{\pi/8}\circ S'\circ R_{\pi/8}^{-1})(z)=e^{i\pi/8}(e^{-i\pi/8}z)^*=e^{i2\pi/8}z^*,
\end{align}
a transformation of the form (\ref{mobiusConj}) with $\alpha=e^{i\pi/8}$, $\beta=0$.

\emph{Transformation $T$.} According to Ref.~[33], $T$ is a $\Mob^-$ transformation which interchanges the origin $z=0$ with a corner of the hyperbolic octagon, e.g., $p_1=2^{-1/4}e^{i\pi/8}$. Since it is a simple interchange, one should have $T^2=e$ where $e$ denotes the identity element in $\Mob^+$. (Note that the composition of two $\Mob^-$ transformations is in $\Mob^+$, and the composition of a $\Mob^+$ transformation and a $\Mob^-$ transformation is a $\Mob^-$ transformation.) Thus we have
\begin{align}
T(z)=\frac{\alpha z^*+\beta}{\beta^*z^*+\alpha^*},
\end{align}
and wish to determine $\alpha$ and $\beta$. Imposing $T(0)=p_1$ implies $\beta=2^{-1/4}e^{i\pi/8}\alpha^*$. Next, since $T^2=e$ one has $T^2(0)=T(T(0))=T(p_1)=0$, which implies $\alpha e^{-i\pi/8}+\alpha^* e^{i\pi/8}=0$. Since $\alpha\neq 0$, writing in polar form $\alpha=s e^{i\theta}$, we find $\theta=5\pi/8$. Finally, requiring that $|\alpha|^2-|\beta|^2=1$, we find $s=2^{1/4}\lambda^{-1}=\sqrt{2+\sqrt{2}}$. Thus we find that $T$ is a transformation of the form (\ref{mobiusConj}) with
\begin{align}
\alpha=e^{i 5\pi/8}\sqrt{2+\sqrt{2}},\hspace{5mm}\beta=e^{-i\pi/2}\sqrt{1+\sqrt{2}}.
\end{align}
Noting by explicit calculation that the composition $\gamma_1\circ\gamma_2$ of two $\Mob^-$ transformations $\gamma_1,\gamma_2$ with $SU(1,1)$ parameters $\alpha_1,\beta_1$ and $\alpha_2,\beta_2$ is a $\Mob^+$ transformation with $SU(1,1)$ parameters $\alpha_{12},\beta_{12}$ given by
\begin{align}\label{compMobminus}
\alpha_{12}=\alpha_1\alpha_2^*+\beta_1\beta_2,\hspace{5mm}
\beta_{12}=\alpha_1\beta_2^*+\beta_1\alpha_2,
\end{align}
we find that the $SU(1,1)$ parameters of $T^2$ are $\alpha_{T^2}=|\alpha|^2+\beta^2=1$ and $\beta_{T^2}=\alpha(\beta+\beta^*)=0$, and thus $T^2=e$.

\emph{Generator relations.} According to Ref.~[33], the automorphism group of the Bolza surface admits a presentation with generators $R,S,T,U$ and relations
\begin{align}
R^8&=S^2=T^2=U^3=(RS)^2=(ST)^2=RTR^3T=e,\label{rel1}\\
UR&=R^7U^2,\\
U^2R&=STU,\\
US&=SU^2,\\
UT&=RSU.\label{rel5}
\end{align}
Using the explicit form of the generators just derived, we verify that all relations hold as identities in $\Mob$, except $RTR^3T=e$. We find
\begin{align}
RTR^3T=\gamma_3\gamma_4^{-1}\gamma_1^{-1}\in\Gamma,
\end{align}
where $\gamma_1,\gamma_2,\gamma_3,\gamma_4$ are the Fuchsian group generators (\ref{BolzaFuchsian}), thus $RTR^3T$ indeed reduces to the identity in $G=\Aut(\Sigma_2)\cong N(\Gamma)/\Gamma$. Similarly, Ref.~[33] states that $Z(G)$, the center of $G$ (i.e., the set of elements that commute with every element of $G$), is isomorphic to $\mathbb{Z}_2$ and generated by $R^4:z\mapsto-z$. To verify this, it is sufficient to verify that $R^4$ commutes with $S,T,U$. Commutation with $S$ holds as an identity in $\Mob$, while for the remaining two generators we find
\begin{align}
\gamma_1 R^4U=UR^4,\hspace{5mm}\gamma_1\gamma_4\gamma_3^{-1}\gamma_2 R^4T=TR^4,
\end{align}
which indeed reduce to $R^4U=UR^4$ and $R^4T=TR^4$ in the quotient $G\cong N(\Gamma)/\Gamma$.

\subsection{Point-group action on hyperbolic \texorpdfstring{$\b{k}$}{k}-space}
\label{app:AutJac}

In the previous section, we explicitly constructed the action of the point group $G$ in real space. In this section, we determine how $G$ acts in four-dimensional $\b{k}$-space, via its action on the automorphic Bloch eigenstates.

Given a general (orientation-preserving or -reversing) transformation $\gamma\in\Mob$, define a linear operator $\c{S}_\gamma$ which performs the corresponding transformation on a wavefunction $\psi(x,y)=\psi(z)$:
\begin{align}\label{Sgamma}
\c{S}_\gamma\psi(z)\equiv\psi(\gamma(z)).
\end{align}
For the derivation that follows, it will be useful to treat $z$ and $z^*$ as independent ``real'' variables, and the transformation (\ref{Sgamma}) will be written as $\c{S}_\gamma\psi(z,z^*)=\psi(\gamma(z),\gamma(z)^*)$.

\emph{Point-group symmetries of the Hamiltonian.} We assume the potential $V$ is not only automorphic with respect to the Fuchsian group $\Gamma$ of hyperbolic lattice translations, but also invariant under point-group transformations,
\begin{align}
\c{S}_h V(z,z^*)\c{S}_h^{-1}=V(h(z),h(z)^*)=V(z,z^*),\hspace{5mm}\forall h\in G\subset\Mob.
\end{align}
In other words, $[\c{S}_h,V]=0$ for all $h\in G$. We also want to show the kinetic term $H_0=-\Delta$ commutes with $\c{S}_h$, where
\begin{align}
\Delta=(1-zz^*)^2\frac{\partial^2}{\partial z\partial z^*},
\end{align}
is the hyperbolic Laplacian [Eq.~(1) in the main text], where we have used $\partial_x=\partial_z+\partial_{z^*}$ and $\partial_y=i(\partial_z-\partial_{z^*})$. We can in fact show that $\Delta$ is invariant under the action of any $\gamma\in\Mob$. First, $\Delta$ is obviously invariant under complex conjugation $z\mapsto z^*$. Since any $\Mob^-$ transformation can be written as the composition of a $\Mob^+$ transformation with complex conjugation, it is sufficient to show that $\Delta$ is invariant under $\Mob^+$ transformations. Consider an arbitrary $\Mob^+$ transformation,
\begin{align}\label{ztow}
z\mapsto w\equiv\gamma(z)=\frac{\alpha z+\beta}{\beta^*z+\alpha^*},\hspace{5mm}
z^*\mapsto w^*\equiv\gamma(z)^*=\frac{\alpha^* z^*+\beta^*}{\beta z^*+\alpha}.
\end{align}
We have
\begin{align}
(\Delta\c{S}_\gamma)\psi(z,z^*)&=(1-zz^*)^2\frac{\partial^2}{\partial z\partial z^*}\psi(w,w^*)\nn\\
&=(1-zz^*)^2\frac{\partial w}{\partial z}\frac{\partial w^*}{\partial z^*}\frac{\partial^2}{\partial w\partial w^*}\psi(w,w^*)\nn\\
&=(1-zz^*)^2\left|\frac{\partial w}{\partial z}\right|^2\frac{\partial^2}{\partial w\partial w^*}\psi(w,w^*).
\end{align}
On the other hand, we have
\begin{align}
(\c{S}_\gamma\Delta)\psi(z,z^*)=\c{S}_\gamma\left((1-zz^*)^2\frac{\partial^2}{\partial z\partial z^*}\psi(z,z^*)\right)=(1-ww^*)^2\frac{\partial^2}{\partial w\partial w^*}\psi(w,w^*).
\end{align}
Using (\ref{ztow}), we have
\begin{align}
1-ww^*=\frac{1-zz^*}{|\beta^* z+\alpha^*|^2}.
\end{align}
On the other hand, we have
\begin{align}
\frac{\partial w}{\partial z}=\frac{1}{(\beta^* z+\alpha^*)^2},
\end{align}
thus $[\c{S}_\gamma,\Delta]=0$ for any $\gamma\in\Mob$, and in particular for $\gamma\in G$. As a result, $[\c{S}_h,H]=0$ for any $h\in G$, with $H=-\Delta+V$ the full Hamiltonian.

\emph{Point-group symmetries of hyperbolic Bloch eigenstates.} We now go back to treating $z=x+iy\in\mathbb{C}$ as a complex coordinate in the Poincar\'e disk. Consider an eigenstate $\psi_\b{k}(z)$ of $H$ with energy $E(\b{k})$, that obeys the four automorphic Bloch conditions (\ref{BolzaPBC}):
\begin{align}\label{BC}
\psi_\b{k}(\gamma_j(z))=e^{ik_j}\psi_\b{k}(z),\hspace{5mm}j=1,\ldots,4.
\end{align}
with $\gamma_1,\ldots,\gamma_4$ in Eq.~(\ref{BolzaFuchsian}). Since $\c{S}_h$ with $h\in G$ commutes with $H$, the state
\begin{align}
\psi_\b{k}^h(z)\equiv\c{S}_h\psi_\b{k}(z)=\psi_\b{k}(h(z)),
\end{align}
for a given $h\in G$ is also an eigenstate of $H$ with the same eigenenergy $E(\b{k})$. However, it does not in general obey the same Bloch conditions as $\psi_\b{k}(z)$. Indeed, first write Eq.~(\ref{BC}) as
\begin{align}
\c{S}_{\gamma_j}\psi_\b{k}(z)=e^{ik_j}\psi_\b{k}(z),\hspace{5mm}j=1,\ldots,4.
\end{align}
Acting with $\c{S}_h$ on both sides and inserting the identity in the form $\c{S}_h^{-1}\c{S}_h=\mathbb{I}$, where the defining action of the inverse operator is
\begin{align}
\c{S}_\gamma^{-1}\psi(z)=\psi(\gamma^{-1}(z)),\hspace{5mm}\gamma\in\Mob,
\end{align}
and $\mathbb{I}$ is the identity operator, we have
\begin{align}
\c{S}_h\c{S}_{\gamma_j}\c{S}_h^{-1}\psi_\b{k}^h(z)=e^{ik_j}\psi_\b{k}^h(z),\hspace{5mm}j=1,\ldots,4.
\end{align}
In other words, $\psi_\b{k}^h$ obeys the modified Bloch conditions
\begin{align}\label{ModBC}
\psi_\b{k}^h((h\gamma_jh^{-1})(z))=e^{ik_j}\psi_\b{k}^h(z),\hspace{5mm}j=1,\ldots,4.
\end{align}

The modified boundary conditions (\ref{ModBC}) involve the conjugation of the Fuchsian group generator $\gamma_j\in\Gamma$ by point-group elements $h\in G$. Given Eqs.~(\ref{FactorGroup}-\ref{normalizer}), $h\gamma_jh^{-1}$ is again necessarily in $\Gamma$. (Although $h$ can be either in $\Mob^+$ or $\Mob^-$, $h\gamma_jh^{-1}$ is necessarily in $\Mob^+$, since the inverse of an element of $\Mob^-$ is also in $\Mob^-$, and only an even number (which could be zero) of $\Mob^-$ transformations appear in $h\gamma_j h^{-1}$. Therefore the boundary conditions obeyed by $\psi_\b{k}^h$ preserve orientation, as those for $\psi_\b{k}$.) From the explicit forms of the four generators $h=R,S,T,U$ of $G$, determined earlier, and the four generators $\gamma_j$ in Eq.~(\ref{BolzaFuchsian}), we can compute $h\gamma_j h^{-1}$. We display the result of these computations in Table~\ref{tab:conjug}, with $h\in\{R,S,T,U\}$ as the row index and $\gamma_j\in\{\gamma_1,\gamma_2,\gamma_3,\gamma_4\}$ as the column index. These expressions in terms of the $\gamma_j$ are not unique, since one can use the relation (\ref{relator}) to obtain different (but equivalent) expressions. Furthermore, since $h\gamma_j^{-1}h^{-1}=(h\gamma_jh^{-1})^{-1}$, the conjugated inverse generators $h\gamma_j^{-1}h^{-1}$ are easily determined from Table~\ref{tab:conjug} as well.

\begin{table}[t]
\begin{tabular}{|l||c|c|c|c|}
\hline 
& $\gamma_1$ & $\gamma_2$ & $\gamma_3$ & $\gamma_4$
 \tabularnewline
\hline 
\hline 
$R$ & $\gamma_2$ & $\gamma_3$ & $\gamma_4$ & $\gamma_1^{-1}$ \\
\hline 
$S$ & $\gamma_2$ & $\gamma_1$ & $\gamma_4^{-1}$ & $\gamma_3^{-1}$ \\
\hline
$T$ & $\gamma_4^{-1}\gamma_3\gamma_2^{-1}$ & $\gamma_3\gamma_4^{-1}\gamma_1^{-1}$ & $\gamma_2\gamma_4^{-1}\gamma_1^{-1}$ & $\gamma_2\gamma_3^{-1}\gamma_1^{-1}$ \\
\hline
$U$ & $\gamma_2\gamma_1^{-1}$ & $\gamma_1^{-1}$ & $\gamma_4^{-1}\gamma_1^{-1}$ & $\gamma_4^{-1}\gamma_3\gamma_2^{-1}$ \\
\hline
\end{tabular}
\caption{Conjugation of Fuchsian group generators by point-group symmetries.}\label{tab:conjug}
\end{table}

Using the results above, we can now show that $\psi_\b{k}^h$ corresponds to a hyperbolic Bloch eigenstate with a transformed wavevector $\b{k}^h$, i.e.,
\begin{align}\label{BlochRotated}
\psi_\b{k}^h(\gamma_j(z))=e^{ik_j^h}\psi_\b{k}^h(z),\hspace{5mm}j=1,\ldots,4.
\end{align}
To show this, we first observe that for a given point-group generator $h$, the relations in Table~\ref{tab:conjug} can be inverted to express any original generator $\gamma_j$ in terms of the conjugated generators $\gamma_j^h\equiv h\gamma_jh^{-1}$. We find
\begin{align}
\gamma_1&=(\gamma_4^R)^{-1}=\gamma_2^S=(\gamma_4^T)^{-1}\gamma_3^T(\gamma_2^T)^{-1}=(\gamma_2^U)^{-1},\\
\gamma_2&=\gamma_1^R=\gamma_1^S=\gamma_3^T(\gamma_4^T)^{-1}(\gamma_1^T)^{-1}=\gamma_1^U(\gamma_2^U)^{-1},\\
\gamma_3&=\gamma_2^R=(\gamma_4^S)^{-1}=\gamma_2^T(\gamma_4^T)^{-1}(\gamma_1^T)^{-1}=\gamma_1^U\gamma_4^U(\gamma_3^U)^{-1},\\
\gamma_4&=\gamma_3^R=(\gamma_3^S)^{-1}=\gamma_2^T(\gamma_3^T)^{-1}(\gamma_1^T)^{-1}=\gamma_2^U(\gamma_3^U)^{-1}.
\end{align}
Writing Eq.~(\ref{ModBC}) as
\begin{align}\label{psigtrans}
\psi_\b{k}^h(\gamma_j^h(z))=e^{ik_j}\psi_\b{k}^h(z),\hspace{5mm}j=1,\ldots,4,
\end{align}
we have, taking $h=R$,
\begin{align}
\psi_\b{k}^R(\gamma_1(z))&=\psi_\b{k}^R((\gamma_4^R)^{-1}(z))=e^{-ik_4}\psi_\b{k}^R(z),\label{BlochR1}\\
\psi_\b{k}^R(\gamma_2(z))&=\psi_\b{k}^R(\gamma_1^R(z))=e^{ik_1}\psi_\b{k}^R(z),\\
\psi_\b{k}^R(\gamma_3(z))&=\psi_\b{k}^R(\gamma_2^R(z))=e^{ik_2}\psi_\b{k}^R(z),\\
\psi_\b{k}^R(\gamma_4(z))&=\psi_\b{k}^R(\gamma_3^R(z))=e^{ik_3}\psi_\b{k}^R(z),\label{BlochR4}
\end{align}
where in Eq.~(\ref{BlochR1}) we have used the relation $\psi_\b{k}^h((\gamma_j^h)^{-1}(z))=e^{-ik_j}\psi_\b{k}^h(z)$, $j=1,\ldots,4$, easily shown by substituting $z\rightarrow(\gamma_j^h)^{-1}(z)$ in Eq.~(\ref{psigtrans}). Thus Eqs.~(\ref{BlochR1}-\ref{BlochR4}) can be written as Eq.~(\ref{BlochRotated}) with
\begin{align}\label{kR}
\b{k}^R=(k_1^R,k_2^R,k_3^R,k_4^R)=(-k_4,k_1,k_2,k_3).
\end{align}
Proceeding similarly for $h=S,T,U$, we find that the $\psi_\b{k}^h$ obey Eq.~(\ref{BlochRotated}) with
\begin{align}
\b{k}^S&=(k_1^S,k_2^S,k_3^S,k_4^S)=(k_2,k_1,-k_4,-k_3),\\
\b{k}^T&=(k_1^T,k_2^T,k_3^T,k_4^T)=(-k_2+k_3-k_4,-k_1+k_3-k_4,-k_1+k_2-k_4,-k_1+k_2-k_3),\\
\b{k}^U&=(k_1^U,k_2^U,k_3^U,k_4^U)=(-k_2,k_1-k_2,k_1-k_3+k_4,k_2-k_3).\label{kU}
\end{align}
The linear relations (\ref{kR}-\ref{kU}) can be written in the matrix form of Eq.~(\ref{kg}),
\begin{align}\label{FourierHyperbolic}
k^h_i=M_{ij}(h)k_j,
\end{align}
with the $4\times 4$ matrices:
\begin{align}\label{GrepJac}
\begin{array}{ll}
M(R)=\left(\begin{array}{cccc}
0 & 0 & 0 & -1 \\
1 & 0 & 0 & 0 \\
0 & 1 & 0 & 0 \\
0 & 0 & 1 & 0
\end{array}\right),&
M(S)=\left(\begin{array}{cccc}
0 & 1 & 0 & 0 \\
1 & 0 & 0 & 0 \\
0 & 0 & 0 & -1 \\
0 & 0 & -1 & 0
\end{array}\right),\\
M(T)=\left(\begin{array}{cccc}
0 & -1 & 1 & -1 \\
-1 & 0 & 1 & -1 \\
-1 & 1 & 0 & -1 \\
-1 & 1 & -1 & 0
\end{array}\right),&
M(U)=\left(\begin{array}{cccc}
0 & -1 & 0 & 0 \\
1 & -1 & 0 & 0 \\
1 & 0 & -1 & 1 \\
0 & 1 & -1 & 0
\end{array}\right).
\end{array}
\end{align}
Equation (\ref{FourierHyperbolic}) can be thought of as the genus-2 analog of the Euclidean relation $k^h_i=h_{ij}k_j$, that follows from straightforward Fourier analysis, where $h\in G\subset O(2)$ is a Euclidean point-group transformation $r_i\rightarrow h_{ij}r_j$ in real space. In fact, we find that the matrices (\ref{GrepJac}) form a linear representation of the automorphism group $G$ of the Bolza surface in four-dimensional hyperbolic $\b{k}$-space, in the sense that those matrices obey the group relations (\ref{rel1}-\ref{rel5}), with $M(h_1h_2)=M(h_1)M(h_2)$ for $h_1,h_2\in G$, and $M(e)$ is the $4\times 4$ identity matrix.

\section{The tight-binding approximation}
\label{sec:TB}

We now discuss how an analog of the tight-binding approximation~[2] can be devised within our general hyperbolic band theory. Consider first the quantum mechanics of an isolated, deep potential well $U(z)$ with compact support in $\c{D}$, e.g., the circular well considered in the main text with $V_0$ large. As with other potentials (e.g., Ref.~[49]), we expect a  number of bound states with discrete eigenenergies $\epsilon_n$ and localized wavefunctions $\phi_n(z)$, orthonormal with respect to the Poincar\'e metric, and satisfying an atomic-like Schr\"odinger equation:
\begin{align}\label{AtomicSchrod}
(-\Delta+U(z))\phi_n(z)=\epsilon_n\phi_n(z),\hspace{5mm}z\in\mathbb{H}.
\end{align}
If the $\phi_n$ are sufficiently localized, they and their $\Gamma$-translates $\phi_n(\gamma(z))$, $\gamma\in\Gamma$, should be good approximations to the true wavefunctions of the full hyperbolic lattice with automorphic potential $V(z)=V(\gamma(z))$ [Eq.~(5) in the main text]. To construct an approximate eigenstate that obeys the automorphic Bloch condition $\psi(\gamma(z))=\chi(\gamma)\psi(z)$ with hyperbolic crystal momentum $\b{k}$, we take an appropriate linear combination of those localized wavefunctions,
\begin{align}\label{LCAO}
\psi_\b{k}(z)\approx\sum_{\gamma\in\Gamma}\sum_n b_{n\b{k}}\chi_\b{k}^*(\gamma)\phi_n(\gamma(z)),
\end{align}
where $n$ ranges over the discrete levels of the atomic problem (\ref{AtomicSchrod}), $b_{n\b{k}}$ is an expansion coefficient, and we explicitly indicated by a subscript the dependence of the Bloch phase factor (\ref{BlochFactorBolza}) on $\b{k}$. Substituting this approximate expansion into the full Schr\"odinger equation $(-\Delta+V)\psi_\b{k}=E(\b{k})\psi_\b{k}$, multiplying from the left on both sides by $\phi_m^*(z)$, and integrating over the entire Poincar\'e disk, we obtain a hyperbolic analog of the standard tight-binding equations~[2]:
\begin{align}
\sum_n\left(\epsilon_m s_{mn}(\b{k})-u_{mn}-t_{mn}(\b{k})\right)b_{n\b{k}}\approx E(\b{k})\sum_n s_{mn}(\b{k})b_{n\b{k}},
\end{align}
a generalized matrix eigenvalue problem whose eigenvalues are approximate hyperbolic band energies $\{E_n(\b{k})\}$ and whose eigenvectors are the expansion coefficients $b_{n\b{k}}$. We define the overlap matrix $s_{mn}(\b{k})$, the on-site potential matrix $u_{mn}$, and the hopping matrix $t_{mn}(\b{k})$ as
\begin{align}
s_{mn}(\b{k})&=\delta_{mn}+\sum_{\gamma\neq e}\chi_\b{k}^*(\gamma)\int_\mathbb{H} d^2z\sqrt{g}\phi_m^*(z)\phi_n(\gamma(z)),\\
u_{mn}&=-\int_\mathbb{H} d^2z\sqrt{g}\phi_m^*(z)\Delta V(z)\phi_n(z),\\
t_{mn}(\b{k})&=-\sum_{\gamma\neq e}\chi_\b{k}^*(\gamma)\int_\mathbb{H} d^2z\sqrt{g}\phi_m^*(z)\Delta V(z)\phi_n(\gamma(z)),
\end{align}
and $\Delta V=\sum_{\gamma\neq e}U(\gamma(z))$ is the sum of $\Gamma$-translates of $U(z)$, excluding the reference cell $\c{D}$.

\section{Hyperbolic Wannier functions and the Bloch transform}

Another key notion of conventional band theory, conceptually related to the tight-binding approximation, is that of Wannier functions~[2]. In the Euclidean context, these are defined as the exact Fourier coefficients $f_n(\b{R},\b{r})$ of a true Bloch eigenstate $\psi_{n\b{k}}(\b{r})$, expanded for each $\b{r}$ as a Fourier series in $\b{k}$, with $\b{R}\in\mathbb{Z}^2$ the sites of the real-space lattice:
\begin{align}\label{WannierEucl}
f_n(\b{r}-\b{R})=\int_{\mathrm{1BZ}}\frac{d^2k}{(2\pi)^2}e^{-i\b{k}\cdot\b{R}}\psi_{n\b{k}}(\b{r}),
\end{align}
where 1BZ denotes the Brillouin zone torus. Since $\psi_{n\b{k}}(\b{r})$ satisfies the Bloch condition, Wannier functions are invariant under simultaneous translations of $\b{r}$ and $\b{R}$ by a given lattice vector $\b{m}\in\mathbb{Z}^2$, and are thus necessarily of the form $f_n(\b{r}-\b{R})$. They can be interpreted as atomic-like wavefunctions associated with site $\b{R}$, analogous to the atomic orbitals. Viewed as a function of $\b{r}\in\mathbb{R}^2$, a Euclidean Wannier function $f_n(\b{r}-\b{R})$ is localized around $\b{R}$, and therefore lives in the space $L^2(\mathbb{R}^2)$ of square-integrable functions over the Euclidean plane $\mathbb{R}^2$. Standard Fourier analysis allows us to invert Eq.~(\ref{WannierEucl}):
\begin{align}\label{EuclBT}
\psi_{n\b{k}}(\b{r})=\sum_{\b{R}}e^{i\b{k}\cdot\b{R}}f_n(\b{r}-\b{R}).
\end{align}
We will refer to Eq.~(\ref{EuclBT}) as the (Euclidean) {\it Bloch transform}, and to Eq.~(\ref{WannierEucl}) as the {\it inverse Bloch transform}.

The tight-binding approximation (\ref{LCAO}) was presented as a means to obtain a delocalized wavefunction $\psi_\b{k}(z)$ obeying the automorphic Bloch condition on $z\in\mathbb{H}$ from a given localized wavefunction $\phi_n(z)$ and its $\Gamma$-translates $\phi_n(\gamma(z))$. However, we may ask whether such a representation can hold {\it exactly}, i.e., whether one can devise a precise hyperbolic analog of the Bloch transform and its inverse. The natural generalization of the Euclidean inverse Bloch transform (\ref{WannierEucl}) would be:
\begin{align}
f_n(\gamma(z))\stackrel{?}{=}\int_{\Jac(\Sigma_g)}\frac{d^{2g}k}{(2\pi)^{2g}}\chi_\b{k}(\gamma)\psi_{n\b{k}}(z),
\end{align}
where we consider the general case of a $\{4g,4g\}$ tessellation with $g\geq 2$. By assumption, $\psi_{n\b{k}}(z)$ obeys the automorphic Bloch condition, thus the putative hyperbolic Wannier function $f_n(\gamma(z))$ depends only on $\gamma(z)$, and not $\gamma$ and $z$ separately. This mirrors the behavior $f_n(\b{R},\b{r})=f_n(\b{r}-\b{R})$ of Euclidean Wannier functions.

However, we find that $f_n(\gamma\gamma^{(1)}(z))=f_n(\gamma(z))$ for any $\gamma^{(1)}\in\Gamma^{(1)}$ where
\begin{align}
\Gamma^{(1)}=[\Gamma,\Gamma]=\langle\gamma_i\gamma_j\gamma^{-1}_i\gamma^{-1}_j\rangle,\hspace{5mm}i,j=1,\ldots,2g,
\end{align}
is the commutator subgroup of $\Gamma\cong\pi_1(\Sigma_g)$, and $\gamma_1,\ldots,\gamma_{2g}$ are the $2g$ generators of $\Gamma$. Loosely speaking, the commutator subgroup of a group measures the extent to which the group is nonabelian (the commutator subgroup of an abelian group is the trivial group with only the identity element). The $\Gamma^{(1)}$-periodicity of $f_n(\gamma(z))$ follows from the fact that $\chi_\b{k}(\gamma\gamma^{(1)})=\chi_\b{k}(\gamma)$, i.e., because $\Gamma^{(1)}$ is in the kernel of the representation $\chi_\b{k}:\Gamma\rightarrow U(1)$ for all $\b{k}\in\Jac(\Sigma_g)$. In other words, the space of functions $\psi_{n\b{k}}$ obeying the automorphic Bloch condition only contains functions that are $\Gamma^{(1)}$-invariant. As a result, any attempt to construct localized Wannier functions $f_n(\gamma(z))$ from such automorphic Bloch functions can only yield functions that are at most $\Gamma^{(1)}$-invariant. Since $\Gamma^{(1)}$ is an infinite group, $\Gamma^{(1)}$-invariant functions cannot be localized and are not square integrable (i.e., they are not part of $L^2(\mathbb{H})$, except the trivial function $f_n=0$). This is problematic for the construction of a naive hyperbolic analog of the Bloch transform (\ref{EuclBT}). Indeed, one would write:
\begin{align}\label{naiveHypBT}
\psi_{n\b{k}}(z)\stackrel{?}{=}\sum_{\gamma\in\Gamma}\chi_\b{k}^*(\gamma)f_n(\gamma(z))
=\left(\mathop{\mathrm{vol}}\Gamma^{(1)}\right)\sum_{[\gamma]\in\Gamma/\Gamma^{(1)}}\chi_\b{k}^*([\gamma])f_n([\gamma](z)),
\end{align}
where $\mathop{\mathrm{vol}}\Gamma^{(1)}$ denotes the volume of the group $\Gamma^{(1)}$, and we have used the fact that $\chi_\b{k}(\gamma)$ and $f_n(\gamma(z))$ are both $\Gamma^{(1)}$-periodic, and thus depend only on the coset $[\gamma]$ of $\Gamma^{(1)}$ in $\Gamma$ to which $\gamma$ belongs (since $\Gamma^{(1)}$ is normal in $\Gamma$, the choice of left or right coset is immaterial). Since $\mathop{\mathrm{vol}}\Gamma^{(1)}$ is infinite, the sum over $\gamma\in\Gamma$ in the naive hyperbolic Bloch transform (\ref{naiveHypBT}) cannot possibly converge.

The appearance of the troublesome infinite factor $\mathop{\mathrm{vol}}\Gamma^{(1)}$ suggests an obvious resolution to the problem, namely that we define hyperbolic Wannier functions as the set $\{f_n([\gamma](z)):[\gamma]\in\Gamma/\Gamma^{(1)}\}$, in which case the Bloch transform becomes:
\begin{align}\label{CosetBT}
\psi_{n\b{k}}(z)=\sum_{[\gamma]\in\Gamma/\Gamma^{(1)}}\chi_{\b{k}}^*([\gamma])f_n([\gamma](z)).
\end{align}
Since $\Gamma/\Gamma^{(1)}$ is nothing but the abelianization of $\Gamma\cong\pi_1(\Sigma_g)$, which is isomorphic to the first homology group $H_1(\Sigma_g,\mathbb{Z})\cong\mathbb{Z}^{2g}$, the sum in Eq.~(\ref{CosetBT}) is in fact over the sites $\b{R}\in\mathbb{Z}^{2g}$ of a $2g$-dimensional Euclidean lattice. We finally recover standard Fourier analysis, albeit in higher dimensions, with the inverse Bloch transform given by:
\begin{align}\label{HypWannierFinal}
f_n([\gamma](z))=\int_{\Jac(\Sigma_g)}\frac{d^{2g}k}{(2\pi)^{2g}}\chi_\b{k}([\gamma])\psi_{n\b{k}}(z).
\end{align}
Indeed, viewed as functions of $\b{k}\in\Jac(\Sigma_g)$, the characters $\chi_\b{k}([\gamma])=e^{i\b{k}\cdot\b{R}}$ are orthonormal:
\begin{align}\label{CharOrthonormal}
\int_{\Jac(\Sigma_g)}\frac{d^{2g}k}{(2\pi)^{2g}}\chi_\b{k}([\gamma])\chi_\b{k}^*([\gamma'])=\delta_{[\gamma],[\gamma']},\hspace{5mm}[\gamma],[\gamma']\in H_1(\Sigma_g,\mathbb{Z})\cong\mathbb{Z}^{2g}.
\end{align}
Note that we also retain the property that the hyperbolic Wannier functions (\ref{HypWannierFinal}) depend only on $[\gamma](z)$, as opposed to depending on $[\gamma]$ and $z$ separately. Viewed from this more abstract point of view, the sum over lattice sites $\b{R}\in\mathbb{Z}^2$ in the Euclidean Bloch transform (\ref{EuclBT}) is in fact over the elements of $H_1(\Sigma_1,\mathbb{Z})\cong\mathbb{Z}^2$, where $\Sigma_1\cong T^2$ is the compactified toroidal unit cell. For a Euclidean lattice, the commutator subgroup is trivial, thus the fundamental group $\pi_1(\Sigma_1)$ and first homology group $H_1(\Sigma_1,\mathbb{Z})$ are isomorphic (and abelian).

When considering the hyperbolic Wannier functions (\ref{HypWannierFinal}) as functions of $z\in\mathbb{H}$, they are still $\Gamma^{(1)}$-periodic, and thus not in $L^2(\mathbb{H})$. However, they become localized on the quotient $\mathbb{H}/\Gamma^{(1)}$, which is an infinite-sheeted abelian cover of $\Sigma_g$. This can be considered as an infinite subset of the original tessellation on which the hyperbolic Bloch states $\psi(\gamma(z))=\chi(\gamma)\psi(z)$ form a complete set; the corresponding Wannier functions $f_n([\gamma](z))$ are then finally in $L^2(\mathbb{H}/\Gamma^{(1)})$. The group of translations on $\mathbb{H}/\Gamma^{(1)}$ is now the abelian group $\Gamma/\Gamma^{(1)}\cong H_1(\Sigma_g,\mathbb{Z})\cong\mathbb{Z}^{2g}$. There exists a family of translation operators $\{T_{[\gamma]}:[\gamma]\in\Gamma/\Gamma^{(1)}\}$ acting on $L^2(\mathbb{H}/\Gamma^{(1)})$ which forms a mutually commuting set that can be simultaneously diagonalized, with eigenvalues $\chi_\b{k}([\gamma])$ corresponding to a complete set of unitary irreducible representations of $\Gamma/\Gamma^{(1)}$.

To complete our discussion of hyperbolic Wannier functions, we show that they obey the expected property~[2] of being orthonormal for different bands $n,n'$ and different homology classes $[\gamma],[\gamma']\in H_1(\Sigma_g,\mathbb{Z})$. The inner product of two hyperbolic Wannier functions on $\mathbb{H}/\Gamma^{(1)}$ is:
\begin{align}\label{WannierOrthog}
\int_{\mathbb{H}/\Gamma^{(1)}}d^2z\sqrt{g}f_n^*([\gamma](z))f_{n'}([\gamma'](z))&=\int_{\Jac(\Sigma_g)}\frac{d^{2g}k}{(2\pi)^{2g}}\int_{\Jac(\Sigma_g)}\frac{d^{2g}k'}{(2\pi)^{2g}}
\chi_\b{k}^*([\gamma])\chi_{\b{k}'}([\gamma'])\nn\\
&\hspace{20mm}\times\int_{\mathbb{H}/\Gamma^{(1)}}d^2z\sqrt{g}\psi_{n\b{k}}^*(z)\psi_{n'\b{k}'}(z).
\end{align}
Here $\sqrt{g}$ denotes the square root of the determinant of the Poincar\'e metric (not to be confused with the genus). For $\b{k}\neq\b{k}'$, the hyperbolic Bloch eigenstates $\psi_{n\b{k}}(z)$ and $\psi_{n'\b{k}'}(z)$, as constructed in our work, obey different boundary conditions and thus formally live in different Hilbert spaces; we cannot directly invoke their orthogonality. To circumvent this problem, we use the invariance of the integration measure $d^2z\sqrt{g}$ under M\"obius transformations and the fact that $\c{D}$ is a fundamental region for $\Gamma$ to write
\begin{align}\label{properlydiscontinuous}
\mathbb{H}/\Gamma^{(1)}=\bigsqcup_{[\gamma]\in\Gamma/\Gamma^{(1)}}[\gamma](\c{D}),
\end{align}
where $\sqcup$ denotes disjoint union, and thus
\begin{align}\label{BlochOrthog}
\int_{\mathbb{H}/\Gamma^{(1)}}d^2z\sqrt{g}\psi_{n\b{k}}^*(z)\psi_{n'\b{k}'}(z)&=\sum_{[\gamma]\in\Gamma/\Gamma^{(1)}}\int_\c{D}d^2z\sqrt{g}\psi_{n\b{k}}^*([\gamma](z))\psi_{n'\b{k}'}([\gamma](z))\nn\\
&=\sum_{[\gamma]\in\Gamma/\Gamma^{(1)}}\chi_\b{k}^*([\gamma])\chi_{\b{k}'}([\gamma])\int_\c{D}d^2z\sqrt{g}\psi_{n\b{k}}^*(z)\psi_{n'\b{k}'}(z),
\end{align}
using the automorphic Bloch condition. Equation~(\ref{properlydiscontinuous}) follows from the fact that the Fuchsian group $\Gamma$ acts properly discontinuously on $\mathbb{H}$~[17], thus $\c{D}\cap\gamma(\c{D})=0$ for any $\gamma\in\Gamma\backslash\{e\}$. The Bloch characters $\chi_\b{k}([\gamma])=e^{i\b{k}\cdot\b{R}}$, $\b{R}\in\Gamma/\Gamma^{(1)}\cong\mathbb{Z}^{2g}$ obey a Schur-type orthogonality relation familiar from Fourier analysis:
\begin{align}
\sum_{[\gamma]\in\Gamma/\Gamma^{(1)}}\chi_\b{k}^*([\gamma])\chi_{\b{k}'}([\gamma])=\sum_{\b{R}\in\mathbb{Z}^{2g}}e^{-i(\b{k}-\b{k}')\cdot\b{R}}=(2\pi)^{2g}\delta(\b{k}-\b{k}').
\end{align}
Substituting into Eq.~(\ref{BlochOrthog}), we have
\begin{align}
\int_{\mathbb{H}/\Gamma^{(1)}}d^2z\sqrt{g}\psi_{n\b{k}}^*(z)\psi_{n'\b{k}'}(z)
=(2\pi)^{2g}\delta(\b{k}-\b{k}')\int_\c{D}d^2z\sqrt{g}\psi_{n\b{k}}^*(z)\psi_{n'\b{k}}(z)
=(2\pi)^{2g}\delta(\b{k}-\b{k}')\delta_{nn'},
\end{align}
using the orthogonality of hyperbolic Bloch eigenstates on $\c{D}$ with the same $\b{k}$. Substituting in Eq.~(\ref{WannierOrthog}), we finally obtain:
\begin{align}
\int_{\mathbb{H}/\Gamma^{(1)}}d^2z\sqrt{g}f_n^*([\gamma](z))f_{n'}([\gamma'](z))
=\delta_{nn'}\int_{\Jac(\Sigma_g)}\frac{d^{2g}k}{(2\pi)^{2g}}\chi_\b{k}^*([\gamma])\chi_\b{k}([\gamma'])
=\delta_{nn'}\delta_{[\gamma],[\gamma']},
\end{align}
making use of Eq.~(\ref{CharOrthonormal}) in the last equality. Thus the hyperbolic Wannier functions behave as Euclidean Wannier functions in $2g$ dimensions.